\documentclass{article}
\usepackage{geometry}
\usepackage{graphicx}
\usepackage{cite}
\usepackage{epsfig}
\usepackage{amsmath}
\usepackage{amsthm}
\usepackage{amssymb}
\usepackage{mathrsfs}
\usepackage{float}
\usepackage{multirow}
\usepackage{verbatim}
\usepackage{fancyhdr}
\usepackage{subcaption}
\usepackage{hyperref}
\usepackage{lineno}
\usepackage{float}
\usepackage{algorithm}
\usepackage[noend]{algorithmic}

\usepackage{color}
\usepackage{xcolor}
\graphicspath{{./}}
\usepackage{mathtools}
\usepackage[normalem]{ulem}
\theoremstyle{plain} 
\newtheorem{theorem}{Theorem}[section]
\theoremstyle{definition}
\newtheorem{remark}{\textbf{Remark}}
\usepackage{threeparttable}
\usepackage{dcolumn}
\usepackage{booktabs}
\usepackage{indentfirst}
\usepackage{setspace}
\usepackage{bm}
\usepackage{enumerate}
\usepackage{diagbox}
\usepackage{authblk}

\newcommand\keywords[1]{\textbf{Keywords}: #1}
\newcounter{example}[section]         % reset every section
\renewcommand{\theexample}{\thesection.\arabic{example}} 
\newenvironment{example}[1][]{
    \refstepcounter{example}%
    \par\noindent\textbf{Example \theexample.} #1 \par
}{\par}
\title{A Morphology-Adaptive Random Feature Method for Inverse Source Problem of the Helmholtz Equation}

\author[a,b]{Xinwei Hu\thanks{huxinwei@mail.ustc.edu.cn}}
\author[a,b,c]{Jingrun Chen\thanks{jingrunchen@ustc.edu.cn}}
\author[d,e]{Haijun Yu\thanks{hyu@lsec.cc.ac.cn(corresponding author)}}
\affil[a]{School of Mathematical Sciences, University of Science and Technology of China, Hefei, China}
\affil[b]{Suzhou Institute for Advanced Research, University of Science and Technology of China, Suzhou, China}
\affil[c]{Suzhou Big Data \& AI Research and Engineering Center, Suzhou, China}
\affil[d]{State Key Laboratory of Mathematical Sciences (SKLMS) \& LSEC, Institute of Computational Mathematics and Scientific/Engineering
Computing, Academy of Mathematics and Systems Science, Chinese Academy of Sciences,
Beijing 100190, China}
\affil[e]{School of Mathematical Sciences, University of Chinese Academy of Sciences, Beijing 100049, China}

\date{} % 去掉日期

\begin{document}
%\linenumbers
\maketitle
\begin{abstract}
The inverse source problem for the Helmholtz equation poses significant challenges, particularly when sources exhibit complex or discontinuous geometries. Traditional numerical methods suffer from prohibitive computational costs, while machine learning-based approaches such as physics-informed neural networks (PINN) and the random feature method (RFM), though computationally efficient for inverse problems, lack the intrinsic machinery to handle the sharp morphological features in such singular problems, leading to inaccurate solutions. To address this issue, we propose the morphology-adaptive random feature method (MA-RFM), a novel two-stage framework designed to adaptively identify critical regions and incorporate morphology-aware activation functions to tackle the multi-frequency inverse source problem with complex geometry. Our framework recasts the ill-posed inverse problem into a well-posed, strictly convex optimization problem by reformulating the governing Helmholtz equation as a Tikhonov-regularized integral equation via its fundamental solution. In the first stage, the integral adaptive RFM (IA-RFM) employs an adaptive algorithm to rapidly localize the source support, thereby reducing computational overhead and accelerating convergence.
 In the second stage, posterior geometric information is progressively integrated into the solver via hybrid basis functions, enabling a precise reconstruction of complex morphologies. The MA-RFM  extends the capabilities of RFM to handle partial differential equations with singular solutions while preserving its mesh-free efficiency.  We demonstrate the superior performance of our approach on ample challenging 2-D and 3-D benchmark problems, including scenarios with limited and noisy measurements, highlighting its robustness and accuracy in reconstructing intricate and discontinuous sources.
\end{abstract}

\keywords{Helmholtz equation, inverse source problem, random feature method, morphology-adaptive basis, iterative adaptive integral mesh, multi-frequency, physics-informed neural networks.}

\section{Introduction}
The inverse source problem of identifying an unknown scalar source in the Helmholtz equation arises in applications such as medical imaging,  antenna synthesis, acoustic tomography, and pollution of the environment \cite{albanese2006medical1,arridge1999medical2,fokas2004medical3,anastasio2007application,devaney2007inverse,ramm1999multidimensional,stefanov2009thermoacoustic,bojarski1973inverse,lin2011helmholtz}.
In this work,  we consider the Helmholtz equation in a homogeneous medium. Let $\Omega$ be a bounded domain in $\mathbb{R}^d$,  with boundary $\Gamma$.  {Assume that the source function $S(\boldsymbol{x})$ is compactly supported with support volume $\tau \subset  \Omega \subset \mathbb{R}^d$ satisfying $\mathrm{dist}(\tau, \Gamma): =\operatorname*{min}\{|\boldsymbol{x}-\boldsymbol{y}|{: } \ \boldsymbol{x}\in\tau, \boldsymbol{y}\in\Gamma\}>0$}. For a radial frequency $\omega$,  the radiating field $u(\boldsymbol{x})$ generated by the source $S(\boldsymbol{x})$  satisfies the following Helmholtz equation with the Sommerfeld radiation condition:
\begin{align}\label{helmholtz}
\left \{
\begin{aligned}
&-\Delta u-k^2u=S,  \quad \text{in} \  \mathbb{R}^d, \\
&\lim_{r\to\infty}r^{\frac{d-1}{2}}\bigl({\partial_r u} -\mathrm{i}ku\bigr)=0, \quad r=|\boldsymbol{x}|, 
\end{aligned}
\right.
\end{align}
where $k={\omega}/{c_0}$ is the wave number,  $c_0$ is the speed of sound.  The Sommerfeld radiation condition ensures that the solution of the forward problem (that is, given $S$, solve \eqref{helmholtz} for $u$) is unique. In the inverse problem, we need to determine $S$ using only the (full or partial) observation of $u$ on the boundary $\Gamma$. 

A classical approach to solving this inverse problem is to formulate it mathematically as an integral equation,  which was originally proposed by Porter \cite{Imageformation, diffraction} and later independently derived in \cite{bojarski1973inverse}. By Green's formula and the radiation condition,  the analytic solution to (\ref{helmholtz}) is given by
\begin{align}\label{analytic}
u(\boldsymbol{x})&=\int_\tau\Phi_k(\boldsymbol{x}, \boldsymbol{y})S(\boldsymbol{y})\mathrm{d}\boldsymbol{y}, \\
\label{Fundamental solution}
\Phi_k(\boldsymbol{x}, \boldsymbol{y})&=\left \{
\begin{aligned}
&\frac{i}{4}H_{0}^{1}(k|\boldsymbol{x}-\boldsymbol{y}|),   &d=2, \\
&\frac{e^{ik|\boldsymbol{x}-\boldsymbol{y}|}}{4\pi|\boldsymbol{x}-\boldsymbol{y}|},  &d=3, \\
&\frac{i}{4}\Bigl(\frac{k}{2\pi|\boldsymbol{x}-\boldsymbol{y}|}\Bigr)^{\frac{d}{2}-1}H_{\frac{d}{2}-1}^{(1)}(k|\boldsymbol{x}-\boldsymbol{y}|), &d\ge4,
\end{aligned}
\right.
\end{align}
where $\Phi_k(\boldsymbol{x}, \boldsymbol{y})$ is the fundamental solution for the Helmholtz equation, and the function $H_0^1$ denotes the Hankel function of the first kind of order zero. 

Bleistein and Cohen \cite{bleistein1977nonuniqueness} investigated the properties of this integral equation at a fixed frequency and demonstrated that its solution is non-unique. Similarly, Isakov established a conditional uniqueness theorem \cite{isakov1990inverse} for the spatial dependence of the source function. In \cite{bao2010multi},  Bao proved that the inverse problem admits a unique solution when multi-frequency data are available. Furthermore,
under suitable regularity assumptions on $S$, stability increases with higher $k_{\text{max}}$,
and the H{\"o}lder type stability estimate can be obtained if $k_{\text{max}}$ is sufficiently large
compared to the size of $\tau$. Traditional solution methods for solving multi-frequency inverse source problems can generally be categorized into iterative and non-iterative approaches. In \cite{bao2011numerical}, Bao, Lin, and Triki proposed a continuation method along the wavenumber,  applying Landweber iteration from low to high frequencies. In contrast,  non-iterative methods avoid the time-consuming iterative process by expanding the source function in terms of specific basis functions and directly solving for the expansion coefficients. Eller and Valdivia proposed a direct method based on eigenfunction expansion of the Laplace operator \cite{eller2009acoustic}, utilizing multi-frequency data corresponding to eigenvalues to recover the expansion coefficients.  In \cite{zhang2015fourier},  Zhang and Guo proposed a Fourier expansion method to compute the source coefficients from data at prescribed wavenumbers. However,  the computational cost is substantial because these methods require discretizing the integral equation on a fine,  uniform grid. Furthermore, their effectiveness is limited when dealing with complex problems.

In recent years, with the rapid development of deep learning, employing neural network models to solve partial differential equations (PDEs) has become an increasingly active research field. Such approaches, such as physics-informed neural networks (PINN~\cite{raissi2019physics}), deep Galerkin method (DGM~\cite{sirignano2018dgm}), deep Ritz method (DRM~\cite{ew2018ritz}), and their variants (see e.g. \cite{jagtap2020extended,kharazmi2021hp, LiaoMing2021DeepNitsche, YuZhang2025NaturalDeep}), need to approximate both $u(\boldsymbol{x})$ and $S(\boldsymbol{x})$, which
face several challenges when applied to inverse source problems: 
\begin{enumerate}
\item \emph{Representational limitations of neural networks:} Standard feedforward neural networks \cite{sheng2024MC-fPINNs} exhibit a well-documented ``spectral bias'', an inherent tendency to learn low-frequency functions more readily than high-frequency ones~\cite{rahaman2019spectral, xu2019frequency}. This makes it difficult to approximate the highly oscillatory solutions $u(\boldsymbol{x})$ of the Helmholtz equation, especially for large wavenumbers. Furthermore, since automatic differentiation computes the required high-order derivatives, it drastically magnifies any initial representation errors in $u(\boldsymbol{x})$. Accurately capturing such oscillatory behavior would necessitate a network with a vast number of parameters, leading to prohibitive computational costs and often yielding poor performance. 
 
\item \emph{Insufficient utilization of physical laws:} These conventional frameworks are difficult to balance the governing PDEs and the Sommerfeld condition, which is considered as soft constraints via penalty terms in the loss function. This represents a superficial use of the underlying physics and does not incorporate the problem's intrinsic mathematical structure. Potent physical priors, such as the fundamental solution, which analytically describes the relationship between the source and the field, are entirely neglected.
\end{enumerate}

To overcome the above limitations, new methodologies have emerged. The boundary integral neural network (BI-Net) \cite{lin2023bi} draws on classical potential theory,  utilizes the fundamental solution of the PDE as a kernel to transform the problem into an integral equation, and then approximates the density function therein with a neural network. Recent advances in solving PDEs have been driven by randomized neural networks, notably the extreme learning machine (ELM)~\cite{dong2021local,wang2024extreme}  and the random feature method (RFM)~\cite{chen2022bridging}. The ELM architecture, which consists of a single hidden layer with randomly assigned and fixed parameters, offers remarkable computational efficiency. By determining the output weights through a simple least-squares fit, it bypasses the costly iterative training required by conventional deep neural networks.  This potential has been further realized in variants such as the physics-informed ELM (PIELM)~\cite{chen2022bridging,dwivedi2020physics}. Building upon this foundation, the RFM proposed by Chen et al. enhances the approach by integrating random feature functions with a partition of unity (PoU)~\cite{chen2022bridging} framework, showing spectral accuracy in many areas with complex domains,  such as interface problems\cite{chi2024random}. This strategy effectively recasts the non-convex optimization problem inherent in many neural network methods into a convex one, thereby ensuring a unique solution and efficient convergence. However, in regions with high gradients or singularities, the fixed random basis functions of RFM may lack the local expressive power necessary to capture the solution accurately. An adaptive feature capture method based on RFM was recently proposed in~\cite{deng2025adaptive}, whose main idea is similar to $r$-refinement. It is able to adaptively adjust the basis functions as well as the collocation points at high gradients to enhance the expressive ability of neural networks. Traditional numerical methods, such as $h$-refinement, $p$-refinement, $hp$-refinement, and $r$-refinement \cite{gui1986h1,gui1986h2,babuska1981p,babuvska1990p,mclean2000elliptic,zegeling1998r}, dynamically concentrate on computational points in critical areas based on adjusting basis functions. Consequently, transferring such powerful adaptive strategies into the context of neural network-based solvers is logical and highly desirable.

Inspired by the above work, we develop in this paper a new framework to solve the inverse source problem of the Helmholtz equation. To this end, we propose a morphology-adaptive random feature method (MA-RFM) and combine it with the idea of the integral equation of BI-Net to obtain efficient approximations. Specifically, we first utilize the fundamental solution of the Helmholtz equation as an integral kernel to transform the original differential equation problem into an integral equation. Subsequently, we employ MA-RFM, a two-phase framework that adaptively locates critical regions and adds morphology basis functions. In its first stage, MA-RFM leverages norms of both the gradient and value of the source function to iteratively update the integral mesh. We call this process integral adaptive RFM (IA-RFM). Building upon this, the second stage expands the approximation space by increasing the basis functions of the corresponding morphology to capture local information, based on posterior information obtained from the rough solution and its gradient.
The innovation of this paper is mainly: 
\begin{enumerate}
\item \emph{Physics-informed integral equation framework}:  The governing differential equation is reformulated into an integral equation with the fundamental solution as its kernel. By replacing high-order differentiation with convolution, this framework embeds physical laws into the model, which enforce the Sommerfeld radiation condition rigorously as a hard constraint and thereby improve numerical stability and robustness.
\item \emph{Dual-criterion iterative adaptive integration strategy}: The RFM with adaptive integration is proposed, driven by a dual criterion: the gradient and value of the solution in the last iteration, which accurately identifies and densely samples critical regions,  significantly reducing computational complexity while maintaining accuracy.
\item \emph{Morphology hybrid-basis enhancement}:  A hybrid-basis enhancement method is designed to achieve high-fidelity reconstruction. This framework introduces auxiliary basis functions capable of capturing local morphological features of the source term, effectively compensating for the deficiency of a global basis in representing intricate details.
\end{enumerate}
The remainder of this paper is structured as follows. Section 2 introduces the multi-frequency inverse source problem, covering the uniqueness theory. Section 3 demonstrates the framework of the MA-RFM, including the IA-RFM, the morphology-adaptive hybrid basis enhancement, and the uniqueness and stability of the Tikhonov regularization solution. Section 4 demonstrates the effectiveness of the proposed method with reconstruction results for challenging 2-D and 3-D inverse source problems, including those with complex geometries and limited aperture data. Section 5 concludes with remarks and future directions.

\section{Mathematical results on multi-frequency inverse source problems}
 {In this section, we present the mathematical foundations of the multi-frequency inverse source problem. We firstly establish the uniqueness results for the continuous problem under certain assumptions. Subsequently, we formulate the corresponding discrete linear system and analyze the stability with the Tikhonov regularization, which provides the theoretical basis for the numerical method proposed in Section 3.}
\subsection{Uniqueness for the continuous inverse source problem}
In this section, we define the multi-frequency inverse source problem for the Helmholtz equation and transform the original differential equations into integral equations by using its fundamental solution. This framework allows the Sommerfeld radiation condition to be used as a hard constraint. Firstly, we introduce the uniqueness theorem for the multi-frequency inverse source problem.

Motivated by uniqueness and stability results \cite{eller2009acoustic,bao2010multi,bleistein1977nonuniqueness}, 
we present a uniqueness result for the multi-frequency inverse source problem under weaker assumptions. It extends the classical result of Bao et al. \cite{bao2010multi},  generalizing the geometry of the domain $\Omega$, and reducing the observation data from the full boundary $\Gamma$ to an arbitrarily small open subset $\Gamma_0$. 
\begin{theorem}
\label{the:Uniqueness of multi-frequency data}
Consider the inverse source problem  {corresponding to the Helmholtz equation \eqref{helmholtz}, i.e. determine $S$ using full or partial observations of $u$ on the boundary $\Gamma$}. Suppose that $S_1$ and $S_2$ be two sources with compact supports $\tau_1,  \tau_2 \ \text{in}\  \Omega$. Let $\Gamma_0$ be an open subset of the boundary $\Gamma = \partial\Omega$. Suppose that  {equation \eqref{helmholtz} holds} for a set of wave numbers $\{k_j\}_{j=1}^{\infty}$ having an accumulation point  {for the source and solution pairs $\{ S_1, u_1(k_j, x)\}$ and $\{ S_2, u_2(k_j, x) \}$}. Then $S_1=S_2$ if the corresponding radiating fields $u_1(k_j,  \boldsymbol{x})$ and $u_2(k_j,  \boldsymbol{x})$ satisfy one of following conditions: 
\begin{align}
 &(a) \ u_1=u_2,  \text{on} \ \Gamma,   \\
 &(b) \ \partial_{\nu} u_1 =\partial_{\nu}u_2,  \ \text{on}\  \Gamma,  \\
&(c) \ u_1=u_2, \quad \text{on} \  \Gamma_{D},  \quad \partial_{\nu} u_1 =\partial_{\nu}u_2,  \ \text{on} \ \Gamma_N,  \  \Gamma=\Gamma_D\cup\Gamma_N, \\
 &(d) \  u_1=u_2,  \quad \partial_{\nu} u_1 =\partial_{\nu}u_2,  \ \text{on} \ \Gamma_0.
\end{align}
\end{theorem}

The proof of Theorem \ref{the:Uniqueness of multi-frequency data} is listed in Appendix A.
\begin{remark}
\label{rem:uniqueness}
 {The inverse source problem is fundamentally a linear inverse problem subject to a compact support constraint. At a single frequency, this linear problem is intrinsically ill-posed due to the existence of non-radiating sources. It is the combination of the compact support constraint and multi-frequency data that eliminates the non-radiating sources. Consequently, Theorem 2.1 establishes uniqueness for multi-frequency problems with Dirichlet, Neumann, mixed boundary, and also the finite aperture data, which provides
a theoretical basis for numerical algorithms.}
\end{remark}

The unbounded domain presents challenges for both traditional numerical methods and neural network methods. A practical exercise is to use an artificially truncated bounded domain together with a proper boundary condition.
To inscribe Sommerfeld conditions on bounded regions, we can make use of the Dirichlet-to-Neumann (DtN) mapping relation. To do this, construct an artificial truncated domain $\Omega_{\rho}$ to be a ball of radius $\rho$, with $\rho$ large enough such that $\Omega \subset \Omega_{\rho}$. Thus, the artificial boundary $\Gamma_{\rho}=\partial \Omega_\rho$ divides $\mathbb{R}^d$ into two parts.
In the outer region $\mathbb{R}^d\setminus\overline{\Omega}_\rho$, there is the Helmholtz outer problem:
\begin{align}\label{out_helmholtz}
\left \{
\begin{aligned}
&-\Delta u-k_j^2u=0,  &\text{in} \  \mathbb{R}^d\setminus\overline{\Omega}_\rho, \\
&\lim_{r\to\infty}r^{\frac{d-1}{2}}\left({\partial_r u}-\mathrm{i}k_ju\right)=0, &\quad r=|\boldsymbol{x}|, \\
&u=g, &\text{on} \ \Gamma_{\rho}.
\end{aligned}
\right.
\end{align}
For this external problem, $u$ can be uniquely determined by the Dirichlet data on its boundary $\Gamma_{\rho}$~\cite{ShenWang2007AnalysisSpectralGalerkin}. 
The DtN mapping $\mathcal{T}$ describes exactly this relationship and maps the Dirichlet data on the boundary $\Gamma_{\rho}$ 
to its corresponding Neumann data $\partial _{\nu}u|_{\Gamma_{\rho}}$. Specifically, the DtN operator $\mathcal{T}$ is defined as follows.
\begin{align}
    \label{DtN}
\mathcal{T}:u|_{\Gamma_\rho}\mapsto\partial_\nu u|_{\Gamma_\rho}.
\end{align}
This operator $\mathcal{T}$ contains exactly the information about the radiation conditions at infinity and can be efficiently calculated (see, e.g.,~\cite{ShenWang2007AnalysisSpectralGalerkin}). By imposing this DtN boundary condition on the artificial boundary $\Gamma_{\rho}$, we can transform the original problem (\ref{helmholtz}) in the unbounded domain into an equivalent problem within the bounded region $\Omega_{\rho}$:
\begin{align}\label{helmholtz-DtN}
\left \{
\begin{aligned}
&-\Delta u-k_{j}^2u=S,  &\text{in} \  \Omega_\rho, \\
& \partial_{\nu} u = \mathcal{T}(u), & \text{on} \ \Gamma_{\rho},\\
& u=u(\boldsymbol{x},k_j), \quad \boldsymbol{x}\in\Gamma_D, \qquad \partial_{\nu} u=\partial_{\nu}u(\boldsymbol{x},k_j), &\boldsymbol{x}\in\Gamma_N.
\end{aligned}
\right.
\end{align}
The above formulation is commonly used in traditional numerical methods. However, it is not efficient in designing neural network methods.
Next, we induce the integral equation framework so that the Sommerfeld condition is naturally satisfied. Based on the fundamental solution of the Helmholtz equations (\ref{Fundamental solution}),  (\ref{helmholtz}) or \eqref{helmholtz-DtN} can be transformed into an integral equation (\ref{analytic}). For a fixed wavenumber $k$,  the radiation operators \( L_k^{(1)} \) and \( L_k^{(2)} \),  mapping from \( L^2(\tau) \) to \( L^2(\Gamma) \),  are defined as follows: 
\begin{align}\label{L1}
L_k^{(1)}(S)(\boldsymbol{x}) &= \int_{\tau} \Phi_k(|\boldsymbol{x} - \boldsymbol{y}|) S(\boldsymbol{y}) \,  \mathrm{d}\boldsymbol{y},  \\
\label{L2}
L_k^{(2)}(S)(\boldsymbol{x}) &= \int_{\tau} {\partial_{\nu(\boldsymbol{x})} \Phi_k(|\boldsymbol{x} -\boldsymbol{y}|)} S(\boldsymbol{y}) \,  \mathrm{d}\boldsymbol{y},  %\quad \boldsymbol{x} \in \Gamma^*, 
\end{align}
where \( \nu(\boldsymbol{x}) \) represents the normal derivative. 
Denote $L_k=(L_k^{(1)}, L_k^{(2)})^\top$,  $g(\boldsymbol{x}, k)=(u(\boldsymbol{x}, k), \partial_{\nu}u(\boldsymbol{x}, k))^\top$,  $\boldsymbol{x} \in \Gamma^*$,  $\Gamma^*\subseteq\Gamma$. Consequently, the inverse source problem can be defined as minimizing the following  objective functional $J(S)$ to reconstruct the optimal $S^*$ by measurement $\{g(\boldsymbol{x},k)\}_{k=k_{\text{min}}}^{k_{\text{max}}}$: 
\begin{align}
    \label{continue_ISP}
    \min_{S} J(S) : = \int_{k_{\text{min}}}^{k_{\text{max}}} \| L_k(S) - g(\cdot,  k) \|_{L^2(\Gamma^*)}^2 \ \mathrm{d}k.
\end{align}

\subsection{ {Discretization and Tikhonov regularization}}

 {In this section, we discuss the discretization of the continuous inverse problem and the stability analysis with the Tikhonov regularization method.}

 {
Define a discrete set of wavenumbers $k_{\text{min}}=k_1 < k_2 < \cdots < k_{N-1} < k_N=k_{\text{max}}$, and denote $U_k$ and ${\partial_\nu U_k}$ as the observational Dirichlet and Neumann data on $\Gamma^*$ for the $k$-th frequency. For the linear representation of the source function
\begin{align}
\label{linear}
    S_{M}(\boldsymbol{x})=\sum_{m=1}^{M} s_{m}\phi_{m}(\boldsymbol{x}),
\end{align}
denote $\boldsymbol{s} = (s_1, s_2, \dots, s_M)^\top \in \mathbb{R}^M$ as the vector of unknown expansion coefficients or trainable weights in the neural network, representing the $M$ degrees of freedom in the source approximation. Define $\Psi_k^{(1)}$ and $\Psi_k^{(2)}$ as the discretization matrices corresponding to operators $L_1^{(k)}$ and $L_2^{(k)}$ acting on the basis functions. Consequently, the discrete form of the loss function in (\ref{continue_ISP}) is given by
\begin{align}
    \label{loss_data}
    \mathcal{L}_{\text{data}}(\boldsymbol{s})&=\sum_{k=k_1}^{k_N} \Bigl(\|\Psi_k^{(1)}\cdot \boldsymbol{s} - U_k\|_2^2+\bigl\|\Psi_k^{(2)}\cdot \boldsymbol{s}  - {\partial_\nu U_k} \bigr\|_2^2\Bigr).
\end{align}
}
To reformulate this summation into a compact linear system, we aggregate the data and operators across all frequencies.  Define the global observation vectors $U_D, U_N$ and the global operator matrices $\Psi_D, \Psi_N$ as follows:
\[
U_D = \begin{bmatrix}
U_{k_1}\\
\vdots \\
U_{k_N}
\end{bmatrix},
\quad
U_N = \begin{bmatrix}
\frac{\partial U_{k_1}}{\partial \nu}\ \\
\vdots \\
\frac{\partial U_{k_N}}{\partial \nu}
\end{bmatrix},
\quad
\Psi_D= \begin{bmatrix}
\Psi_{k_1}^{(1)}\\
\vdots \\
\Psi_{k_N}^{(1)}
\end{bmatrix},
\quad
\Psi_N= \begin{bmatrix}
\Psi_{k_1}^{(2)}\\
\vdots \\
\Psi_{k_N}^{(2)}
\end{bmatrix}.
\]
We construct the final real-valued system by separating the real and imaginary parts. Define the global data vector $U$ and the system matrix $\Psi_M$ as:
\[
U = \begin{bmatrix}
\text{Re}(U_D)\\
\text{Im}(U_D)\\
\text{Re}(U_N)\\
\text{Im}(U_N)\\
\end{bmatrix}\in \mathbb{R}^{n\times 1}, \quad
\Psi_M= \begin{bmatrix}
\text{Re}(\Psi_D)\\
\text{Im}(\Psi_D)\\
\text{Re}(\Psi_N)\\
\text{Im}(\Psi_N)\\
\end{bmatrix} \in \mathbb{R}^{n\times M}.
\]
Thus, the discrete inverse problem can be equivalently expressed as the linear system:
\begin{align}\label{Ax=b}
    \Psi_M \boldsymbol{s} = U.
\end{align}
Typically,  such problems are ill-posed, as $\Psi_\text{M}$ has a large condition number,  making the solution highly sensitive to noise in the data. Assume that the measured data $U^{\delta}$ relates to the true,  noise-free data as 
$(U^{\delta})_i=(1+\delta \epsilon)\cdot(U_{\text{true}})_i$, $\epsilon\sim \text{Uniform}(-1,1)$, or $U^{\delta} = U_{\text{true}} + e$,  where $e$ represents measurement noise. A direct inversion, as shown through singular value decomposition (SVD) $\Psi_\text{M} = Q_1\Sigma Q_2^\top$, where $\Sigma = \text{diag}(\sigma_1, \ldots, \sigma_M)$, 
$Q_1=(\hat{u}_1, \hat{u}_2, \ldots, \hat{u}_n)$, $Q_2=(\hat{v}_1, \hat{v}_2, \ldots, \hat{v}_M)$,
would yield a solution: 
\begin{align*}
   \boldsymbol{s}^{\delta}=\sum_{i=1}^M\frac{\hat{u}_i^* U^{\delta}}{\sigma_i}\hat{v}_i=\sum_{i=1}^M\left(\frac{\hat{u}_i^* U_{\text{true}}}{\sigma_i}+\frac{\hat{u}_i^* e}{\sigma_i}\right)\hat{v}_i. 
\end{align*}
The term $\hat{u}_i^* e / \sigma_i$ demonstrates that small singular values $\sigma_i$ can drastically amplify the noise component, corrupting the solution. To counteract this effect,  we employ Tikhonov regularization,  which seeks to find a solution by minimizing a composite objective function: 
\begin{equation}\label{eq:T-reg-loss}
  \mathcal{L}_{\text{reg}}(\boldsymbol{s}^{\delta})= \|\Psi_\text{M} \boldsymbol{s}^{\delta} - U^{\delta}\|_2^2 + \lambda_{\text{reg}}^2 \|\boldsymbol{s}^{\delta}\|_2^2.  
\end{equation}
The regularization parameter $\lambda_{\text{reg}} > 0$ balances the trade-off between fitting the data and controlling the norm of the solution. The solution becomes: 
\begin{align}\label{T-SVD}
\boldsymbol{s}^{\delta}=\sum_{i=1}^M f_i \left(\frac{\hat{u}_i^* U_{\text{true}}}{\sigma_i}+\frac{\hat{u}_i^* e}{\sigma_i}\right)\hat{v}_i,  \quad  \quad f_i = \frac{\sigma_i^2}{\sigma_i^2+\lambda_{\text{reg}}^2}.
\end{align}
The filter factors $f_i$ suppress the influence of noise associated with small singular values ($\sigma_i \ll \lambda_{\text{reg}}$),  thus stabilizing the solution. The choice of $\lambda_{\text{reg}}$ is critical: a value too large introduces excessive bias,  while a value too small fails to adequately suppress noise. The following theorem provides the uniqueness and stability of the Tikhonov regularization solution.
\begin{theorem}[Uniqueness and stability of Tikhonov regularization solution]
\label{Existence and Uniqueness}
\leavevmode\\[-8pt] % 在换行的同时，减少了6pt的垂直间距
\begin{enumerate}[(a)]
    \item \label{thm:Tikhonov_uniqueness} For any regularization parameter $\lambda_{\text{reg}}>0$, objective function $\mathcal{L}_{\text{reg}}(\boldsymbol{s}^{\delta})= \|\Psi_\text{M} \boldsymbol{s}^{\delta} - U^{\delta}\|_2^2 + \lambda_{\text{reg}}^2 \|\boldsymbol{s}^{\delta}\|_2^2$, is strictly convex. Therefore, the minimization problem has a unique solution.
    \item \label{thm:Tikhonov_stability} Let $S_M^*(\boldsymbol{x}) = \sum s_m^* \phi_m(\boldsymbol{x})$ be the best approximation of the true source in the finite space ${\rm{span}}\{\phi_m(\boldsymbol{x}), m=1,\ldots, M\}$,  with coefficient vector $\boldsymbol{s}^{*}$. Let $S_M^{\delta}$ be the regularized solution obtained from data $U^{\delta}$ with a coefficient vector $\boldsymbol{s}^{\delta} = (\Psi_\text{M}^\top\Psi_\text{M}+\lambda_{\text{reg}}^2 I)^{-1}\Psi_\text{M}^{T}U^{\delta}$. Assume the noise level $\|U_{\text{true}} - U^{\delta}\|_2 \le \delta \|U_{\text{true}}\|_2 \le \delta_{\text{all}}$ and the model inconsistency is $\eta_\text{M} = \|\Psi_\text{M} \boldsymbol{s}^{*} - U_{\text{true}}\|_2$. If the source condition $\boldsymbol{s}^{*} = (\Psi_\text{M}^\top \Psi_\text{M})^\nu w$ holds for some vector $w$,  $0<\nu\le1$,  then the error is bounded by
\begin{equation}
    \|\boldsymbol{s}^{\delta}-\boldsymbol{s}^{*}\|_2 \le \frac{1}{2 \lambda_{\text{reg}}}(\delta_{\text{all}}+\eta_\text{M})+ C_{\nu} \lambda_{\text{reg}}^{2\nu}\|w\|_2.
\end{equation}
Furthermore,  choosing the optimal parameter $\lambda_{\text{reg}}^2=\left(\frac{\delta_{\text{all}}+\eta_\text{M}}{4\nu C_{\nu}\|w\|_2}\right)^{\frac{2}{(2\nu+1)}}$ yields the minimum error bound: 
\begin{equation}
    \|\boldsymbol{s}^{\delta}-\boldsymbol{s}^{*}\|_2 \le \left[ (2\nu+1) \cdot (4\nu)^{-\frac{2\nu}{2\nu+1}} \cdot (C_{\nu}\|w\|_2)^{\frac{1}{2\nu+1}} \right] (\delta_{\text{all}}+\eta_\text{M})^{\frac{2\nu}{2\nu+1}}.
\end{equation}
where $C_{\nu}$ is a positive constant that depends only on $\nu$.
\end{enumerate}
\end{theorem}
The proof of Theorem \ref{Existence and Uniqueness} is given in Appendix B.
\begin{remark}
     Statement (\ref{thm:Tikhonov_uniqueness}) demonstrates a key benefit of Tikhonov regularization: it formulates the inverse problem as a strictly convex optimization, which ensures both the existence and uniqueness of the solution, thereby establishing a solid foundation for the subsequent numerical algorithm. 
      Statement (\ref{thm:Tikhonov_stability}) establishes an optimal $\mathcal{O}((\delta_{\text{all}}+\eta_\text{M})^{\frac{2\nu}{2\nu+1}})$ convergence rate for the coefficients and provides an a priori estimate for the optimal regularization parameter $\lambda_{\text{reg}}$.
\end{remark}

\section{The morphology-adaptive random feature method}
\subsection{Integral adaptive random feature method}
To describe our method and the baseline methods, we first consider the inverse problem (\ref{helmholtz-DtN}). 
The standard PINN and RFM for this problem is to approximate the unknown scalar field $u(\boldsymbol{x})$ and $S(\boldsymbol{x})$ at the same time.
Let $u_{M}(\boldsymbol{x})$ and $S_{M}(\boldsymbol{x})$ be the output approximated by the fully connected neural network:
\begin{align}\label{approximation function}
    u_{M}(\boldsymbol{x})&=\sum_{m=1}^{M_u} u_{m} \phi_{m,1}(\boldsymbol{x}), \quad    S_{M}(\boldsymbol{x})=\sum_{m=1}^{M_S} s_{m}\phi_{m,2}(\boldsymbol{x}).
\end{align}
In RFM, each random feature function $\phi_{m,1}, \phi_{m,2}$ is constructed by
\begin{align}
\label{activation funtion}
\phi_{m}(\boldsymbol{x})= \sigma(\boldsymbol{k}_m\cdot \boldsymbol{x}+b_m),
\end{align}
where $\sigma$ is the activation function,   {$\boldsymbol{k}_m$, $b_m$ are initialized randomly, i.e.  $\boldsymbol{k}_m \sim \text{Uniform}(-R_m, R_m)^d$,  $b_m \sim \text{Uniform}(-R_m, R_m)$, and remain fixed throughout the training process}.  {Therefore,  only the outer parameters $\{u_m\}_{m=1}^{M_u}$, $\{s_m\}_{m=1}^{M_S}$ } need to be determined. In RFM, the domain $\Omega$ is partitioned into $N_p$ subdomains $\{ \Omega_n, n=1,\ldots, N_p \}$, each centered at $\boldsymbol{x}_n$, such that $\Omega=\cup_{n=1}^{N_p}\Omega_n$. For each $\Omega_n$, taking cubic-type subdomains as examples, RFM applies a standardization:
\begin{align}
\tilde{\boldsymbol{x}}^{(n)}= {\left(\frac{x_1-x_{n1}}{r_{n1}},\frac{x_2-x_{n2}}{r_{n2}},\ldots,\frac{x_d-x_{nd}}{r_{nd}}\right)^\top},\quad n=1,\cdots,N_p,
\end{align}
to map  {$\Omega_n = \prod_{k=1}^d [x_{nk}-r_{nk}, x_{nk}+r_{nk}]$ into $[-1,1]^d$, where $\boldsymbol{r}_n$ represents the radius of $\Omega_n$}. We take the following PoU functions in the one-dimensional case:
\begin{align}
\label{PoU}
\begin{aligned}
&\psi(x)=\mathbb{I}_{\left[-\frac{5}{4},-\frac{3}{4}\right]}\left(x\right)\cdot\frac{1+\sin\left(2\pi x\right)}{2}+\mathbb{I}_{\left[-\frac{3}{4},\frac{3}{4}\right]}\left(x\right)+\mathbb{I}_{\left[\frac{3}{4},\frac{5}{4}\right]}\left(x\right)\cdot\frac{1-\sin\left(2\pi x\right)}{2}.\end{aligned}
\end{align}
For $d \ge 1$, the PoU function $\psi_n(\boldsymbol{x})$ is defined as $\psi_n(\boldsymbol{x})=\Pi_{i=1}^{d}\psi_n(x_{i})$, where $\psi_n(x_i) = \psi(\tilde{x}_i^{(n)})$.
Then,
\begin{align} \label{eq:u-S-approx}
u_{M}(\boldsymbol{x})=\sum_{n=1}^{N_p}\psi_{n,1}(\boldsymbol{x})\sum_{j=1}^{J_{n,1}}u_{nj}\phi_{nj,1}(\boldsymbol{x}), \quad S_{M}(\boldsymbol{x})=\sum_{n=1}^{N_p}\psi_{n,2}(\boldsymbol{x})\sum_{j=1}^{J_{n,2}}s_{nj}\phi_{nj,2}(\boldsymbol{x}).
\end{align}
 {Here, the subscripts $1$ and $2$ to indicate that different PoUs and basis functions can be used for $u$ and $S$, where $J_{n,1}$ and $J_{n,2}$ denote the number of basis functions used on the $n$-th partition for approximating $u$ and $S$, respectively.} 
Let $C_I$, $C_B$, $C_D$, and $C_N$ represent inner points, DtN boundary points, Dirichlet and Neumann observation points, respectively. 
 The loss function for problem (\ref{helmholtz-DtN}) is
\begin{align}
    \label{RFM_LOSS_inverse_source}
\mathcal{L}_{\text{RFM}}&=\sum_{k=k_1}^{k_N}
\biggl( \lambda_   I\sum_{\boldsymbol{x}_i \in C_I}\left|-(\Delta+k^2) u_{M}^{k}(\boldsymbol{x}_i)-S_{M}(\boldsymbol{x}_i)\right|^2
+\lambda_B\sum_{\boldsymbol{x}_j\in C_B}\left|\partial_{\nu} u^{k}_{M}(\boldsymbol{x}_j)-\mathcal{T}^{k}(u^{k}_{M}(\boldsymbol{x}_j))\right|^2 \notag\\
&{}\quad +\lambda_{D}\sum_{\boldsymbol{\xi}_j\in C_D}\left|u^{k}_{M}(\boldsymbol{\xi}_j)-u^{k}(\boldsymbol{\xi}_j)\right|^2
+\lambda_{N}\sum_{\boldsymbol{\xi}_j\in C_N}\left|\partial _{\nu}u^{k}_{M}(\boldsymbol{\xi}_j)-\partial_{\nu}u^{k}(\boldsymbol{\xi}_j)\right|^2 \biggr).
\end{align}
This approach needs $N+1$ neural networks to approximate the field and the source $\{\{u^{k}_{M}\}_{k_\text{min}}^{k_\text{ max}}, S_{M}\}$. The number of parameters is huge, and the boundary condition is a kind of soft constraint. Moreover, one has to tune $\lambda_I, \lambda_B$, $\lambda_D$, and $\lambda_N$ to balance the PDE, the boundary condition, and the observation data, which is not an easy task.

To this end, we propose the integral random feature method (IRFM) based on (\ref{L1})-(\ref{L2}), which naturally satisfies the Sommerfeld condition and requires only one network to approximate the source $S$ in (\ref{L1})--(\ref{L2}), which takes the form:  
\begin{align} \label{eq:S-approx}
S_{M}(\boldsymbol{x})=\sum_{n=1}^{N_p}\psi_{n}(\boldsymbol{x})\sum_{j=1}^{J_{n}}s_{nj}\phi_{nj}(\boldsymbol{x}),
\end{align}
 {where $J_{n}$ denotes the number of basis functions used on the $n$-th partition for approximating $S$.}
The next step concentrates on the integral discretization. 
Since the support $\tau$ is not known, to make things simple, we do numerical integration inside a larger box (cube for 3-D) denoted by $V_0$ that contains $\tau$.
IRFM employs a tensor-product Gauss--Legendre quadrature to integrate \eqref{L1}-\eqref{L2}, and then solve for $\boldsymbol{s}$ by minimizing 
\begin{align}
    \label{IRFM_LOSS}
\mathcal{L}_{\text{IRFM}} = \sum_{k=k_1}^{k_N} \Bigl(
\lambda_{D}\!\!\sum_{\boldsymbol{\xi}_j\in C_D}\bigl|\hat{L}_k^{(1)}(S_M)(\boldsymbol{\xi}_j)-u^k (\boldsymbol{\xi}_j)\bigr|^2
+ \lambda_{N}\!\!\sum_{\boldsymbol{\xi}_j\in C_N}\bigl|\hat{L}_k^{(2)}(S_M)(\boldsymbol{\xi}_j)-\partial_\nu u^k (\boldsymbol{\xi}_j)\bigr|^2 \Bigr),
\end{align}
which is a discretized version of \eqref{continue_ISP}. Here, $\hat{L}_k^{(1)}$ and $\hat{L}_k^{(2)}$ denote the operators defined in  \eqref{L1}-\eqref{L2} with the integrations approximated by numerical quadratures.

However, the tensor-product Gaussian integration deteriorates for non-smooth functions.
To effectively reduce computational costs,  {we employ an adaptive mesh refinement strategy. During the $n$-th iteration, for the $i$-th quadrature cell $C_i$, we calculate
\begin{align} \label{eq:quad-indicator}
 I_{i}^{\text{abs}} = \Bigl(%\sum_{j=1}^{2^d} 
 \sum_{\ell}  
 \bigl|S_{\text{M}}^{(n)}(\boldsymbol{x}^{i}_\ell)\bigr|^2 \omega^{i}_\ell\Bigr)^{1/2}, 
\quad
I_{i}^{\nabla} &= \Bigl(\sum_\ell 
 \bigl|\nabla S_{\text{M}}^{(n)}(\boldsymbol{x}^i_\ell)\bigr|^2\omega^i_\ell\Bigr)^{1/2},  
 \end{align}
where $\{ \boldsymbol{x}^i_\ell, \omega^i_\ell \}$ are the Gaussian quadrature nodes and weights on quadrature cell $C_i$.
A cell is marked for refinement if it meets either of the following criteria:
\begin{itemize}
\item \emph{Intensity criterion:} To adequately resolve the source support, we refine regions with high local magnitude: $I_{i}^{\text{abs}}/\sum_i I_{i}^{\text{abs}}  > {\eta_{\text{abs}}}$.
\item \emph{Gradient criterion:} To accurately capture boundaries and sharp transitions, we refine regions with high local gradient: 
$ I_{i}^{\nabla}/\sum_i I_{i}^{\nabla} > \eta_{\text{grad}}$. 
\end{itemize}
 The thresholds $\eta_{\text{abs}}$ and $\eta_{\text{grad}}$ are inversely proportional to the number of quadrature cells. %See Figure \ref{process: two-phase} for a sketch of adaptive integration. 
 The details of the iterative adaptive random feature method (IA-RFM) are shown in Algorithm \ref{alg: IA}.}

\begin{algorithm}[H]
\caption{The Iterative Adaptive Random Feature Method (IA-RFM)}
\label{alg: IA}
\begin{algorithmic}[1]
\REQUIRE Quadrature domain $V_0$,  {initial network $\mathcal{N}_0$}, initial mesh  {set} $\mathcal{C}_0$, $n^d$ Gauss points. Thresholds  $\eta_{\text{grad}}$, $\eta_{\text{abs}}$, tolerance $\varepsilon_{\text{IA-RFM}}$, max refine level {\tt max\_{level}}, max iterations {\tt max\_iter}.

\ENSURE Final adaptive mesh  $\mathcal{C}_{\text{final}}$  and numerical solution $S_{M}^{\text{final}}$.

\STATE \emph{Step 1: } Compute initial solution $S_{M}^{(0)}$ on mesh $\mathcal{C}_0$  {based on $\mathcal{N}_0$} using IRFM. Set $n=0$.

\STATE \emph{Step 2: } Initialize $\mathcal{C}_{n+1} = \emptyset$, then:
    \STATE \quad \emph{2.1:} Calculate indicators $I_{i}^{\text{abs}}$, $I_{i}^{\nabla}$ for each $C_i \in \mathcal{C}_k$ by \eqref{eq:quad-indicator}.

    \STATE \quad \emph{2.2:} For each cell $C_i \in \mathcal{C}_k$, mark $C_i$  if  {$\text{level}(C_i) < \texttt{max\_level}$} and the following is satisfied:
    \begin{equation}
        I_{i}^{\text{abs}}/\sum_i I_{i}^{\text{abs}}  > {\eta_{\text{abs}}}
        \quad \text{or}
        \quad 
        I_{i}^{\nabla}/\sum_i I_{i}^{\nabla} > \eta_{\text{grad}}.
    \end{equation}
    If marked, divide $C_i$ into sub-cells, then add all sub-cells to $\mathcal{C}_{n+1}$; otherwise, add $C_i$ to $\mathcal{C}_{n+1}$.

\STATE \emph{Step 3: } Compute refined solution $S_{M}^{(n+1)}$ on $\mathcal{C}_{n+1}$ using IRFM. 

\STATE \emph{Step 4: } Calculate iterative relative error $\Delta S_{M}^{(n)} = {\|S_{M}^{(n+1)} - S_{M}^{(n)}\|_2}/{\|S_{M}^{(n)}\|_2}$. \\
\ \ \ If $\Delta S_{M}^{(n)} < \varepsilon_{\text{IA-RFM}}$ or $n \geq$ {\tt max\_iter}, jump to Step 5;\\
\ \ \ Else set $n \leftarrow n+1$, and repeat Steps 2-4.
\STATE \emph{Step 5: } Set $\mathcal{C}_{\text{final}} = \mathcal{C}_{n+1}$ and  $S_{M} = S_{M}^{(n+1)}$, then stop.
\end{algorithmic}
\end{algorithm}

\begin{remark} 
{
\label{rem:partition_strategy}
Although the general RFM framework allows for a static domain decomposition into $N_p$ subdomains (as defined in Eq. \eqref{eq:S-approx}), the framework proposed in this work is designed with a dynamic refinement strategy. 
For simplicity, we initialize the solver with a single global representation (setting $N_p=1$ and $\Omega_1 = \Omega$, the extension to $N_p>1$ is straightforward). Consequently, the parameter $J_{n}$ corresponds to the  total number of basis functions to approximate $S$, denoted as $M_0$ in our numerical experiments. 
\begin{align} \label{eq:S-approx-fix}
S_{M}(\boldsymbol{x})=\sum_{j=1}^{M_0}s_{j}\phi_{j}(\boldsymbol{x}).
\end{align}
As subsequently detailed in Section \ref{sec:MA-bases}, instead of increasing $N_p$ geometrically, the method adaptively expands the approximation space by adding morphology-aware basis functions, resulting in a total basis count of $M_{\text{total}} = M_0 + \sum_{i=1}^{I_{\text{max}}} \sum_{{k=1}}^{K_i} M_k^{(i)}$.
}
\end{remark}

 {Substituting the global representation (\ref{eq:S-approx-fix}) into (\ref{L1}) and (\ref{L2}), and applying the quadrature rule from Algorithm \ref{alg: IA}, the integral operators (\ref{L1}), (\ref{L2}) can be discretized as}
\begin{align}
    \label{matrix1}
    &L_k^{(1)}(S_M)(\boldsymbol{X})\approx \hat{L}_k^{(1)}(s)(\boldsymbol{X})=\Psi_k ^{(1)} \cdot \boldsymbol{s},  
    \quad 
    &L_k^{(2)}(S_M)(\boldsymbol{X})\approx \hat{L}_k^{(2)}(s)(\boldsymbol{X})=\Psi_k ^{(2)} \cdot \boldsymbol{s},
\end{align}
where
\begin{align}
(\Psi_k^{(1)})_{n,m} &=\sum_{ {C_i \in \mathcal{C}_{\text{final}}}}\sum_{\mathbf{j}\in  {Z_n^d}}\Phi_k\left(|\mathbf{x}_n-\mathbf{y}_\mathbf{j}^i|\right)\phi_m(\mathbf{y}_\mathbf{j}^i)\omega_\mathbf{j}^i,\\
(\Psi_k^{(2)})_{n,m} &=\sum_{ {C_i \in \mathcal{C}_{\text{final}}}}\sum_{\mathbf{j}\in  {Z_n^d}} 
{\partial_{\nu(\mathbf{x}_n)}\Phi_k\bigl(|\mathbf{x}_n-\mathbf{y}_\mathbf{j}^i|\bigr)}\phi_m(\mathbf{y}_\mathbf{j}^i)\omega_\mathbf{j}^i.
\end{align}
 {Here the index set $Z_n^d:=\{1,\cdots,n\}^d$ refers to all $d$-dimensional multi-indices i.e., $j = (j_1,\dots,j_d)^\top$ with each $j_k \in Z_n :=\{1,\dots,n\}$.}  {$\{ \mathbf{y_j}^i \}$, $\{ w^i_\mathbf{j} \}$ denote the quadrature points and associated quadrature weights  in $C_i \in \mathcal{C}_{\text{final}}^i$ generated by Algorithm \ref{alg: IA}}.  {With the unknown coefficient vector defined as $\boldsymbol{s}=(s_1, s_2, \cdots, s_M)^\top$ and these discretized operators, we can explicitly formulate the $\mathcal{L}_{\text{data}}(\boldsymbol{s})$  previously defined in (\ref{loss_data}).}

\begin{figure}[htbp]
    \centering
    \begin{subfigure}[b]{0.9\textwidth}
        \includegraphics[width=\textwidth]{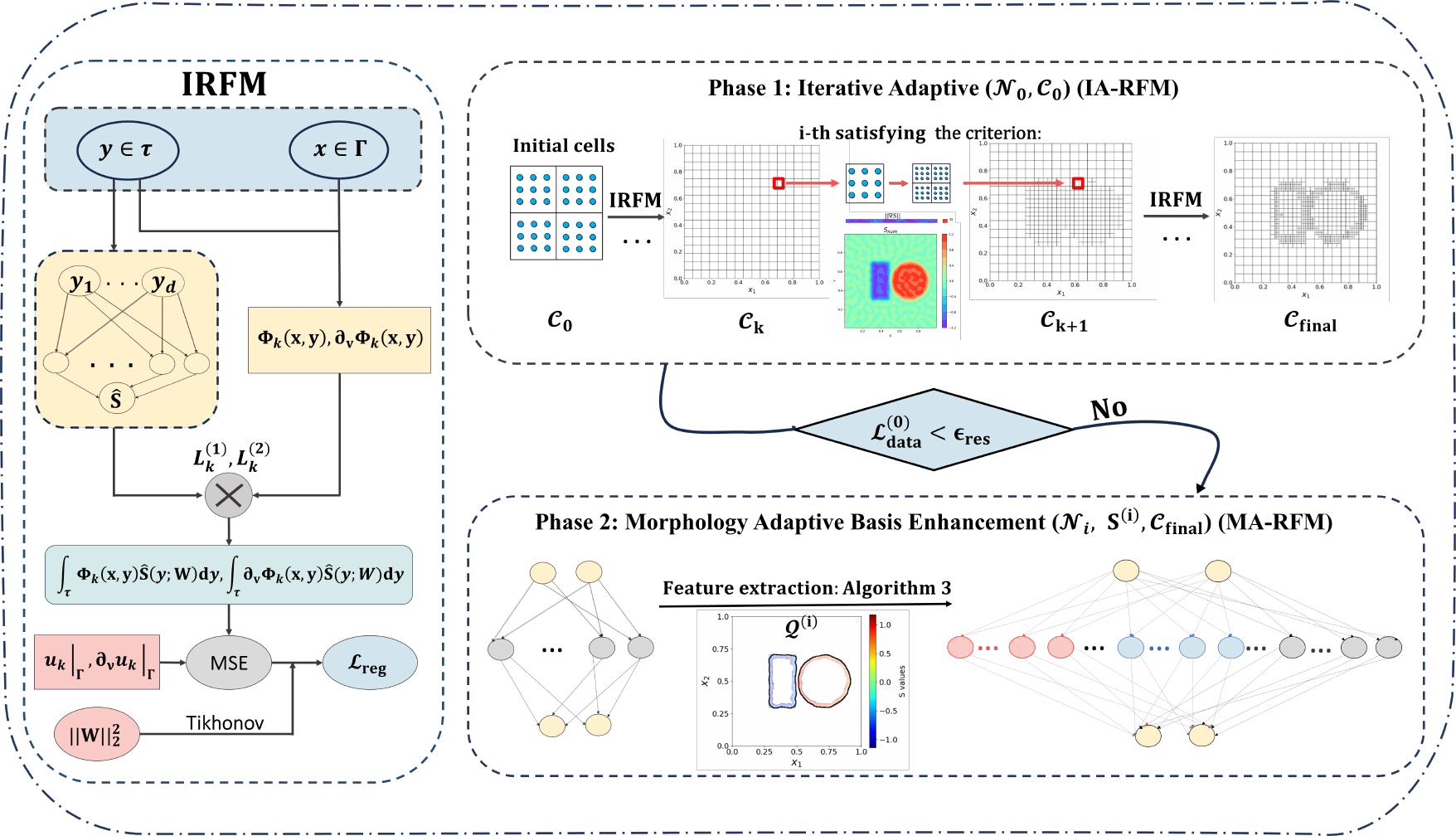}
    \end{subfigure}
    \caption{Integral Random Feature method (IRFM), Integral Adaptive Random Feature Method (IA-RFM), Morphology-Adaptive Random Feature Method (MA-RFM).}
    \label{process: two-phase}
\end{figure}

\subsection{Morphological adaptive hybrid basis enhancement}\label{sec:MA-bases}
The numerical method proposed in this section centers on adaptively enhancing the representational capacity of RFM. The hidden layer of RFM can be viewed as a function basis set. The entire RFM defines an approximation space $V$ spanned by these basis functions
\begin{equation}\label{space}
    V=\text{span}\{\phi_1(\boldsymbol{x}), \phi_2(\boldsymbol{x}), \cdots, \phi_M(\boldsymbol{x})\}.
\end{equation}
From this perspective,  RFM uses the activation functions from neural networks as basis functions, combining the advantages of neural networks and spectral methods. To overcome the limitation of the standard RFM,  whose fixed basis functions struggle to capture complex local features, we propose an adaptive basis functions enhancement strategy based on posterior information, such as the initial solution and its gradient, to locate local features, critical regions in the problem, and then add new, morphological basis functions.

For discontinuous features along specific boundaries or curves,  the geometric regions where these features reside can be implicitly defined with level set functions $\psi_\alpha(\boldsymbol{x})$, which divide the space into two regions (inside and outside the interface): 
\begin{align}
\label{area}
\Omega_\alpha &: = \{ \boldsymbol{x} \in \mathbb{R}^2 \mid \psi_\alpha(\boldsymbol{x}) \le 0 \}, \ \alpha=1, \ldots,  {K},
\end{align}
 {where 
$K$ is the number of non-overlapping regions}. Based on this,  the idea is to design a family of basis functions that can simulate jump behavior while being smooth everywhere. Its general form is as follows: 
\begin{align}
\label{jump}
    \phi_\alpha^{(j)}(\boldsymbol{x};K_\alpha^{(j)},\boldsymbol{\theta}_\alpha^{(j)}) = \sigma_{\alpha}\left(K_\alpha^{(j)} \cdot d_{\mathcal{T}_\alpha}^{(j)}(\boldsymbol{x};\boldsymbol{\theta}_\alpha^{(j)}) \right),  \ \alpha=1, \ldots,  {K},\  \ j=1, \ldots, M_\alpha. 
\end{align}
$\phi^{(j)}_\alpha$ represents the $j$-th basis function added to the $\alpha$-th group. $\boldsymbol{\theta}$ are uniformly sampling fixed shape parameters. The signed distance function (SDF) $d_{\mathcal{T}_\alpha}(\boldsymbol{x})$, which is closely related to $\psi_\alpha(\boldsymbol{x})$, represents the distance to the interface and $\mathcal{T}_\alpha$ specifies the geometric type.  $\sigma_\alpha$ typically is chosen as {\it tanh},  {\it sigmoid},  or similar S-shaped functions to simulate the interface. The soft boundary parameters $K_\alpha^{(j)}$ are used to control the ``hardness'' of the transition across the boundary. Ultimately,  we use a smooth and differentiable ``soft'' function to effectively approximate a discontinuous ``hard'' geometric boundary, allowing it to integrate into gradient-based optimization algorithms,  which traditional hard boundary models cannot achieve.  

To determine which $\mathcal{T}_\alpha$ and $\sigma_\alpha$ to apply and to estimate the corresponding shape parameters $\boldsymbol{\theta}_\alpha$, we employ a data-driven strategy. The process relies on the point cloud $\mathcal{Q}_{\text{grad}}$ or $\mathcal{Q}_{\text{abs}}$, defined as
\begin{align}
 \mathcal{Q}_{\text{abs}}(S_M)=\Bigl\{\boldsymbol{x} \in P_{\text{test}}, \frac{|S_M(\boldsymbol{x})|}{\|S_M\|_\infty} \geq t_{\text{abs}} \Bigr\},
 \ \mathcal{Q}_{\text{grad}}(S_M)=\Bigl\{\boldsymbol{x} \in P_{\text{test}},\  \frac{|\nabla S_M(\boldsymbol{x})|}{\|\nabla S_M\|_\infty} \geq t_{\text{grad}}\Bigr\},
\end{align}
where $t_{\text{abs}}, t_{\text{grad}} \in (0,1)$ are hyperparameters. 
We firstly employ the DBSCAN algorithm~\cite{EsterKX1996DensitybasedAlgorithm} to partition the candidate point set $\mathcal{Q} \subseteq \mathcal{Q}_{\text{grad}} \cup \mathcal{Q}_{\text{abs}}$ into $K$ disjoint clusters {with adaptive neighborhood radius
  \[
  \varepsilon_{\rm clu}=c_\varepsilon h,\qquad
  h=\max_i(\Delta x_i),
  \]
  and minimum core-point size $N_{\min}^{\rm clu}$.}
 These clusters serve as the distinct support regions for the $K$ sources,
$$\mathcal{Q}=\bigcup_{\alpha=1}^{K} C_\alpha, \quad C_i\cap C_j=\emptyset, \quad \forall i \neq j.$$
We categorize the identified clusters based on a prior shape library, which classifies geometries into three primary families: \emph{ellipsoidal} (including circle/ellipse-like shapes), \emph{rectangular} (cubic/box-like shapes), and \emph{general} (irregular shapes). To implement the classification, we firstly extract a set of geometric descriptors and fitting metrics from the cluster $C_\alpha$ and its boundary $\hat{\Gamma}_\alpha = \{\boldsymbol{x}^{(j)}\}_{j=1}^{|\hat{\Gamma}_\alpha|}$. Starting with the spatial extent and radial distribution, we define the centroid $\boldsymbol{c}_\alpha$, half axis-aligned lengths $\boldsymbol{L}_\alpha$.

\begin{equation}
\label{eq: geometric_basic}
\begin{aligned}
    & \boldsymbol{c}_\alpha = \frac{1}{|C_\alpha|}\sum_{\boldsymbol{x}\in C_\alpha}\boldsymbol{x}, 
    \quad 
    L_{\alpha,i} = \frac{1}{2} \bigl( \max_{\boldsymbol{x} \in C_\alpha}(x_i) - \min_{\boldsymbol{x} \in C_\alpha}(x_i) \bigr), \ i=1,\dots,d. 
\end{aligned}
\end{equation}

 {Finally, to distinguish between the \emph{rectangular} and \emph{ellipsoidal} families, we compute the fitting residuals $\mathcal{E}$, which measure the deviation of the boundary points from an ideal box or ellipsoid parameterized by $(\boldsymbol{c}_\alpha, L_{\alpha,i})$:
\begin{equation}
\label{eq: fitting_residuals}
\begin{aligned}
    \mathcal{E}_{\mathrm{rect}}^{(\alpha)} = \frac{1}{|\hat{\Gamma}_\alpha|} \sum_{j=1}^{|\hat{\Gamma}_\alpha|}\left|\max_{\ell \in\{ 1,...,d\}} \bar{x}_l^{(j)} -1\right|, \quad
    \mathcal{E}_{\mathrm{ellip}}^{(\alpha)} = \frac{1}{|\hat{\Gamma}_\alpha|}\sum_{j=1}^{|\hat{\Gamma}_\alpha|}\left|\sqrt{\sum_{\ell=1}^{d}\bigl( \bar{x}_l^{(j)} \bigr)^2}-1 \right|,
    \quad \bar{x}_l^{(j)} = \frac{x^{(j)}_{\ell} - c_{\alpha,\ell}}{L_{\alpha,\ell}}.
\end{aligned}
\end{equation}}
  
\begin{remark}
 {
The prior library is topology-aware but not completely fixed a priori: it only specifies the admissible morphology classes for the enrichment space. In two dimensions, we take
$\mathcal{T}_{\alpha}\in\{\mathrm{rectangle},\mathrm{ellipsoid},\mathrm{general}\}$,
while in three dimensions we enlarge the candidate set to
$\mathcal{T}_{\alpha}\in\{\mathrm{rectangle},\mathrm{ellipsoid},\mathrm{torus},\mathrm{general}\}$.
This library may be extended at the solver-design level when the stage-one coarse reconstruction provides stable evidence of an additional regular morphology. Such an extension concerns only the admissible basis families, whereas the actual label $\mathcal{T}_{\alpha}$ and its geometric parameters are still inferred from the stage-one coarse reconstruction through clustering, morphological post-processing, and residual-based shape comparison. In particular, for three-dimensional clusters with toroidal topology, we additionally evaluate a torus-type fitting residual together with the rectangle- and ellipsoid-type residuals; see Appendix~\ref{app:torus_residual} for its explicit definition.}
\end{remark}
  
 {Next, we classify the basis function construction into two distinct scenarios.
The first scenario addresses canonical geometries, allowing for efficient parametric fitting. The second scenario addresses general, irregular geometries where no prior shape knowledge is assumed, necessitating a robust, data-driven numerical level set approach.}

\subsubsection{ {Scenario I: canonical geometries}}
 {For sources exhibiting canonical geometries (e.g., spheres, cubes, or tori), we construct the basis functions by fitting geometric primitives.}

 {For example,  for the indicator function of a circular boundary $\mathbb{I}_{\boldsymbol{x} \in B(c, r)}$,  choose 
$$\psi_\alpha^{(j)}(\boldsymbol{x};\boldsymbol{\theta}_\alpha^{(j)}) = -d_{\text{circle}}^{(j)}(\boldsymbol{x},\boldsymbol{\theta}_\alpha^{(j)}) = |\boldsymbol{x} - \boldsymbol{c}_\alpha^{(j)}|^2 - (r_\alpha^{(j)})^2,\ j=1,\cdots,M_\alpha,$$
where $\boldsymbol{\theta}_\alpha^{(j)}=(\boldsymbol{c}_\alpha^{(j)},r_\alpha^{(j)})$. Use the sigmoid function for approximation: 
$$\phi_\alpha^{(j)}(\boldsymbol{x};K_\alpha^{(j)},\boldsymbol{\theta}_\alpha^{(j)}) = \sigma_\alpha\Bigl(K_\alpha^{(j)} \cdot \bigl((r_\alpha^{(j)})^2 - |\boldsymbol{x} - \boldsymbol{c}_\alpha^{(j)}|^2\bigr)\Bigr),  \quad \boldsymbol{c}_\alpha^{(j)}\in \mathbb{R}^d ,$$
 {or equivalently, 
$$\phi_\alpha^{(j)}(\boldsymbol{x};K_\alpha^{(j)},\boldsymbol{\theta}_\alpha^{(j)}) = \sigma_\alpha\left(\tilde{K}_\alpha^{(j)} \cdot \left(1- \left|\frac{\boldsymbol{x} - \boldsymbol{c}_\alpha^{(j)}}{r_\alpha^{(j)}}\right|^2\right)\right),  \quad \boldsymbol{c}_\alpha^{(j)}\in \mathbb{R}^d ,$$
where $\tilde{K}_{\alpha}^{(j)}=(r^{(j)}_{\alpha})^2K_{\alpha}^{(j)}$, and the shape parameters are sampled from uniform distributions around the detected estimates:
$K_\alpha^{(j)} \sim \text{Uniform}(K_{\alpha, \text{min}}, K_{\alpha, \text{max}})$, 
$r_\alpha^{(j)} \sim \text{Uniform}\left((1-\epsilon_r)\hat{r}_{\alpha}, (1+\epsilon_r)\hat{r}_{\alpha}\right)$, \\ 
$c_{\alpha,l}^{(j)} \sim \text{Uniform}\left(\hat{c}_{\alpha,l} - \epsilon_c |\hat{c}_{\alpha,l}|, \hat{c}_{\alpha,l} + \epsilon_c |\hat{c}_{\alpha,l}|\right)$.
Here, $l=1,2,\cdots,d$, is the $l$-th component, and $\hat{r}_\alpha$, $\hat{\boldsymbol{c}}_\alpha$ denote the posterior shape parameters derived from the geometry detection (\ref{eq: geometric_basic}). The terms $\epsilon_r$ and $\epsilon_c$ serve as perturbation tolerances, introduced to account for potential inaccuracies in the geometric estimation and to enhance the robustness of the basis set. More generally, it can be generalized to the form of an ellipsoid and rectangle,
\begin{align}
    \label{ellip}
    \phi_\alpha^{(j)}(\boldsymbol{x};K_\alpha^{(j)},\boldsymbol{\theta}_\alpha^{(j)}) &= \sigma_\alpha\left(\tilde{K}_\alpha^{(j)} \cdot \left(1- \sum_{i=1}^d\left|\frac{x_i - c_{\alpha,i}^{(j)}}{L_{\alpha,i}^{(j)}}\right|^2\right)\right),  \quad \boldsymbol{c}_\alpha^{(j)}\in \mathbb{R}^d ,\\
    \label{rec}
    \phi_\alpha^{(j)}(\boldsymbol{x};K_\alpha^{(j)},\boldsymbol{\theta}_\alpha^{(j)})
    &=
    \sigma_\alpha\!\left(
    \hat{K}_\alpha^{(j)}
    \left(
    1-\max_{i\in\{1,\dots,d\}}
    \left|\frac{x_i-c_{\alpha,i}^{(j)}}{L_{\alpha,i}^{(j)}}\right|
    \right)
    \right),
    \qquad
    \boldsymbol{c}_\alpha^{(j)}\in\mathbb{R}^d.
    % \phi_\alpha^{(j)}(\boldsymbol{x};K_\alpha^{(j)},\boldsymbol{\theta}_\alpha^{(j)}) &= \sigma_\alpha\left(\hat{K}_\alpha^{(j)} \cdot
    % \max_{i\in\{1,\dots,d\}}\{L_{{\alpha},i}^{(j)}-(x_i-c_{\alpha}^{(j)}) \}\right)  \quad \boldsymbol{c}_{\alpha^{(j)},i}\in \mathbb{R}^d ,
\end{align}
where $\tilde{K}_{\alpha}^{(j)}=(\prod_{i=1}^dL_{\alpha,i})^{2/d}K_{\alpha}^{(j)}$, $\hat{K}_{\alpha}^{(j)}=(\prod_{i=1}^dL_{\alpha,i})^{1/d}K_{\alpha}^{(j)}$.}}

\begin{figure}[H]
    \centering
    \captionsetup[subfigure]{justification=centering, labelformat=parens}

    \begin{subfigure}[b]{0.24\textwidth}
        \includegraphics[width=\textwidth]{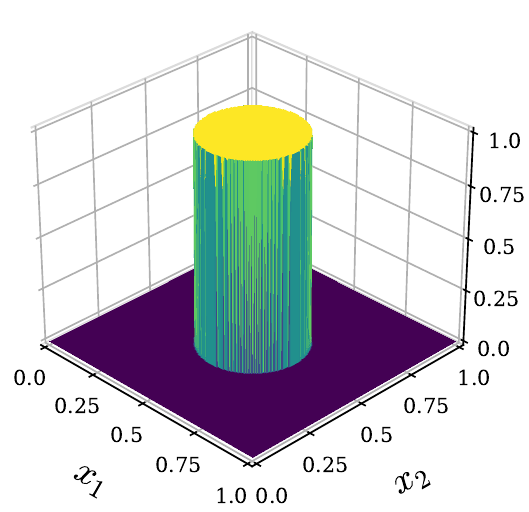}
        \caption{ }
        \label{fig:morph_circle}
    \end{subfigure}
    \begin{subfigure}[b]{0.24\textwidth}
        \includegraphics[width=\textwidth]{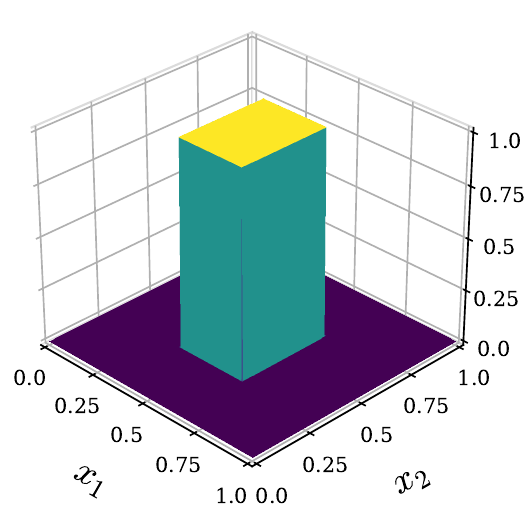}
\caption{ }
        \label{fig:morph_rect}
    \end{subfigure}
    \begin{subfigure}[b]{0.24\textwidth}
        \includegraphics[width=\textwidth]{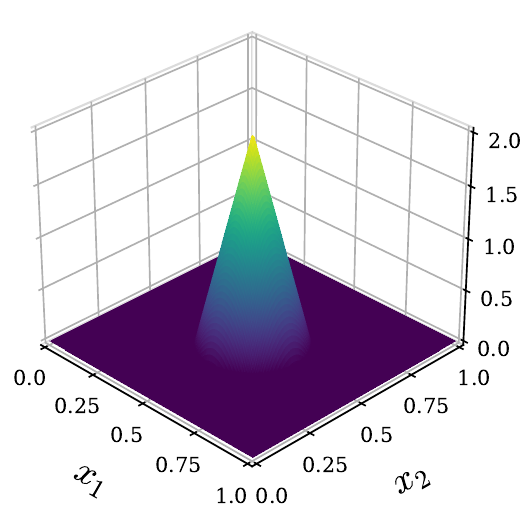}
        \caption{ }
        \label{fig:morph_cone}
    \end{subfigure}
    \begin{subfigure}[b]{0.24\textwidth}
        \includegraphics[width=\textwidth]{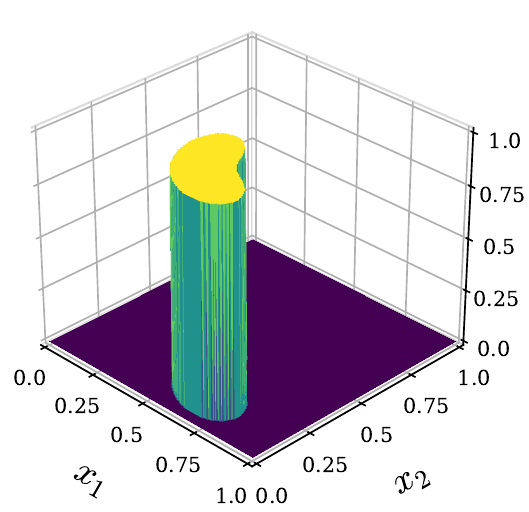}
        \caption{ }
        \label{fig:morph_bean}
    \end{subfigure}   
    \caption{ {Visualization of morphological basis functions. (a) $\text{sigmoid}\left(K\cdot( r^2-\|\boldsymbol{x}-\boldsymbol{c}\|_2^2)\right)$. Circular domain with $K=5\times10^4,r=0.2,\boldsymbol{c}=(0.5,0.5)^\top$. (b) $\text{sigmoid}\big(K\cdot \max\limits_{l=1,2}\{L_L-(x_l-c_l)\}\big)$. Rectangular domain with $K=5\times10^4,\boldsymbol{c}=(0.5,0.5)^\top, \boldsymbol{L}=(\text{$L_1$,$L_2$})^\top=(0.15,0.2)^\top$. (c) $ \text{ReLU}\left(K\cdot( r-\|\boldsymbol{x}-\boldsymbol{c}\|_2)\right)$. Conical with $K=2,r=0.2,\boldsymbol{c}=(0.5,0.5)^\top$. (d) $\text{sigmoid}\left(K\cdot d(\boldsymbol{x};\boldsymbol{\theta})\right)$. Bean domain using SDF in (\ref{SDF}) with $K=5\times10^4$.}}
    \label{fig:morphology_function}
\end{figure}

 {This adaptive idea is equally applicable to other types of local features,  such as Gaussian-type source terms with $\boldsymbol{\theta}_\alpha^{(j)}=(\boldsymbol{c}_\alpha^{(j)},v_\alpha^{(j)})$:
\begin{align}
    \label{Gauss}
    \phi_\alpha^{(j)}(\boldsymbol{x};\boldsymbol{\theta}_\alpha^{(j)})= \exp\left(-\tilde{v}_\alpha^{(j)} \cdot \sum_{i=1}^d\left|\frac{x_i - c_{\alpha,i}^{(j)}}{L_{\alpha,i}^{(j)}}\right|^2\right),
\end{align}
where $\tilde{v}^{(j)}_\alpha=(\prod_{i=1}^dL_{\alpha,i})^{2/d}v^{(j)}_\alpha$. The center $\boldsymbol{c}^{(j)}_\alpha$ and the decay rate $v^{(j)}_{\alpha}$, can be determined through posterior information and random sampling. See Figure \ref{fig:morphology_function} for diverse morphological basis functions.}

\subsubsection{ {Scenario II: general, irregular geometries}}
 {For complex geometries exhibiting non-convex features, we extract the discrete numerical boundary: $\hat{\Gamma}_\alpha$ corresponding to the $\alpha$-th source cluster $C_\alpha$ from $\mathcal{Q}$, establishing this boundary as the level set of the implicit function $\psi_\alpha$:
$$\hat{\Gamma}_\alpha = \partial C_\alpha = \{ \boldsymbol{x} \mid \psi_\alpha(\boldsymbol{x}) = 0 \}.$$
Then generate a family of perturbed interfaces by expanding or contracting $\hat{\Gamma}_\alpha$ along its unit outward normal $\boldsymbol{n}(\boldsymbol{x})$,}
\begin{align}
\label{general}\boldsymbol{x}^{(j)}=\boldsymbol{x}+ \rho ^{(j)}_\alpha \boldsymbol{n}(\boldsymbol{x}) ,  \ \forall \ \boldsymbol{x} \in  {\hat{\Gamma}_{\alpha}},
\end{align}
where $\rho^{(j)}_\alpha \sim U(\rho^{(j)}_{\min},  \rho^{(j)}_{\max})$.  {In practice, when the subsequent screening set is chosen as $\mathcal{Q}=\mathcal{Q}_{\mathrm{grad}}\cap \mathcal{Q}_{\text{abs}}$, we take $\rho^{(j)}_{\min}<0<\rho^{(j)}_{\max}$ so that the detected curve can be perturbed both inward and outward. The intersection reduces the effective local region, and bidirectional offsets help preserve a sufficiently informative neighborhood of the interface. If only one-sided $\mathcal{Q}_{\mathrm{grad}}$ or $\mathcal{Q}_{\mathrm{abs}}$ is used, we set $\rho^{(j)}_{\min}=0$, so that the offset is restricted to a single direction.}

Each $\rho^{(j)}$ defines a new,  slightly deformed boundary,  thereby generating a family of new level set functions $\psi_\alpha^{(j)}(\boldsymbol{x};\rho_\alpha^{(j)})$ and $\hat{\Gamma}^{(j)}_\alpha=\{\boldsymbol{x} \ |\ \psi_\alpha^{(j)}(\boldsymbol{x};\rho_\alpha^{(j)})=0\}$.  {Next, we explicitly define the region polarity using a simple approximated SDF
\begin{align}
\label{SDF}
d_{\text{general}}^{(j)}(\boldsymbol{x};\rho^{(j)})=\left \{
\begin{aligned}
&1,  \quad &\psi_\alpha^{(j)}(\boldsymbol{x};\rho^{(j)})\le 0 , \\
&-1,  \quad &\psi_\alpha^{(j)}(\boldsymbol{x};\rho^{(j)})> 0.
\end{aligned}
\right.
\end{align}
When we select a larger $K$, the basis function (\ref{jump}) $\phi_\alpha^{(j)}(\boldsymbol{x};K_\alpha^{(j)},\boldsymbol{\theta}_\alpha^{(j)})$ can clearly distinguish between the interior and exterior of the interface, thereby functioning as a local basis function. Figure \ref{fig:morph_bean} shows the morphological basis functions of non-convex regions.}  {In the numerical experiments, local circular basis functions are introduced to compensate for the limited interface-capturing capability. 
$$
\phi^{(j)}_{\mathrm{circular}}(\boldsymbol{x};K^{(j)},\boldsymbol{\theta}^{(j)})
=\mathrm{sigmoid}\!\left(
K^{(j)}\cdot\left((r^{(j)})^2-\|\boldsymbol{x}-\boldsymbol{c}^{(j)}\|_2^2\right)
\right), \quad j=1,\cdots,M_{\mathrm{circular}}.
$$
Their centers $\boldsymbol{c}^{(j)}$ are randomly sampled from a set of candidate points $\{\boldsymbol{x}^{(j)}\}$, and the radius $r^{(j)}$ are determined by the mesh scale.}

 Based on the descriptors defined above and the two scenarios, we perform shape classification and basis initialization through the following steps:
  \begin{enumerate}
      \item  {\emph{Shape classification and SDF selection ($d_{\mathcal{T}_\alpha}$):}
      We use the threshold $t_{\text{reg}}$ as a residual tolerance to determine whether the detected cluster admits a sufficiently accurate regular-shape fit.}
       {\begin{itemize}
          \item \emph{Scenario 1:} If $\min\bigl(\mathcal{E}_{\mathrm{rect}}^{(\alpha)},\mathcal{E}_{\mathrm{ellip}}^{(\alpha)}\bigr)\le t_{\text{reg}}$,
          then the boundary is classified as \emph{regular}. In this case, we compare the two residuals: if $\mathcal{E}_{\mathrm{rect}}^{(\alpha)}<\mathcal{E}_{\mathrm{ellip}}^{(\alpha)}$, we select the rectangular SDF; otherwise, we select the ellipsoidal SDF.
          \item \emph{Scenario 2:} If
$\min\bigl(\mathcal{E}_{\mathrm{rect}}^{(\alpha)},\mathcal{E}_{\mathrm{ellip}}^{(\alpha)}\bigr)> t_{\text{reg}}$,
          then the boundary is classified as \emph{general}, and we construct a numerical SDF directly from $\hat{\Gamma}_\alpha$.
      \end{itemize}}

   \item  {\emph{Basis type selection ($\sigma_\alpha$):}
  For the $\alpha$-th detected component, let $\widehat{\tau}_\alpha\subset V_0$ denote the
  interior of the detected geometry.
  For regular shapes, $\widehat{\tau}_\alpha$ is the interior of the fitted rectangle, ellipsoid,
  or torus; for a general component, it is the region enclosed by the detected numerical boundary.
  Let $\mathcal{G}_h=\{\mathbf{x}_m\}_{m=1}^{N_h}$ be the evaluation grid, and define
  \[
  \mathcal{I}_\alpha:=\{m:\mathbf{x}_m\in \widehat{\tau}_\alpha\}.
  \]
  We denote by
  \[
  S_{M,\alpha}:=\{\,S_{M}(\mathbf{x}_m):m\in\mathcal{I}_\alpha\,\}
  \]
  the numerical source values inside the detected region.
  Then the coefficient-of-variation-type indicator is defined by
  \[
  \text{CV}_\alpha
  :=
  \frac{\operatorname{std}(S_{M,\alpha})}
  {\operatorname{mean}(|S_{M,\alpha}|)}.
  \]
  A small $\text{CV}_\alpha$ indicates a relatively flat-top profile, while a large $\text{CV}_\alpha$ indicates
  a more concentrated peak-like profile.
  Therefore, if $\text{CV}_\alpha<t_{\text{CV}}$, we choose the \emph{sigmoid} activation; otherwise, we choose a
  peak-type activation such as \emph{exponential} (or \emph{ReLU}).}
      \item \emph{Basis assembly:}
      With the determined shape type $\mathcal{T}_\alpha$ and basis type $\sigma_\alpha$, the final basis functions are initialized as
      \[
      \phi_\alpha^{(j)}(\boldsymbol{x};K_\alpha^{(j)},\boldsymbol{\theta}_\alpha^{(j)})
      =
      \sigma_{\alpha}\bigl(
      K_\alpha^{(j)}\, d_{\mathcal{T}_\alpha}^{(j)}(\boldsymbol{x};\boldsymbol{\theta}_\alpha^{(j)})
      \bigr),
      \qquad j=1,\dots,M_\alpha.
      \]
  \end{enumerate}
  The detailed implementation and the threshold choices used in this automatic generation procedure are summarized in \emph{Algorithm~\ref{alg: shape_detection}}.

\begin{algorithm}[H]
\caption{Shape detection and basis initialization}
\label{alg: shape_detection}
\begin{algorithmic}[1]
\REQUIRE  Approximate solution $S_M$, $\mathcal{Q}_{\text{grad}}$, $\mathcal{Q}_{\text{abs}}$,  {cluster parameters $\varepsilon_{\rm clu}=5h, N_{\min}^{\rm clu}=20$}, thresholds of CV  $t_{\text{CV}} = {0.5}$,  {regular shape $t_{\text{reg}}=0.05$}.
\ENSURE Set of new basis functions $\mathcal{B}_{\text{new}}$ with initialized parameters.
\STATE Partition $\mathcal{Q}\subseteq \mathcal{Q}_{\text{grad}}\cap \mathcal{Q}_{\text{abs}}$ into disjoint clusters $\{C_\alpha\}_{\alpha=1}^K$ using DBSCAN. $\mathcal{B}_{\text{new}} \gets \emptyset$.
\FOR{each cluster $C_\alpha$ in $\{C_1, \dots, C_K\}$}
    \STATE Compute center $\boldsymbol{c}_\alpha$, half axis-aligned lengths $L_{\alpha,i}$, aspect ratio $\text{AR}_\alpha$.
    %, distance $D_{\alpha}(\boldsymbol{x})$.
    
    \STATE \textbf{if} $\min \bigl(\mathcal{E}_{\mathrm{rect}}^{(\alpha)},\mathcal{E}_{\mathrm{ellip}}^{(\alpha)}\bigr) > t_{\mathrm{reg}}$ \textbf{then} $\mathcal{T}_\alpha \gets \text{general}$, \ $\boldsymbol{\theta}_\alpha \gets \{\boldsymbol{c}_\alpha\}$.
    
    \STATE \textbf{else if} $\mathcal{E}_{\mathrm{rect}}^{(\alpha)} \le \mathcal{E}_{\mathrm{ellip}}^{(\alpha)}$ \textbf{then}  $\mathcal{T}_\alpha \gets \text{Rectangle}$, $\boldsymbol{\theta}_\alpha \gets \{ \boldsymbol{c}_\alpha, \{L_{\alpha,i}\}_{i=1}^d \}$.

    \STATE \textbf{else} $\mathcal{T}_\alpha \gets \text{Ellipsoid}$, $\boldsymbol{\theta}_\alpha \gets \{ \boldsymbol{c}_\alpha, \{L_{\alpha,i}\}_{i=1}^d \}$.
    
    \STATE Calculate Coefficient of Variation: $\text{CV}_\alpha  $.
    \STATE \textbf{if} $\text{CV}_\alpha \ge t_{\text{CV}}$ \textbf{then} $\sigma_\alpha \gets \text{Exp}$ (or ReLU) \textbf{else} $\sigma_\alpha \gets \text{Sigmoid}$.
    
    \STATE \textbf{if} $\mathcal{T}_\alpha == \text{general}$ \textbf{then} \emph{Scenario 2:} Construct SDF $d_{\text{general}}^{(j)}$ from $\hat{\Gamma}_\alpha^{(j)}$ (Eq. \ref{SDF}).
    \STATE \textbf{else} \emph{Scenario 1:} Sample $M_\alpha$ parameter sets $\boldsymbol{\theta}_\alpha^{(j)}=(\theta_{\alpha,1}^{(j)},\cdots,\theta_{\alpha,d_{\boldsymbol{\theta}}}^{(j)})^\top$ uniformly around $\hat{\boldsymbol{\theta}}_\alpha$ with tolerances $\epsilon_l$:
    \[
    \boldsymbol{\theta}_\alpha^{(j)} \in \prod_{l=1}^{d_{\boldsymbol{\theta}}} U\left( \hat{\theta}_{\alpha,l}^{(j)}-\epsilon_l |\hat{\theta}_{\alpha,l}^{(j)}|, \hat{\theta}_{\alpha,l}^{(j)}+\epsilon_l|\hat{\theta}_{\alpha,l}^{(j)}| \right).
    \]
    \STATE Update $\mathcal{B}_{\text{new}} \gets \mathcal{B}_{\text{new}} \cup \left\{ \sigma_{\alpha}\left(K_\alpha^{(j)} \cdot d_{\mathcal{T}_\alpha}^{(j)}(\boldsymbol{x};\boldsymbol{\theta}_\alpha^{(j)}) \right)\right\}_{j=1}^{M_\alpha}$.
\ENDFOR
\STATE \emph{Return} $\mathcal{B}_{\text{new}}$.
\end{algorithmic}
\end{algorithm}
In summary, MA-RFM acts as a two-stage adaptive enhancement process. Firstly, an initial coarse solution is obtained in the baseline space $V$ using IA-RFM. A posterior analysis is then performed on this solution and its gradient field to identify geometric features. Based on these features, specialized morphological basis functions are generated to expand the approximation space from $V$ to a more robust space 
$V' = V_0 \oplus V_{\text{expansion}}$, where
 {\begin{align}
\label{expansion_V}
&V_0 = \text{span}\bigl\{ \phi_0^{(j)}(\boldsymbol{x}; \boldsymbol{k}_0, b_0),\ j=1,\ldots, M_0 \bigr\}, \notag \\
&V_{\text{expansion}} = \text{span}\bigl\{ \phi_{\alpha}^{(j)}(\boldsymbol{x}; K_\alpha^{(j)}, \boldsymbol{\theta}_{\alpha}^{(j)}) \mid \alpha=1,\dots,K, \, j=1,\dots,M_\alpha \bigr\}.
\end{align}
The final system matrix $\Psi_M:=\Psi_{M_{\text{total}}}$ is constructed by augmenting the initial basis matrix $\Psi_{M_0}$ with the $K$ adaptive block matrices corresponding to the detected regions, and the vector of unknown coefficients is also expanded to align with this block structure.
Consequently, the final reconstruction is achieved by solving the Tikhonov regularization problem governed by $\mathcal{L}_{\text{reg}}$ within this enriched space $V'$.} To save computational resources,  a stopping criterion is added,  which will terminate the iteration if $\mathcal{L}_{\text{data}}$ is less than a given threshold $\epsilon_{\text{res}}$. 
The complete two-phase MA-RFM algorithm is outlined in Figure \ref{process: two-phase} and Algorithm \ref{alg: two_phase}.

\begin{algorithm}[htbp]
\caption{Morphology-Adaptive Random Feature Method (MA-RFM)}
\label{alg: two_phase}
\begin{algorithmic}[1]
\REQUIRE Quadrature domain $V_0$, initial mesh $\mathcal{C}_0$, initial network $\mathcal{N}_0$, initial number of basis functions $M_0$, $n^d$ Gauss points in the reference cell,  refinement threshold $\eta_{\text{grad}}=1/300$, $\eta_{\text{abs}}=1/100$, max division level {\tt max\_level} , max iteration {\tt max\_iter},  number of measurement points $N_s$,   {cluster parameter $\varepsilon_{\rm clu}=5h, N_{\min}^{\rm clu}=20$}, threshold  $t_{\text{grad}}$, $t_{\text{abs}}=1/3$,  {$t_{\text{reg}}=0.05$,\ $t_{\text{CV}} = 0.5$},  test points $P_{\text{test}}$, 
max enhancement iteration $I_{\text{max}}$, 
residual tolerance $\varepsilon_{\text{res}}$.
\ENSURE Final network $\mathcal{N}_{\text{final}}$ and solution $S_{\text{final}}$.
\STATE Train $\mathcal{N}_0$ by IA-RFM to obtain coefficients $s^{(0)}$, initial hidden layer $H^{(0)}$, final number of integral points $n_{\text{integral}}$, final integral points $\mathcal{P}_{\text{final}}$ and $\mathcal{C}_{\text{final}}$. Compute $S^{(0)}$ and $\nabla S^{(0)}$ on $P_{\text{test}}$. Calculate data loss $\mathcal{L}^{(0)}_{\text{data}}(s^{(0)}, \mathcal{P}_{\text{final}})$.
\STATE \textbf{if} $\mathcal{L}_{\text{data}}^{(0)} < \varepsilon_{\text{res}}$ \textbf{then} $i=0$. \textbf{else}
\FOR{$i = 1$ to $I_{\text{max}}$}
\STATE  {Filter out the point sets $\mathcal{Q}\subseteq\mathcal{Q}_{\text{grad}}^{(i)}\cup\mathcal{Q}_{\text{abs}}^{(i)}$ with $t_{\text{grad}}, t_{\text{abs}}.$}
\STATE  {Get disjoint cluster $\{C_\alpha^{(i)}\}_{\alpha=1}^K$ and new basis functions using Algorithm \ref{alg: shape_detection},
\hspace*{\fill}%
$\phi_\alpha^{(i,j)}(\boldsymbol{x};K^{(i,j)},\boldsymbol{\theta}_\alpha^{(i,j)})
=
\left\{
\sigma_{\alpha}\!\left(
K_\alpha^{(j)} d_{\mathcal{T}_\alpha}^{(j)}(\boldsymbol{x};\boldsymbol{\theta}_\alpha^{(i,j)})
\right)
\right\}_{j=1}^{M_\alpha^{(i)}}$%
\hspace*{\fill}} 
\vspace{0.6em}
\STATE Construct the enhanced network $\mathcal{N}_{i}$ by incorporating new basis functions $\{\phi_\alpha^{(i,j)}\}_{\alpha=1, j=1}^{K^{(i)}, M_\alpha^{(i)}}$: 
\hspace*{\fill}%
$H^{(i)} = H^{(i-1)}+
\{\phi_\alpha^{(i,j)}\}_{\alpha=1, j=1}^{K^{(i)}, M_\alpha^{(i)}}
, \quad \mathcal{N}_{i}(\boldsymbol{x}) =  {f_{\text{out}} \circ H^{(i)}(\boldsymbol{x})}.$
\hspace*{\fill}%
\vspace{0.6em}
\STATE Retrain to obtain ${s}^{(i)}$ and ${S}^{(i)}$ through $\mathcal{N}_{i}$. 
\STATE Calculate the data loss $\mathcal{L}_{\text{data}}^{(i)}(s^{(i)},  \mathcal{P}_{\text{final}})$.
\STATE \textbf{if} $\mathcal{L}_{\text{data}}^{(i)} < \varepsilon_{\text{res}}$ \textbf{then} \emph{break}.
    \ENDFOR
\STATE $\mathcal{N}_{\text{final}} \gets \mathcal{N}_{i}$,  $S_{\text{final}} \gets S^{(i)}$.
\STATE \emph{Return} $\mathcal{N}_{\text{final}},  S_{\text{final}}$.
\end{algorithmic}
\end{algorithm}

\section{Numerical experiments}
In this section, we present several 2-D and 3-D numerical examples, including continuous sources, discontinuous sources, and complex geometric sources, to evaluate the robustness and flexibility of the proposed algorithm.\\

\noindent\emph{Baseline Models.} To show that we can better reconstruct the source with our framework, we set up baseline models. In addition to the IA-RFM, MA-RFM we derived before, we consider the IRFM which uses fixed training integral points in $V_0$, as well as the PINN which approximates $\{\{u^{k}_{\theta}\}_{k_\text{min}}^{k_\text{ max}}, S_{\theta}\}$ based on the differential equation (\ref{helmholtz-DtN}).\\

\noindent{\emph{Date Generation.}} The artificial data can be generated by evaluating the observation operators \eqref{L1}-\eqref{L2} on given source. 
 Random noise is added to the artificial data in the magnitude and phase angle of the radiation field,
\begin{align}\label{u_noise}
    u^{\delta} _k: =u_k+\delta \epsilon_1 |u_k|e^{i\pi\epsilon_2 }, \     
    {\partial_\nu u^{\delta}_k}:={\partial_\nu u_k} +\delta \epsilon_1 \bigl|{\partial_\nu u_k}\bigr|e^{i\pi\epsilon_2 },  \ 
    \epsilon_1,  \epsilon_2 \sim \text{Uniform}(-1, 1), 
\end{align}
where $\delta$ is the noise level. The observation data $\{u_k, {\partial_\nu u_k}\}$ are acquired on $\Gamma$ for $k\in [k_{\text{min}}, k_{\text{max}}]$. In the following numerical examples,  unless otherwise specified,  we set $k_{\text{min}}=1$,  $k_{i+1}=k_i+ k_\delta$ for $i=1, \cdots, N-1$ with a uniform increment $k_\delta = 4$. By default, observations are performed on a rectangular boundary. Let $\Omega=[a, b]\times[c, d]$, and the sample step size is $h_x=\frac{b-a}{N_s-1}$, $h_y=\frac{d-c}{N_s-1}$, 
\begin{align*}
\mathcal{P}_{\text{mea}}= \bigcup_{i=0}^{N_s - 1}
\{(a, y)\cup(b, y)\mid y=c+i h_y\}
\cup\{(x, c)\cup(x, d)\mid x=a+ih_x \},\\
\bigl\{u(k_i, x_s)\mid k_i\in\{k_1, \ldots, k_N\}, x_s\in \mathcal{P}_{\text{mea}} \bigr\}, \quad 
\Bigl\{{\partial_\nu u(k_i, x_s)} \mid k_i\in\{k_1, \ldots, k_N\}, x_s\in \mathcal{P}_{\text{mea}}\Bigr\}.
\end{align*}
The unspecified $N_s$ refers to the number of observation points on one side of a rectangle or 1/4 of an arc.  {The IA-RFM threshold and max division level , max iteration  is fixed as $\epsilon_{\mathrm{IA\text{-}RFM}} = 1\times10^{-3}$, {\tt max\_level}=4, {\tt max\_iter}=5 for all experiments.
The accuracy of the reconstruction is evaluated on a test set $P_{\text{test}}$, which consists of a $300 \times 300$ uniform grid on $V_0$. }  The relative $\|\cdot\|_2$ error and  {pointwise absolute error} of $S$ are given by: 
\begin{align}\label{error}
E_{l^2}(S_M)=\frac{\sqrt{\frac{1}{n}\sum_{i=1}^{n}[S_M(\mathbf{x}_{i})-S_{\text{ex}}(\mathbf{x}_{i})]^{2}}}{\sqrt{\frac{1}{n}\sum_{i=1}^{n}S_{\text{ex}}^{2}(\mathbf{x}_{i})}}, \quad  {e(\mathbf{x}_i)=|S_M(\mathbf{x_i})-S_{\text{ex}}(\mathbf{x}_i)|,\quad \text{for} \ \mathbf{x}_i \in P_{\text{test}}.}
\end{align}
 {In the following experiments, we denote by $E_{l^2}(S^{(0)})$ and $E_{l^2}(S_{\text{final}})$ the relative errors of the solutions obtained from the IA-RFM phase and the MA-RFM phase, respectively.} To select $\lambda_{\text{reg}}$, we adopt the posterior L-curve \cite{hansen1992L-curve} method, which was initially proposed by Lawson and extended by Hansen, and the optimal regularization parameter corresponds to the corner of the curve.
 \begin{remark}
  It is worth noting that, although Algorithm~\ref{alg: two_phase} allows for multiple enhancement iterations ($I_{\text{max}}>1$), the stopping criterion is satisfied after only one
  iteration in all numerical examples reported in this work. This efficiency is largely due to the robust feature extraction and basis-initialization procedure described in
  Algorithm~\ref{alg: shape_detection}.%, which accurately selects the support candidates via $\mathcal{Q}_{\text{grad}}$ and $\mathcal{Q}_{\text{abs}}$. 
\end{remark}

\begin{figure}[ht]
    \centering
    \includegraphics[width=1\textwidth]{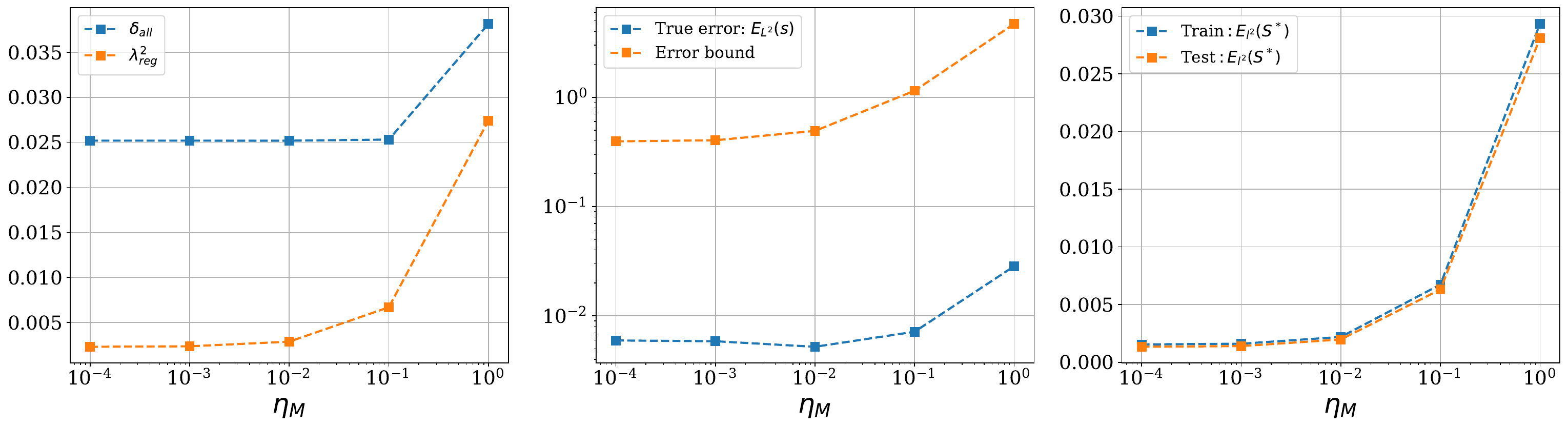}
    \caption{Example~\ref{ex:Prior parameters}: {\bf Left}) Prior parameters:  $\eta_\text{M}$ versus $\delta_{\text{all}}$; {\bf Middle}) $\lambda_{\text{reg}}^2$, $\|\boldsymbol{s}^{\delta}-\boldsymbol{s}^{*}\|_2$ and the theoretical error bounds; {\bf Right}) $\|\boldsymbol{s}^{\delta}-S^*_M\|_2$ for the training set versus the test set.}
    \label{fig:theory}
\end{figure}

\begin{figure}[ht]
    \centering
    \begin{subfigure}[b]{0.32\textwidth}
        \includegraphics[width=\textwidth]{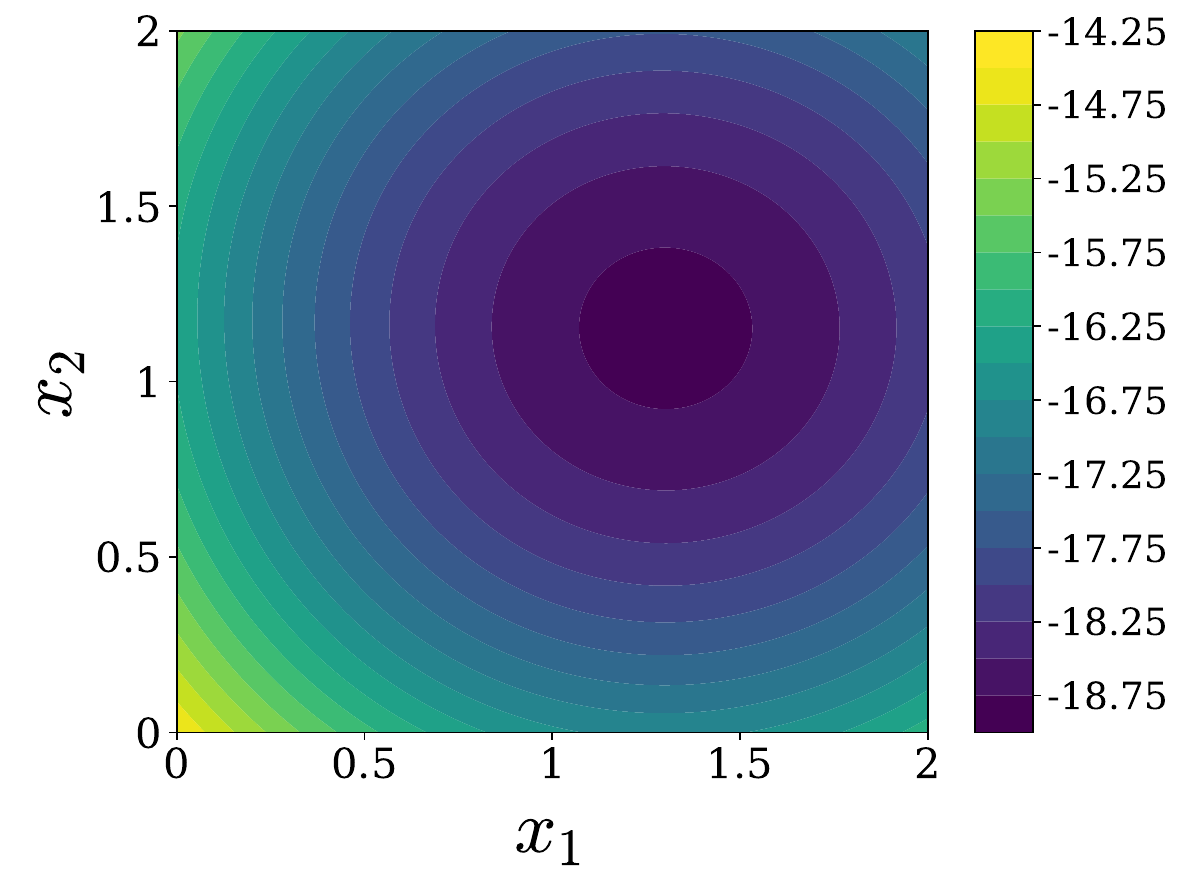}
        \caption{ $S_M^*(\boldsymbol{x})$ }
    \end{subfigure}
    \hfill
    \begin{subfigure}[b]{0.32\textwidth}
        \includegraphics[width=\textwidth]{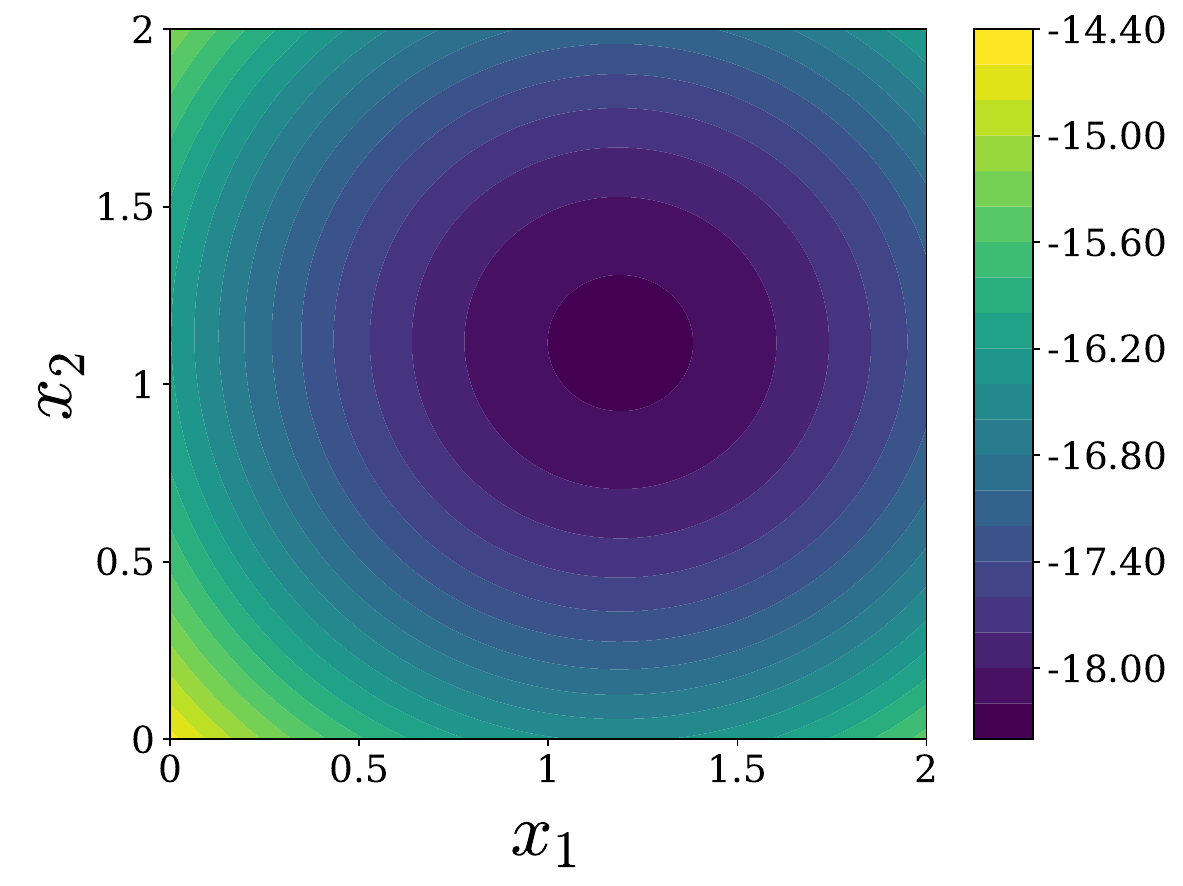}
        \caption{${S}_M^{\delta}(\boldsymbol{x})$}
    \end{subfigure}
    \hfill
    \begin{subfigure}[b]{0.32\textwidth}
        \includegraphics[width=\textwidth]{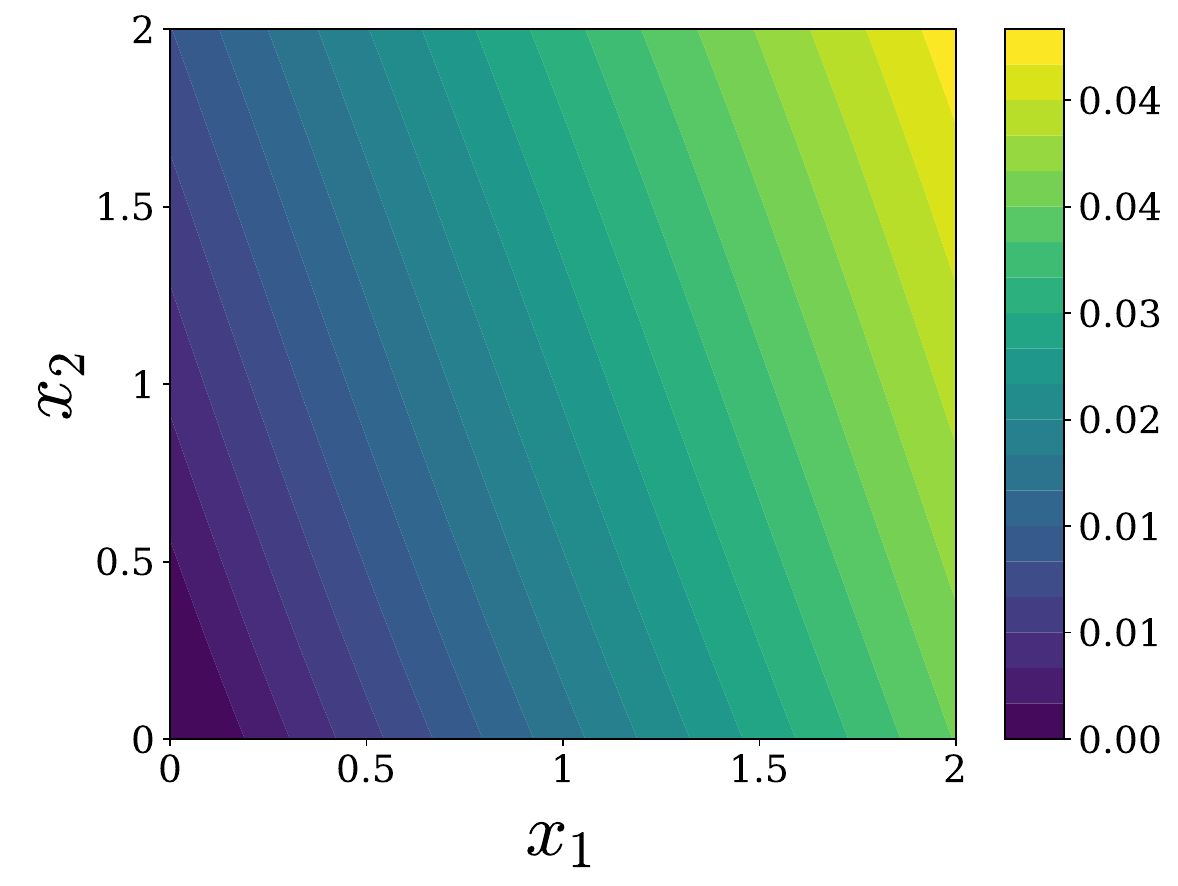}
        \caption{ {Pointwise relative error}}
    \end{subfigure}
    \hfill
    \caption{Example~\ref{ex:Prior parameters}: Prior parameters: IRFM results with $\eta_\text{M}=1$, $\nu=1$, $\delta=5\%$, $\delta_{\text{all}}=3.1\times10^{-2}$,  $M_0=3200$, $N_s=15$,  yield $3.01\%$ $E_{l^2}(S)$.}
    \label{fig:S_last}
\end{figure}

\begin{example}[\emph{Prior parameters}]\label{ex:Prior parameters}
In this example, we construct the source to validate Theorem \ref{Existence and Uniqueness}.

Firstly, we create a random vector $w$, and directly compute $\boldsymbol{s}^{*}=(\Psi_\text{M}^\top\Psi_\text{M})^{\nu} w$, $S^*_M(\boldsymbol{x})=\sum s_m^* \phi_m(\boldsymbol{x})$ after generating a random feature space that defines the operator $\Psi_\text{M}$. 
Secondly, we generate $U_{\text{true}}$ with a precisely controlled model inconsistency $\eta_\text{M}$.
The idea is based on the orthogonal decomposition of a vector space. $\forall  \ U\in\mathbb{R}^n$ can be decomposed into the direct sum of the column space of the operator $\Psi_\text{M}$, denoted as $\text{col}(\Psi_\text{M})$, and its left null space $\text{ker}(\Psi_\text{M}^\top)$.
\begin{align}
\mathbb{R}^n = \text{col}(\Psi_\text{M}) \oplus \ker(\Psi_\text{M}^\top).
\end{align}
Utilizing SVD, we have $\Psi_\text{M}=U\Sigma V^\top$. If $\text{rank}(\Psi_\text{M})=r$, the first $r$ columns of  $U\in \mathbb{R}^{n\times n}$ form the orthonormal basis for $\text{col}(\Psi_\text{M})$, while the last $n-r$ columns, denoted as $U_{\text{null}}$ form the orthonormal basis for $\text{ker}(\Psi_\text{M}^\top)$. $\boldsymbol{\eta_{\text{vec}}}$ is constructed as a random linear combination of the basis for the left null space: $\boldsymbol{\eta_{\text{vec}}}=U_{\text{null}} \cdot \boldsymbol{c}$, where $\mathbf{c}\in \mathbb{R}^{n-r}$ is a vector of random coefficients. To precisely control its norm to a target error value $\eta_\text{M}$ we normalize and scale it:
 \begin{align}
\boldsymbol{\eta}_{\text{vec}} \leftarrow \frac{\boldsymbol{\eta}_{\text{vec}}}{\|\boldsymbol{\eta}_{\text{vec}}\|_2} \times \eta_\text{M},
\end{align}
Finally, the synthetic data is generated by adding the explainable part $\Psi_\text{M} \cdot \boldsymbol{s}^{*}$
$$
U_{\text{true}}=\Psi_\text{M} \cdot \boldsymbol{s}^* + \boldsymbol{\eta}_{\text{vec}}.
$$
\noindent\emph{Experimental setup:}\ $V_0=[0, 2]^2$,  $\Omega=[-0.5,2.5]^2$. Randomly generate vector $w$, $N_s=15$, $\nu=1$, $C_{\nu}=1$, $\eta_{\text{M}} = 10^{-4}, 10^{-3}, 10^{-2}, 10^{-1}, 1$, $\delta=5\%$, $k_{\text{min}}=1, k_{\text{max}}=89$.\\ 
\noindent\emph{Hyperparameter settings:}\ The activation function used for IRFM is $\sin$ , $R_m=20$, $M_0=3200$,  initial integral mesh $ {N_{x_1}}= {N_{x_2}}=1$  with  $ {n_{x_1}}= {n_{x_2}}=100$ Gauss points per cell. \\

The selection of the regularization parameter is guided by Theorem \ref{Existence and Uniqueness}. Figure \ref{fig:theory} illustrates that both the total error $\delta_{\text{all}}$ and regularization term $\lambda_{\text{reg}}^2$ increase with $\eta_\text{M}$. Meanwhile, the actual reconstruction error $\|\boldsymbol{s}^{\delta}-\boldsymbol{s}^{*}\|_2$
is approximately one order of magnitude below the theoretical upper bound. The close correspondence between the training and test set errors demonstrates the framework's strong generalization capability. Figure \ref{fig:S_last} shows the reconstructed solution when $\eta_\text{M}=1$ with $E_{l^2}(S)=3.01\%$.
\end{example}

\begin{example}[\emph{Mountain shape source function}]\label{ex:Mountain shape source function}
In the example, we aim to reconstruct a mountain-shaped source function
$$
S(x_1, x_2)=1.1e^{-200((x_1-0.01)^2)+(x_2-0.12)^2}-100(x_2^2-x_1^2)e^{-90(x_1^2+x_2^2)}.
$$
We adopt the data generation procedure which generates Cauchy data on $\rho>R$  based on Dirichlet observations on $x=R$ based on an exterior extension from  \cite{zhang2015fourier} to compare. 
\begin{align}
    \label{artificial}
    \nu^{\delta}(k, x)=\sum_{n\in\mathbb{Z}}\frac{H_{n}^{(1)}\left(k\rho\right)}{H_{n}^{(1)}\left(kR\right)}\hat{u}_{k, n}^{\delta}\mathrm{e}^{\mathrm{i}n\theta}, 
    \quad
    \partial_{\nu}\nu^{\delta}(k, x)=\sum_{n\in\mathbb{Z}}k\frac{H_{n}^{(1)^{\prime}}\left(k\rho\right)}{H_{n}^{(1)}\left(kR\right)}\hat{u}_{k, n}^{\delta}\mathrm{e}^{\mathrm{i}n\theta}, \quad x\in\Gamma_{\rho}, \ \rho>R.
\end{align}
$\hat{u}_{k, n}^{\delta}$ is the Fourier coefficient of the data $u^{\delta}(k, R, \theta)$. All other parameters,  such as wavenumbers $K_N \cup \{k^*\} $ and the truncation number $N=2[\delta ^{-1/3}]$, where [X] denotes the largest integer that is smaller than X + 1,  are chosen to be identical to those in \cite{zhang2015fourier} (Thm 3.3,  Rem 3.1). Here we use $N_R$, $N_{\rho}$ to denote the total number of observations on $\Gamma_R$, and the total number of data points generated on $\Gamma_\rho$  {via the exterior extension 
(\ref{artificial})}, respectively.  \\

\noindent\emph{Experimental setup:}\ $V_0=[-0.3, 0.3]^2$,  $\Omega=B_{0.5
    }(0, 0)$, $\rho=0.6$.  The observation region is the circular arc $\Gamma_{R}=\partial \Omega$, and the generated region is $\Gamma_{\rho}$, $\delta=5\%$, truncation number $N=6$.\\
\noindent\emph{Hyperparameter settings:}\  The activation function used for IRFM and IA-RFM is {\it sine}, $R_m=20$, $M_0=3200$,  $ {N_{\rho}}=400$,  initial integral mesh $ {N_{x_1}}= {N_{x_2}}=5$  with  $ {n_{x_1}}= {n_{x_2}}=5$ per cell, $\epsilon_{\text{res}}=0.5\delta$.\\

In order to understand the reason for the failure of PINN, we analyze its prediction for the inverse source problem with a loss function similar to (\ref{RFM_LOSS_inverse_source}). To simplify the Loss and reduce the difficulty caused by the penalty term, set the observation position and the artificial boundary to be $\Gamma_{\rho}$, and then use the paradigm triangulation inequality to get $\mathcal{L}_{\text{PINN}}$ as follows: 
\begin{align}
    \label{PINN_LOSS_inverse_source}
\mathcal{L}_{\text{PINN}}&=\sum_{k=k_1}^{k_N}
\Bigl( \lambda_ {I} \sum_{\boldsymbol{x}_i \in V_0}\left|-(\Delta+k^2) u_\text{NN}^{k}(\boldsymbol{x}_i)-S_\text{NN}(\boldsymbol{x}_i)\right|^2
+ \lambda_{D}\sum_{\boldsymbol{x}_j\in \Gamma_{\rho}}\left|u^{k}_\text{NN}(\boldsymbol{x}_j)-u^{k}(\boldsymbol{x}_j)\right|^2 
\notag\\
&{}\qquad\quad +\lambda_{B}\sum_{\boldsymbol{x}_j\in \Gamma_{\rho}}\left|\mathcal{T}^{k}(u^{k}_\text{NN}(\boldsymbol{x}_j))-\partial_{\nu}u^{k}(\boldsymbol{x}_j)\right|^2 \Bigr).
\end{align}
The structures of $u_\text{NN}$ and $S_\text{NN}$ are [2,\ 50,\ 50,\ 2] and [2,\ 50,\ 50,\ 1], respectively. Set $n_{\text{pde}}=100^2$, $\lambda_{I}=1$, $\lambda_{B}=10$, $\lambda_{D}=10$, and we perform noise-free experiments, with the other parameters consistent with the settings above. Training is first performed using 30,000 ADAM iterations, then the subsequent training is performed using 5,000 L-BFGS iterations. From results shown in  {Figure \ref{fig:PINN_continuous_source_10}}, we observe that PINN fails to reconstruct the source. Since in this framework, not only do we need $N+1$ networks to approximate the scattered fields $u$ and $S$, which is a very large number of parameters and difficult to balance with a multitude of penalties, but also $u$ is more oscillatory as $k$ gets larger, yet neural networks have difficulty in approximating high frequency.
\begin{figure}[ht]
    \centering
    \begin{subfigure}[b]{0.35\textwidth}
        \includegraphics[width=\textwidth]{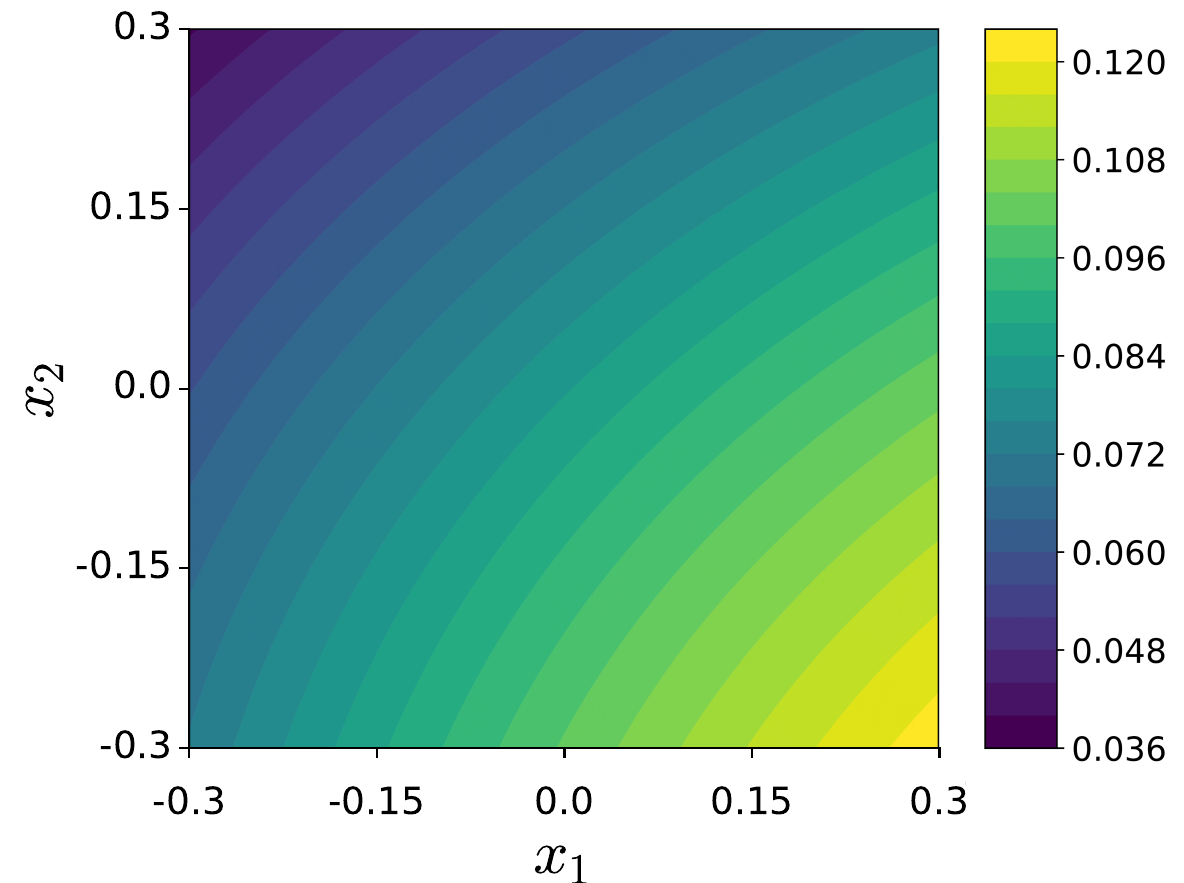}
        \caption{ }
    \end{subfigure}
    \hspace{5ex}
    \begin{subfigure}[b]{0.34\textwidth}
        \includegraphics[width=\textwidth]{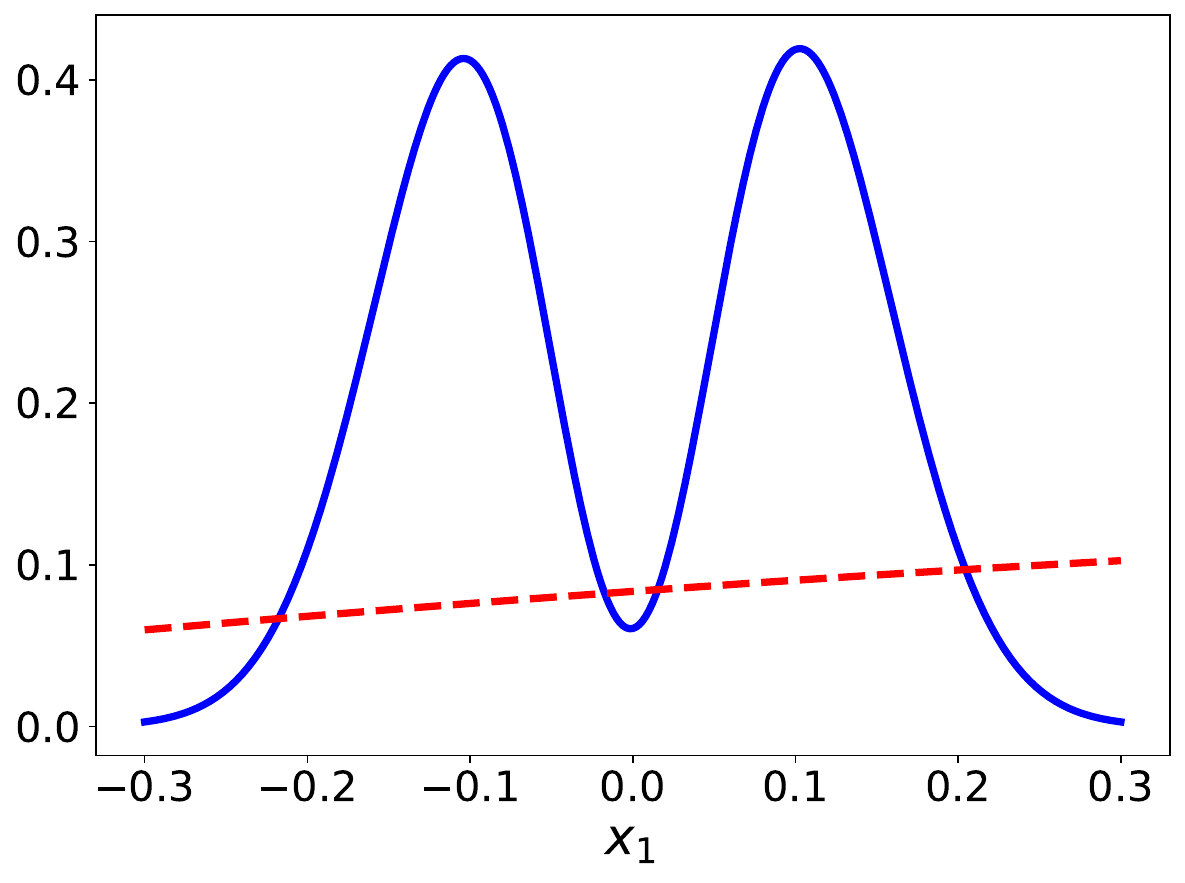}
        \caption{ }
    \end{subfigure}
    \caption{Example \ref{ex:Mountain shape source function}: Mountain shape source function: (a). PINN solution with $u_\text{NN}$: [2,\ 50,\ 50,\ 2]  and $S_\text{NN}$: [2,\ 50,\ 50,\ 1]. (b). The true solution (\textcolor{blue}{\rule[0.5ex]{1.2em}{0.5pt}}) and PINN solution (\textcolor{red}{\rule[0.5ex]{0.6em}{0.8pt}\hspace{0.2em}\rule[0.5ex]{0.6em}{0.8pt}}) at($x_1$,0).}
    \label{fig:PINN_continuous_source_10}
\end{figure}

According to Table \ref{tab:FM, IRFM, IA-RFM},  IA-RFM shows an overwhelming advantage in computational efficiency compared with the conventional Fourier method (FM) and IRFM. The number of generated points $N_{\rho}$ and integration points $n_{\text{integral}}$ are reduced by one and two orders of magnitude, respectively. Through its adaptive integration,  it saves approximately 80\% of the integration computations compared with IRFM,  while simultaneously achieving a comparable or even superior $l^2$ error. Subsequently,  with noise level $\delta=10\%$ and $N_{\rho}$=100,  and all other parameters remaining the same,  the $E_{l^2}(S)$ is 1.38\% with IA-RFM, which is shown in Figure \ref{fig:continuous_source_10}.

In addition,  it is worth mentioning that this experiment satisfies the convergence condition ($\epsilon_{\text{res}}<\delta$) in the first stage of the algorithm, and does not enable the basis function enhancement in the second stage, which suggests that IA-RFM is sufficient for solving this problem.\\
\begin{table}[htbp]
    \centering
    \caption{Comparison of FM,  IRFM,  and IA-RFM with $N=6$, $\delta = 5\%$ for mountain shape source.}
    \begin{tabular}{ccccc}
        \toprule
        Method & $N_R$& $N_{\rho}$ & $n_{\text{integral}}$ &  $E_{l^2}(S)$ \\
        \midrule
        \multirow{4}{*}{FM}
        &50&5000&$800^2$&1.620\% \\
        &100 &5000 &$800^2$ &1.150\%\\
        &200&5000&$800^2$ &0.824\%\\
        &400&5000 &$800^2$ &0.629\% \\
        \hline
        \multirow{4}{*}{IRFM}
        &50 &400 &$100^2$& 0.735\% \\
        &100& 400&$100^2$ &0.638\% \\
        &200 &400 &$100^2$&0.580\% \\
        &400&400 &$100^2$ &0.560\% \\
        \bottomrule
        \multirow{4}{*}{IA-RFM}
        &50 &400 &2500& 0.726\% \\
        &100& 400&2500 &0.803\% \\
        &200 &400 &2500 &0.562\% \\
        &400&400 &2500 &0.550\% \\
        \bottomrule
    \end{tabular}
    \label{tab:FM, IRFM, IA-RFM}
\end{table}

\begin{figure}[ht]
    \centering
    \begin{subfigure}[b]{0.32\textwidth}
        \includegraphics[width=\textwidth]{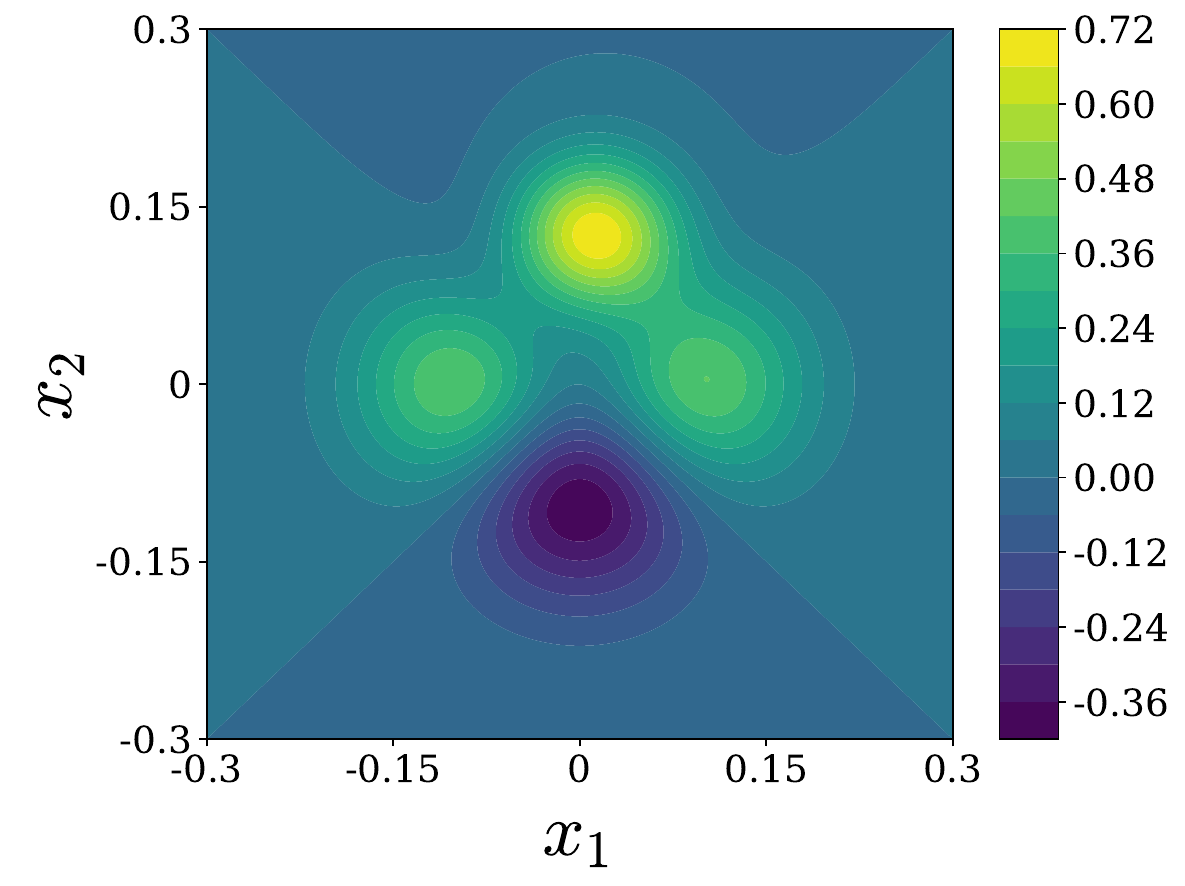}
        \caption{ True source }
       %\label{S_ex}
    \end{subfigure}
    \hfill
    \begin{subfigure}[b]{0.32\textwidth}
        \includegraphics[width=\textwidth]{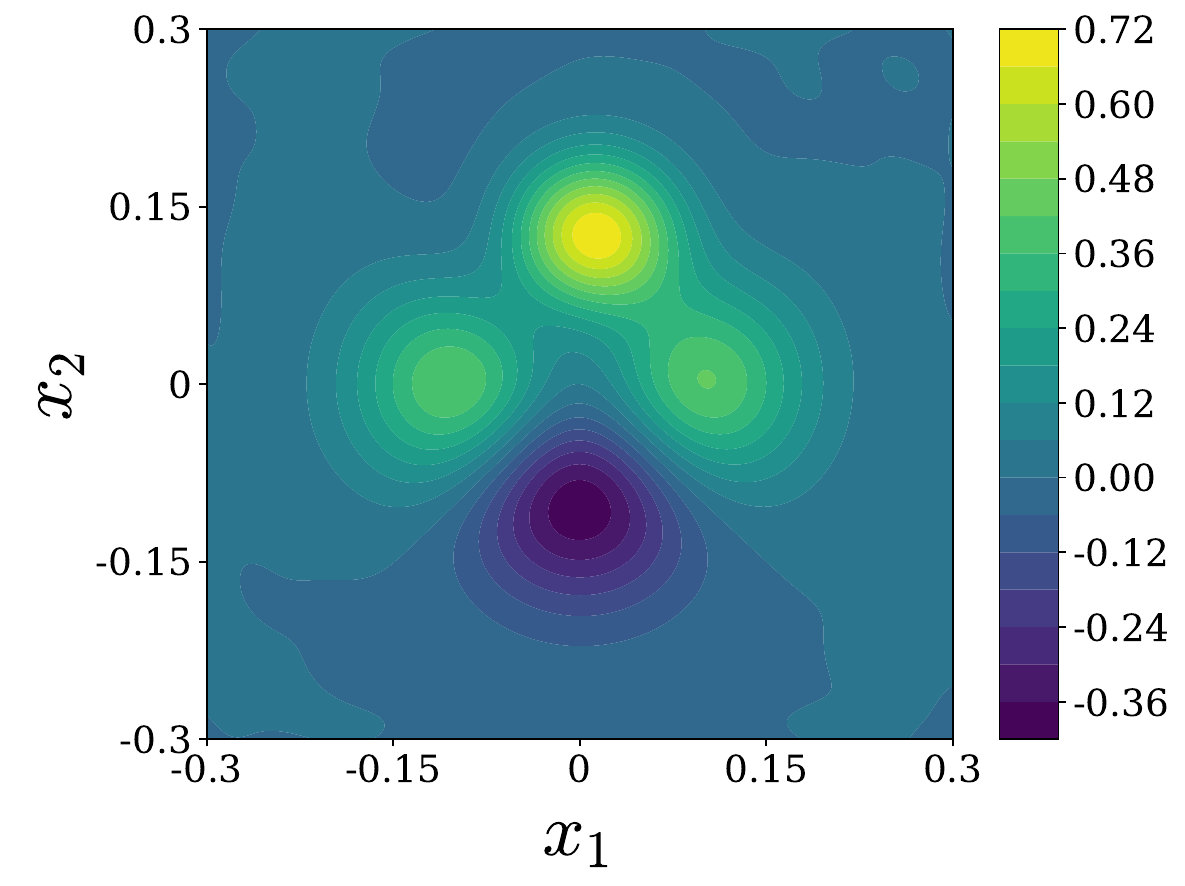}
        \caption{Reconstructed source  }
        %\label{S_num}
    \end{subfigure}
    \hfill
    \begin{subfigure}[b]{0.32\textwidth}
        \includegraphics[width=\textwidth]{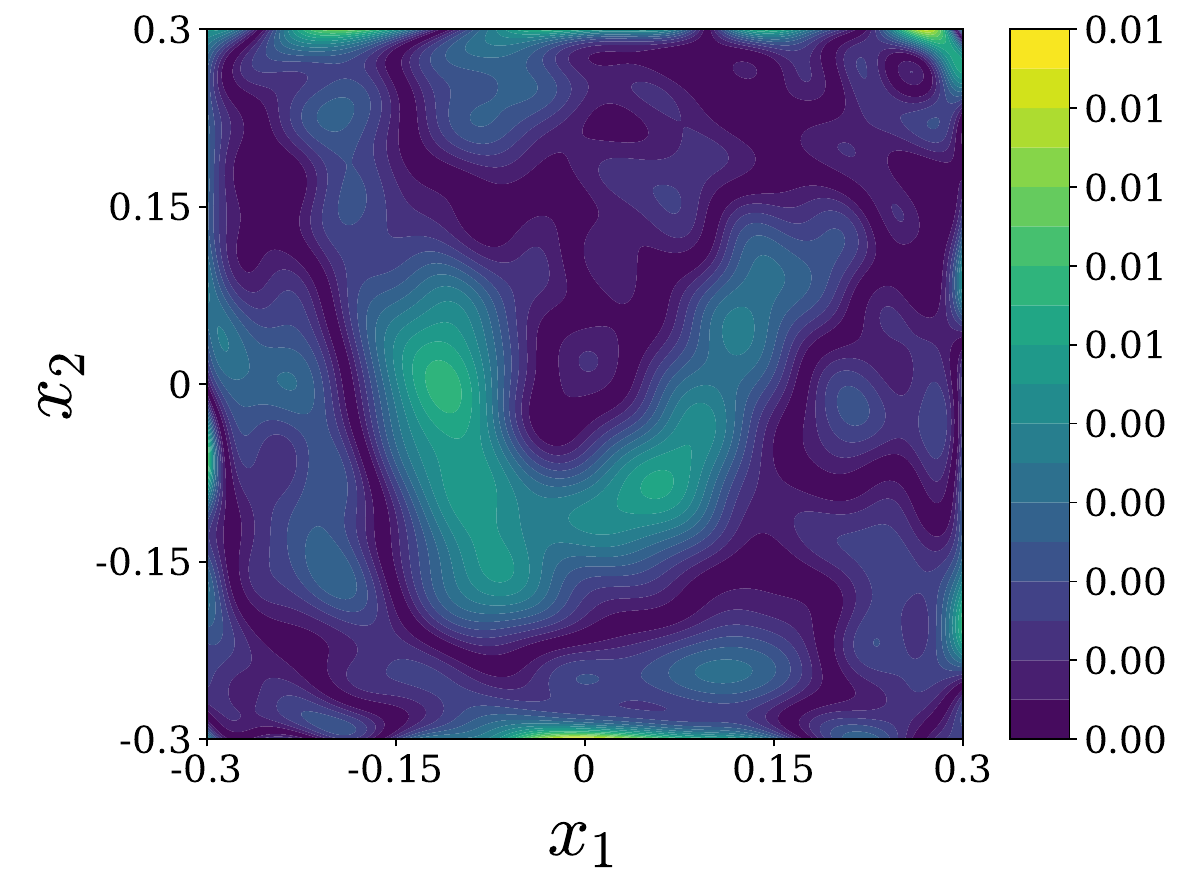}
        \caption{ {Pointwise absolute error}}
        \label{continous_source}
    \end{subfigure}
    \hfill
    \caption{Example \ref{ex:Mountain shape source function}: Mountain shape source function: IA-RFM results with $\delta=10\%$,  $M_0=3200$,  $N_s=50$, $N_{\rho}$=100, $\lambda_{\text{reg}}^2$ = 1.00e-2 yield 1.38\% $E_{l^2}(S)$.}
    \label{fig:continuous_source_10}
\end{figure}
\end{example}
 {Before presenting the MA-RFM hyperparameter settings for Examples 4.3--4.9, we summarize the outputs of Algorithm \ref{alg: shape_detection} in Table \ref{tab:extracted_parameters}. The table reports the detected geometric parameters, residual indicators, morphology label $\mathcal{T}_k$, and basis profile $\sigma_k$ obtained from the first stage coarse reconstruction. For completeness, Table \ref{tab:Comparison of IA-RFM and MA-RFM} provides the corresponding  comparison between the IA-RFM stage and the MA-RFM stage.}\\
\begin{example}[\emph{Discontinuous source}]\label{ex:Gibbs phenomenon}
In the following numerical experiments,  a more challenging scenario is considered where the support of the source $\tau$,  is a proper subset of $V_0$,  $\tau \subsetneqq V_0$,  which means the source is discontinuous within the region $V_0$.
$$S(x_1, x_2)=\mathcal{X}_{B_r}, \quad  B_r(0.5, 0.5)=\{(x_1, x_2)| (x_1-0.5)^2+(x_2-0.5)^2 \leq r^2 ,  r=0.2\}.$$
\noindent \emph{Experimental setup:}\ $V_0=[0, 1]^2$, $\Omega=[-0.5, 1.5]^2$. $k_{\text{min}}$=1,  $k_{\text{max}}=89$,  $N_s$=15.

 \noindent \emph{Hyperparameter settings:}  {The activation function Tanh is used for IRFM and IA-RFM. In MA-RFM, the shape type $\mathcal{T}_1$ and the basis profile $\sigma_1$ are selected by Algorithm~\ref{alg:
  shape_detection}. In this example, one cluster is detected and the corresponding morphology-aware basis is chosen as
  \begin{align}
  \label{one_circle}
      \phi_1^{(j)}(\boldsymbol{x};K_1^{(j)},\boldsymbol{\theta}_1^{(j)})
      =
      \text{sigmoid}\left(
      \tilde{K}_1^{(j)}
      \cdot
      \left(
      1-\sum_{i=1}^2\left|\frac{x_i-c_{1,i}^{(j)}}{L_{1,i}^{(j)}}\right|^2
      \right)
      \right),
      \qquad j=1,2,\cdots,M_1.
  \end{align}
  We take $R_m=20$, $M_1=M_0$, and $M_{\text{total}}=M_0+M_1$. The initial integral mesh is $N_{x_1}=N_{x_2}=4$ with $n_{x_1}=n_{x_2}=3$, and the parameter ranges are set by $K_1^{(j)}\sim U(1000,20000)$, $\epsilon_c=5\%$, $
  \epsilon_L=5\%-10\%$, and $\epsilon_{\text{res}}=0.5\delta$.}\\

 {Table \ref{tab:compare_one_circle} presents a quantitative comparison among IRFM, IA-RFM, and MA-RFM.
 With the same number of basis functions, IA-RFM achieves a comparable error level while using much fewer integration points than IRFM.
However, both single-phase methods eventually exhibit error saturation around 19\%, confirming that mesh refinement alone cannot capture the strong discontinuity. 
In contrast, the proposed MA-RFM breaks this accuracy barrier, achieving an error of around 13.50\%. Notably, MA-RFM with $M=800$ outperforms the finest IRFM result ($M=6400$) in accuracy, while requiring computational time (8.5s vs. 35.9s), demonstrating the necessity and efficiency of the two-phase strategy.}

  To further validate robustness, we conducted a noise-resistance experiment with parameters set as follows: $M_0=1600$, $M_1=1600$, $\epsilon_c=5\%$, and $\epsilon_r=5\%-10\%$. As shown in Table~\ref{tab:one_circle_noise}, the
  inclusion of morphology-adaptive basis functions leads to a relative accuracy improvement of 40\% compared to the IA-RFM baseline. Visual details are provided in Figure~\ref{fig:one_circle_MA-RFM}, which show that the final
  adaptive mesh $\mathcal{C}_{\text{final}}$ and the gradient field $\mathcal{Q}_{\text{grad}}$ align well with the true boundary of the source, while the L-curve confirms the appropriateness of the selected regularization
  parameter. Moreover, the MA-RFM results in Figure~\ref{fig:one_circle_IA-RFM} under different noise levels $\delta$ show that, even at the relatively high noise level $\delta=10\%$, the relative error is only 13.96\%.  {For the
  case $\delta=5\%$, the quantitative comparison between the IA-RFM and MA-RFM stages is summarized in Table \ref{tab:Comparison of IA-RFM and MA-RFM}, while the extracted geometric parameters together with the corresponding enriched basis information are reported in
  Table \ref{tab:extracted_parameters} in Appendix C.}

\begin{table}[htbp]
    \centering
    \caption{Example \ref{ex:Gibbs phenomenon}: Comparison for discontinuous source with $\delta=5\%$ with $t_{\text{grad}}=t_{\text{abs}}=1/3$.}
    \begin{tabular}{cccccc}
        \toprule
        Method & $M_{\text{total}}$ & $n_{\text{integral}}$ &  $\lambda^2_{\text{reg}}$ & $E_{l^2}(S)$ & Time(second) \\
        \midrule
        \multirow{4}{*}{IRFM}
        &800 &$100^2$ &1.00e-4 &22.29\% &5.2 \\
        &1600&$100^2$ &1.00e-4 &20.56\% &8.1\\
        &3200 &$100^2$ &1.00e-4 &19.26\% &15.8 \\
        &6400&$100^2$ &1.00e-4 &18.55\% &35.9\\
        \hline
        \multirow{4}{*}{IA-RFM}
        &400 &$6948$ &1.00e-4&26.67\%&5.9 \\
        &800&$6678$ &1.00e-4 &22.45\% &10.2 \\
        &1600 &$6273$ &1.00e-4 &20.00\% &18.4\\
        &3200&$6165$ &1.00e-4 &19.10\%&38 \\
        \hline
        \multirow{4}{*}{MA-RFM}
        &800 &$6948$ &1.00e-4 &13.53\% & 8.5\\
        &1600&$6678$ &1.00e-5 &13.51\% &20.7 \\ 
        &3200 &$6273$ &1.00e-5 &13.24\% &41.7 \\
        &6400&$6165$ &1.00e-4&13.48\% &120.2 \\
        \bottomrule
    \end{tabular}
    \label{tab:compare_one_circle}
\end{table}
  \begin{table}[htbp]
    \setlength{\tabcolsep}{3pt}
      \centering
      \caption{Example \ref{ex:Gibbs phenomenon}: The shape parameters and reconstruction errors with $t_{\text{grad}}=t_{\text{abs}}=1/3$.}
      \begin{tabular}{cccccc}
          \toprule
          $\delta$ & 0.5\% & 1\% & 5\% & 10\% & 20\% \\
          \midrule
          $n_{\text{integral}}$ & 6327 & 6408 & 6165 & 6192 & 6165 \\
$\hat{\boldsymbol c}$ & (5.00e-1,5.00e-1) & (5.00e-1,5.00e-1) & (5.00e-1,5.00e-1) & (5.00e-1,5.00e-1) & (5.00e-1,5.01e-1) \\
$\hat{\boldsymbol L}$ & (2.05e-1,2.05e-1) & (2.03e-1,2.03e-1) & (2.03e-1,2.03e-1) & (2.03e-1,2.05e-1) & (2.03e-1,2.03e-1) \\
          $E_{l^2}(S^{(0)})$ & 20.38\% & 20.03\% & 20.03\% & 20.28\% & 19.94\% \\
          $\lambda^2_{\text{MA-RFM}}$ & 1.00e-5 & 1.00e-4 & 1.00e-3 & 1.00e-3 & 1.00e-2 \\
          $E_{l^2}(S_{\text{final}})$ & 13.43\% & 13.61\% & 14.02\% & 13.96\% & 14.50\% \\
          \bottomrule
      \end{tabular}
      \label{tab:one_circle_noise}
  \end{table}
\begin{figure}[ht]
    \centering
    \begin{subfigure}[b]{0.275\textwidth}
        \includegraphics[width=\textwidth]{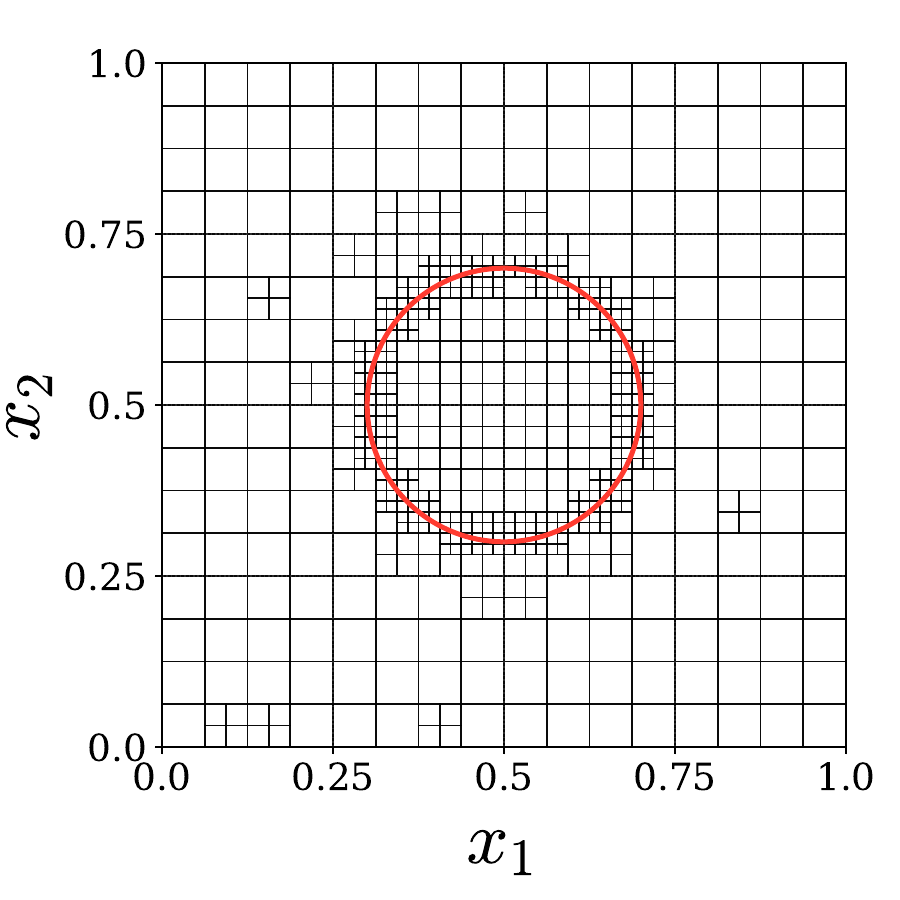}
        \caption{$\mathcal{C}_{\text{final}}$}
        %\label{S_ex}
    \end{subfigure}
    \hfill
    \begin{subfigure}[b]{0.33\textwidth}
        \includegraphics[width=\textwidth]{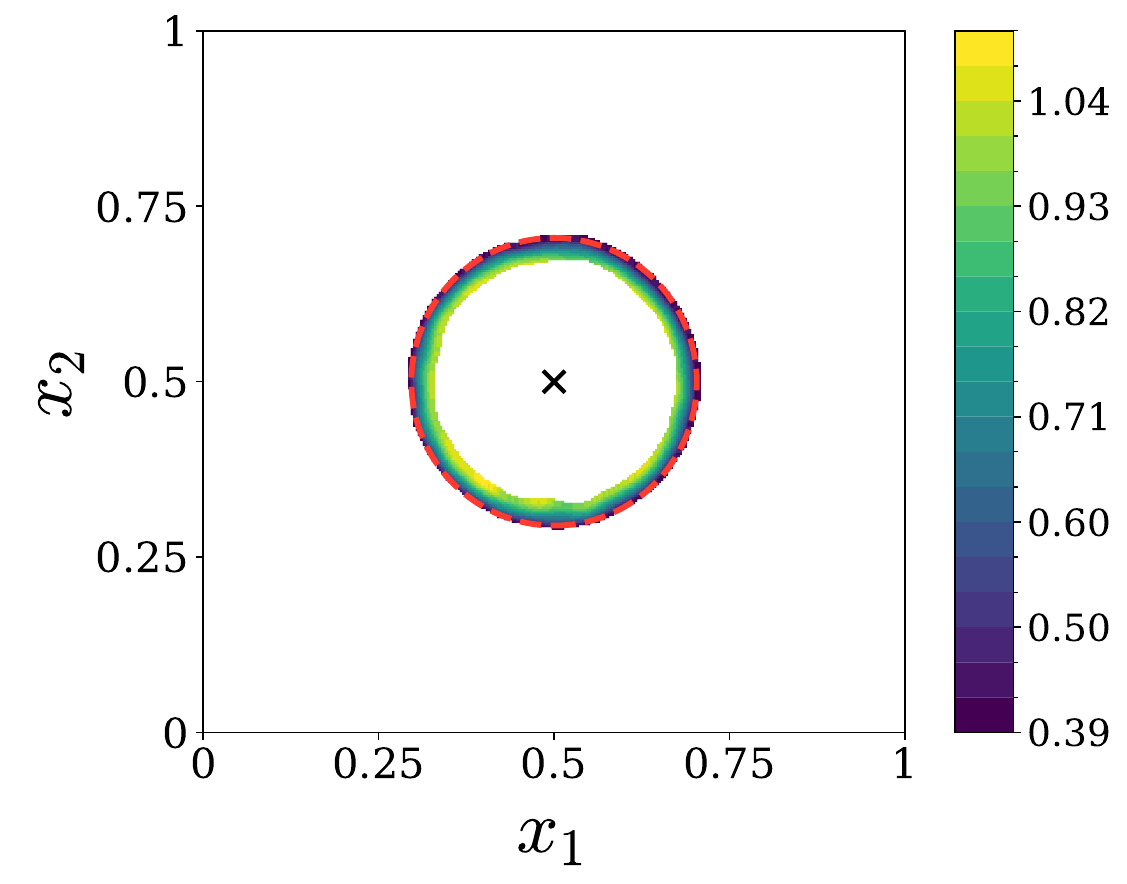}
\caption{$\mathcal{Q}_{\text{grad}}$}
        %\label{S_num}
    \end{subfigure}
    \hfill
    \begin{subfigure}[b]{0.345\textwidth}
        \includegraphics[width=\textwidth]{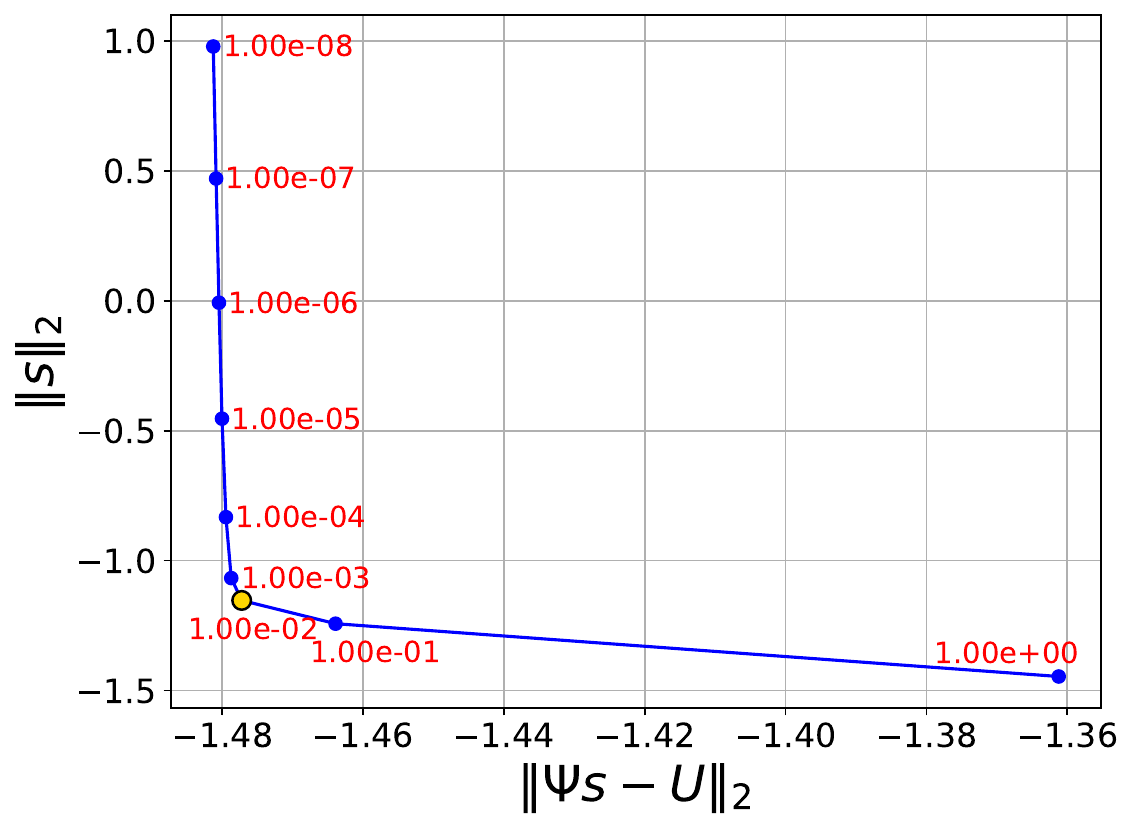}
         \caption{L-curve}
    \end{subfigure}
    \caption{Example \ref{ex:Gibbs phenomenon}: Discontinuous source: (a) shows the $\mathcal{C}_{\text{final}}$ with IA-RFM, and the red line represents the true interface. (b) displays the $\mathcal{Q}_{\text{grad}}$ with  {$t_{\text{grad}}=t_{\text{abs}}=1/3$}. The red dashed lines indicate the detected boundary, while the “×” symbols denote the detected center obtained by Algorithm \ref{alg: shape_detection}. (c) records the L-curve  $\text{log}_{10}(\|\Psi \cdot \boldsymbol{s}-U\|_2)$-$\text{log}_{10}(\|\boldsymbol{s}\|_2)$ with different $\lambda_{\text{reg}}^2$.}
    \label{fig:one_circle_IA-RFM}
\end{figure}
\begin{figure}[ht]
    \centering
    \begin{subfigure}[b]{0.32\textwidth}
\includegraphics[width=\textwidth]{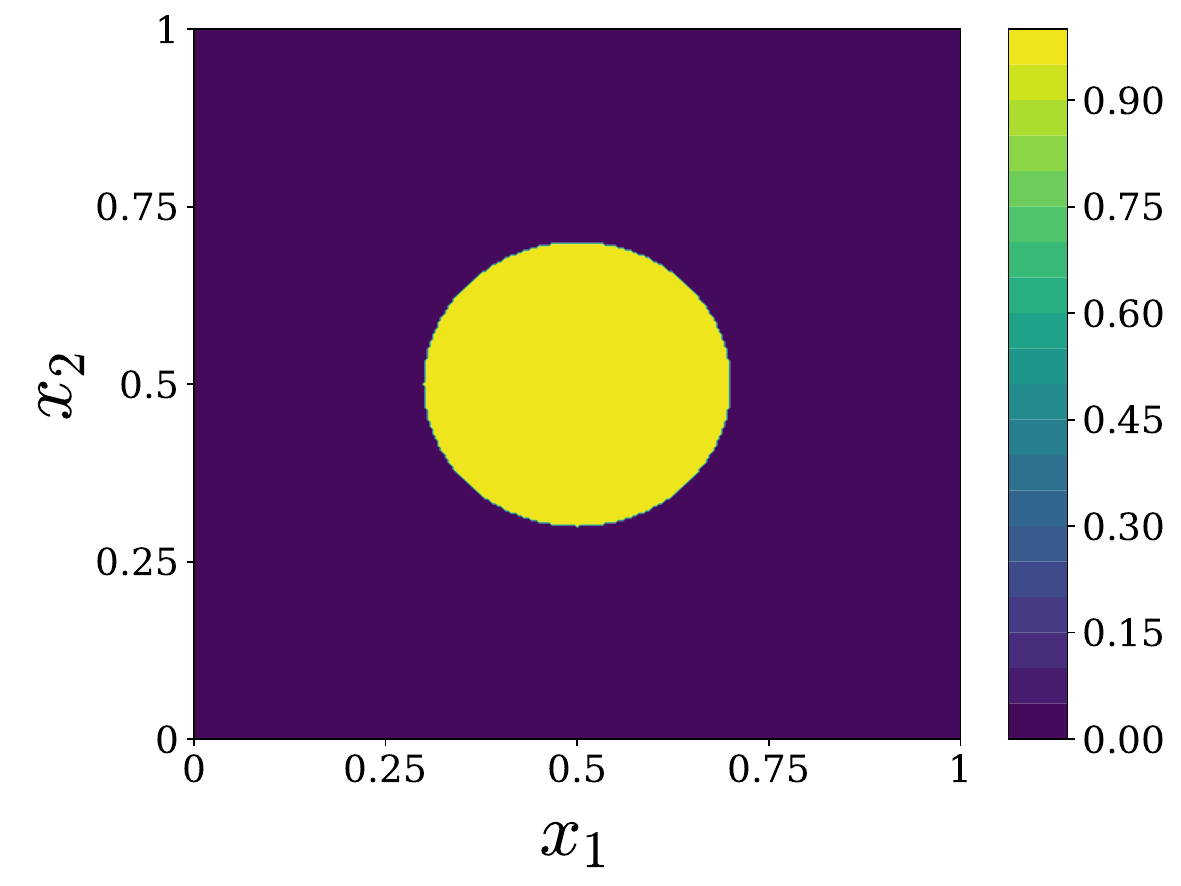}
        \caption{ True source }
        %\label{S_ex}
    \end{subfigure}
    \hfill
    \begin{subfigure}[b]{0.32\textwidth}
        \includegraphics[width=\textwidth]{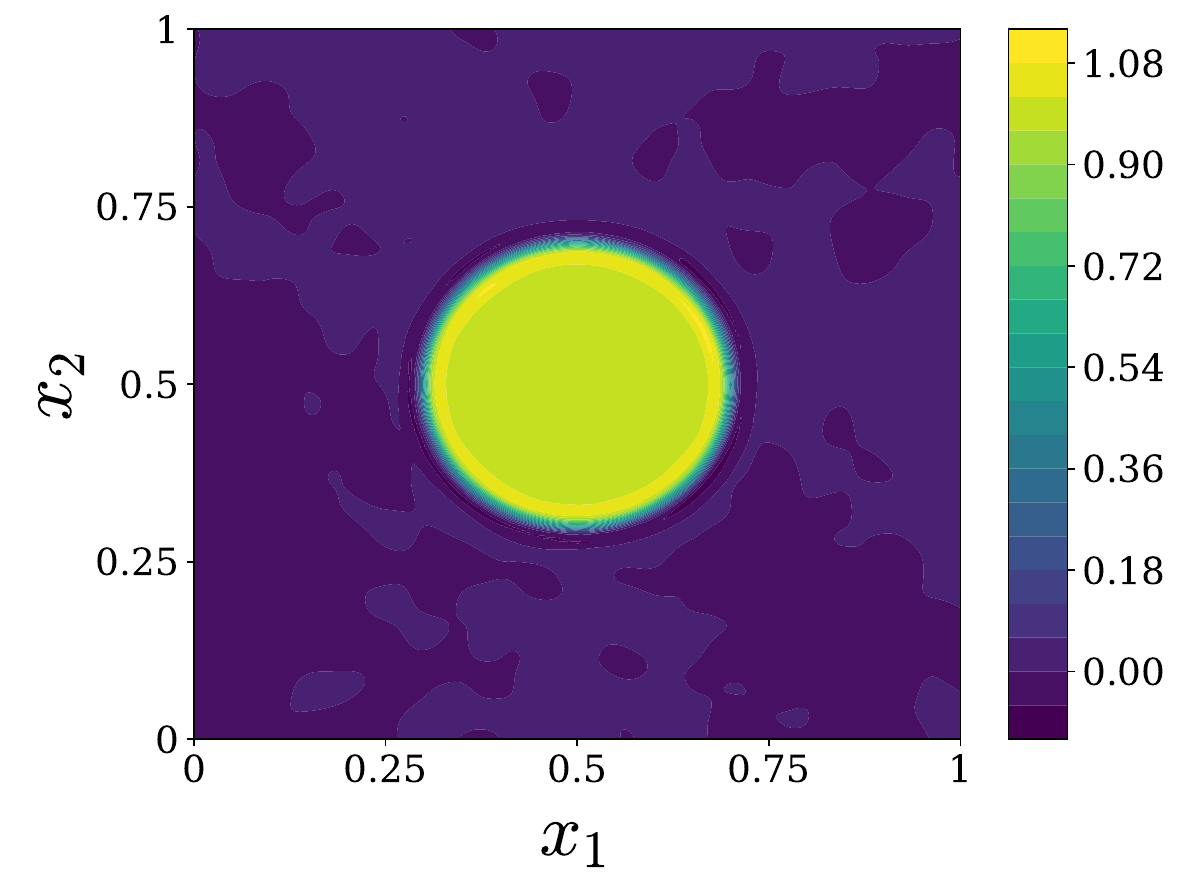}
            \caption{Reconstructed source  }
        %\label{S_num}
    \end{subfigure}
    \hfill
    \begin{subfigure}[b]{0.32\textwidth}
        \includegraphics[width=\textwidth]{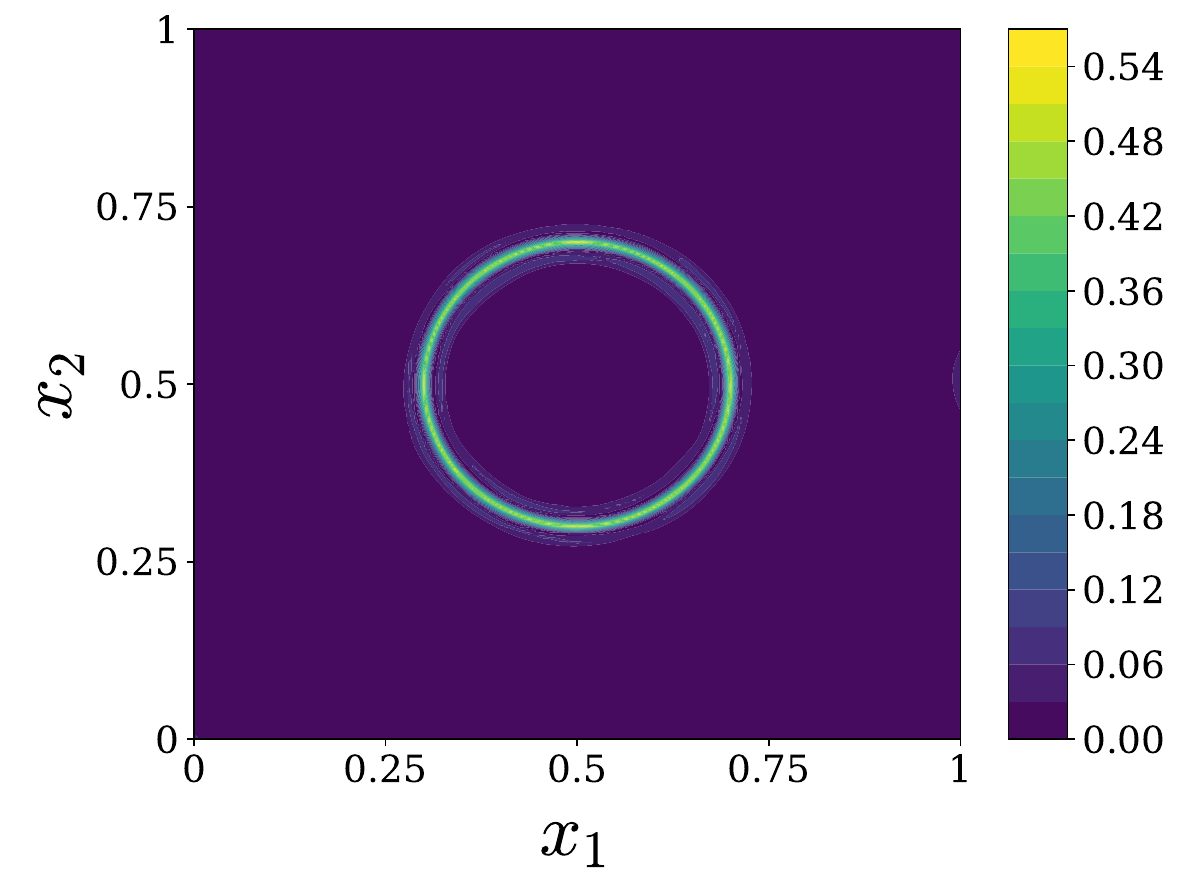}
         \caption{ {Pointwise absolute error}}
    \end{subfigure}
    \caption{Example \ref{ex:Gibbs phenomenon}: Discontinuous source: MA-RFM results with $\delta=10\%$, $M_{\text{total}}=3200$, $N_s=15$, $\lambda_{\text{reg}}^2 $= 1.00e-4 yield 13.96\% $E_{l^2}(S)$.}
    \label{fig:one_circle_MA-RFM}
\end{figure}
\end{example}
\begin{example}[\emph{Two circle sources}]\label{ex:Two circle sources}
 Consider the reconstruction of the discontinuous source function defined in the rectangular domain $V_0$.
 \begin{align*}
    S(x_1, x_2)=\left\{\begin{array}{ccc}0.5\exp(-550r_1^2),&\quad\mathrm{in}\ B_r(\hat{x}_0, 0),\\0.5\exp(-550r_2^2),&\quad\mathrm{in}\ B_r(\bar{x}_0, 0),\\0,\quad\text{elsewhere.}\end{array}\right.
\end{align*}
$r_1=\sqrt{(x_1-\hat{x}_0)^2+x_2^2},\   r_2=\sqrt{(x_1-\bar{x_0})^2+x_2^2}$ and $B_r(x_0, 0)$=$\{\boldsymbol{x}:  (x_1-x_0)^2+x_2^2 \leq r^2 \}$.\\

\noindent \emph{Experimental setup:}\ $V_0=[-0.30, 0.30]^2$, $\Omega=[-0.35, 0.35]^2$, $\hat{x}_0=-0.06$, $\bar{x}_0=0.08$, $r=0.06$, $N_s=20$,  $k_{\text{min}}=1$,  $k_{\text{max}}=77$.

\noindent \emph{Hyperparameter settings:}\ The initial activation function for MA-RFM is Tanh with $R_m=5$.  {Two clusters are detected and the corresponding morphology-aware basis functions are chosen as two families of truncated Gaussian basis function}.
    \begin{align*}
    \phi_{\alpha}^{(j)}(\boldsymbol{x};K^{(j)}_\alpha,\boldsymbol{\theta}_\alpha^{(j)})&=\text{sigmoid}\left(\tilde{K}_\alpha^{(j)} \cdot \left(1- \sum_{i=1}^2\left|\frac{x_i - c_{\alpha,i}^{(j)}}{L_{\alpha,i}^{(j)}}\right|^2\right)\right)
    \cdot \exp\left(-\tilde{v}_\alpha^{(j)} \cdot \sum_{i=1}^2\left|\frac{x_i - c_{\alpha,i}^{(j)}}{L_{\alpha,i}^{(j)}}\right|^2\right) \\
    &\alpha=1, 2,  \ j=1, 2, \cdots, M_\alpha.
    \end{align*}
    The exponential component shapes the morphology of the source,  with the parameter $v$ controlling its decay rate. $K_\alpha^{(j)} \sim U(1000,20000)$,  $v_\alpha^{(j)} \sim U(0,1000)$. $M_0=2400$,  $M_1=2000$,  $M_2=2000$,  initial integral mesh $N_{x_1}=N_{x_2}=4$ with $n_{x_1}=n_{x_2}=3$, $\epsilon_c=3\%$,  $\epsilon_{L}=3-10\%$.
\\

 {Table \ref{tab:two_circle_non_constant_transposed} summarizes the estimated shape parameters, the L-curve-selected regularization parameter, and the relavtive error. Figure \ref{Cell:two_circle_nonconstant} depicts the mesh evolution of IA-RFM, highlighting adaptive refinement near jumps and high solution magnitudes. The final reconstruction performance is visualized as heatmaps in Figure \ref{fig:two_circle_nonconstant}, which shows that MA-RFM maintains a low relative error $E_{l^2}(S)=7.08\%$ even with $\delta=20\%$.}  {For the
  case $\delta=5\%$, the quantitative comparison between the IA-RFM and MA-RFM stages is summarized in Table \ref{tab:Comparison of IA-RFM and MA-RFM}, while the extracted geometric parameters together with the corresponding enriched basis information are reported in
  Table \ref{tab:extracted_parameters} in Appendix C.}\\
  \begin{table}[htbp]
  \setlength{\tabcolsep}{2pt}
      \centering
      \caption{The shape parameters and reconstruction errors for two circle sources.}
      \begin{tabular}{c|ccccc}
          \toprule
          $\delta$ & 0.5\% & 1\% & 5\% & 10\% & 20\% \\
          \midrule
          $n_{\text{integral}}$ & 2844 & 2952 & 3006 & 3006 & 3870 \\
$\hat{\boldsymbol c}_1$ & (-6.08e-2,1.47e-5) & (-6.04e-2,8.27e-5) & (-6.06e-2,1.24e-4) & (-6.04e-2,4.29e-5) & (-6.03e-2,1.09e-4) \\
$\hat{\boldsymbol c}_2$ & (8.04e-2,-4.06e-5) & (8.06e-2,-2.28e-4) & (8.04e-2,9.14e-6) & (8.07e-2, 3.59e-4) & (8.04e-2,-3.53e-4) \\
$\hat{\boldsymbol L}_1$ & (6.40e-2,6.60e-2) & (6.40e-2,6.60e-2) & (6.40e-2,6.60e-2) & (6.40e-2,6.60e-2) & (6.40e-2,6.40e-2) \\
$\hat{ \boldsymbol L}_2$ & (6.40e-2,6.60e-2) & (6.40e-2,6.60e-2) & (6.40e-2,6.60e-2) & (6.40e-2,6.50e-2) & (6.40e-2,6.60e-2) \\
          \midrule
          $E_{l^2}(S^{(0)})$ & 9.92\% & 9.77\% & 10.06\% & 10.54\% & 12.89\% \\
          $\lambda_{\text{MA-RFM}}^2$ & 1.00e-6 & 1.00e-5 & 1.00e-4 & 1.00e-3 & 1.00e-3 \\
          $E_{l^2}(S_{\text{final}})$ & 4.72\% & 5.42\% & 6.04\% & 6.44\% & 7.08\% \\
          \bottomrule
      \end{tabular}
\label{tab:two_circle_non_constant_transposed}
  \end{table}

\begin{figure}[ht]
    \centering
\includegraphics[width=1\textwidth]{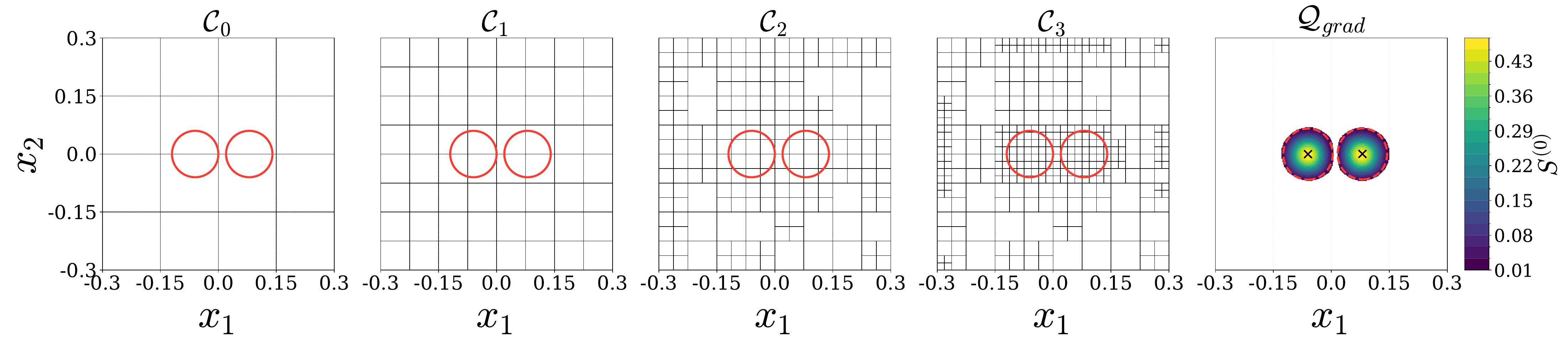}
    \caption{Example \ref{ex:Two circle sources}:\ Two circle sources: grid division diagram for IA-RFM with $\delta=10\%$, $M_0=2400$, $\text{Tanh}$, $R_m=5$. The last one shows the $\mathcal{Q}_{\text{grad}}$ with  {$t_{\text{grad}}$=1/3}. The red dashed lines indicate the detected boundary, while the “×” symbols denote the detected center.}
    \label{Cell:two_circle_nonconstant}
\end{figure}

\begin{figure}[H]
    \centering
    \begin{subfigure}[b]{0.32\textwidth}
        \includegraphics[width=\textwidth]{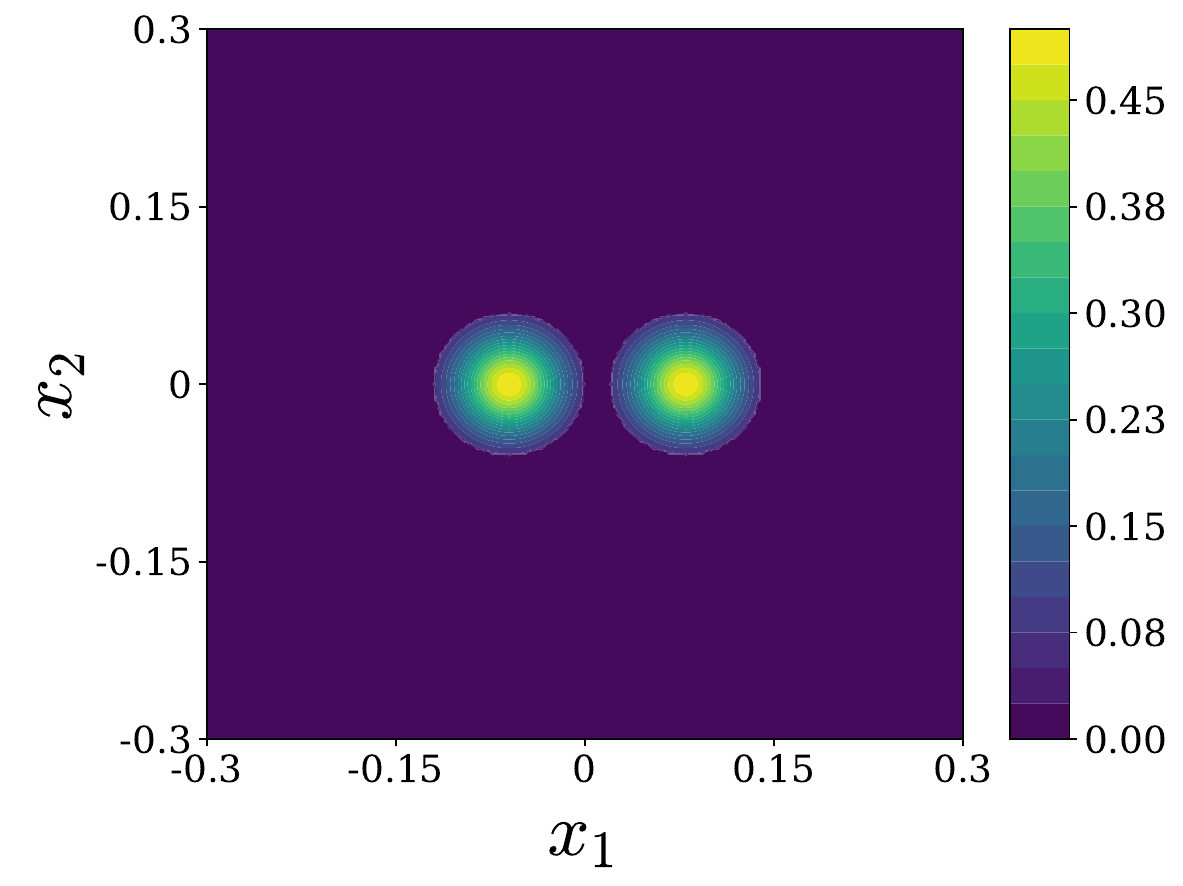}
        \caption{True source}
        %\label{S_ex}
    \end{subfigure}
    \begin{subfigure}[b]{0.32\textwidth}
        \includegraphics[width=\textwidth]{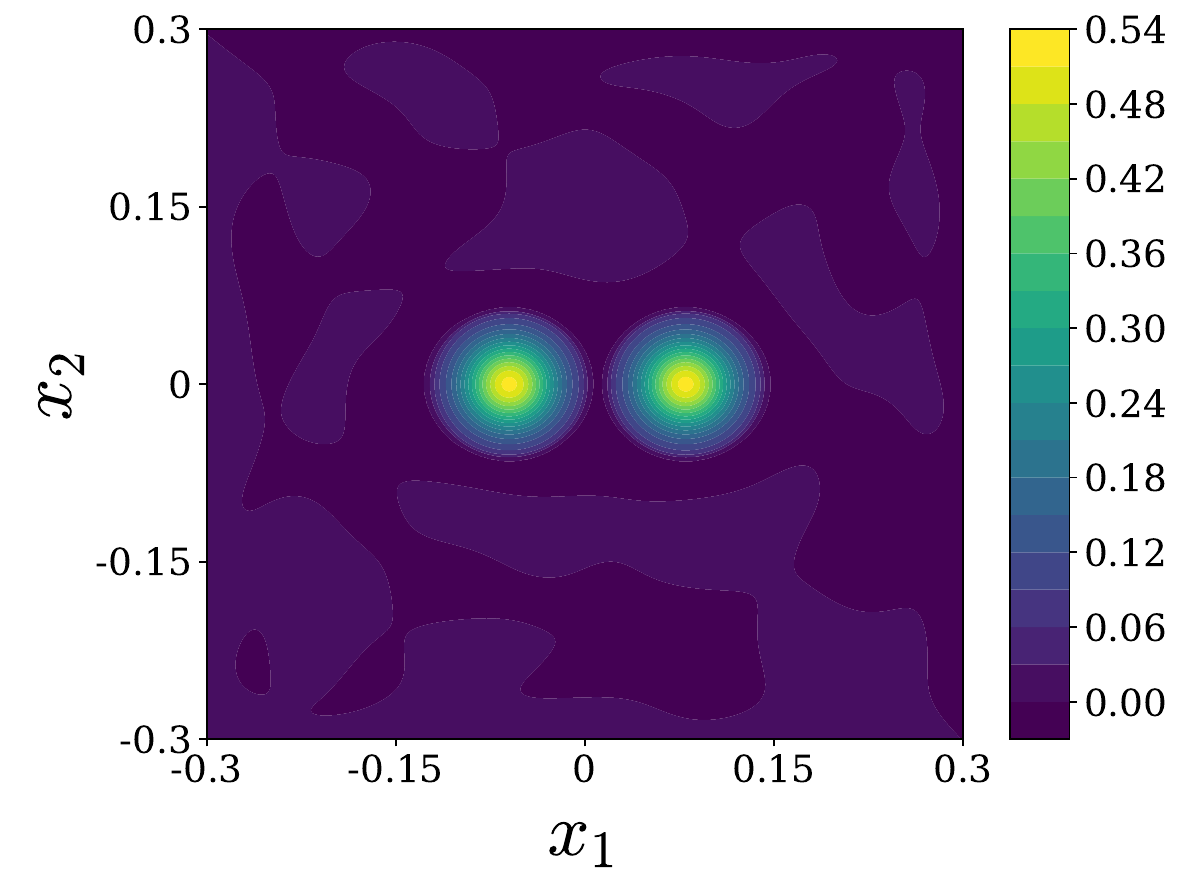}
        \caption{Reconstructed source}
        % \caption{Numerical source  }
        %\label{fig:one_circle_one_rec_com}
    \end{subfigure}
    \begin{subfigure}[b]{0.32\textwidth}
        \includegraphics[width=\textwidth]{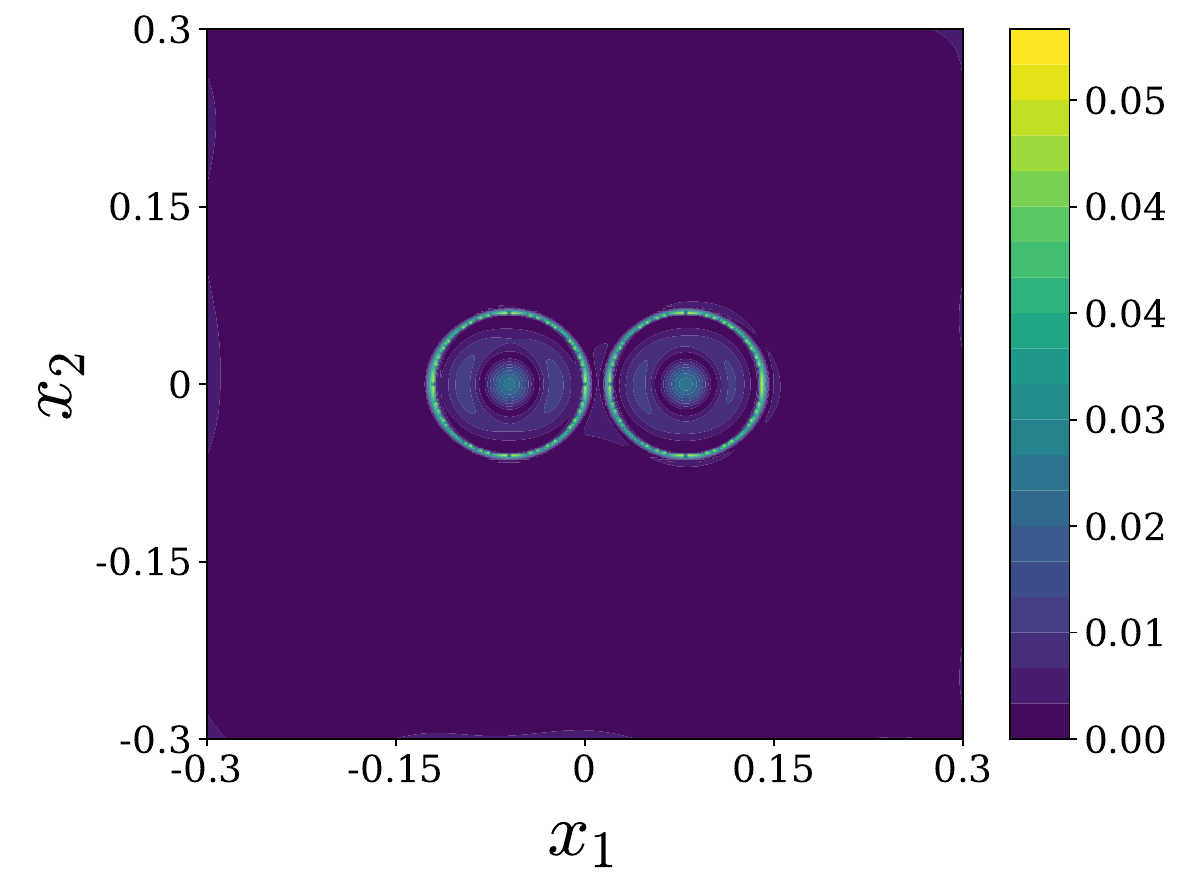}
        \caption{ {Pointwise absolute error}}
    \end{subfigure}
    \caption{Example \ref{ex:Two circle sources}:\ Two circle sources: MA-RFM results with $\delta=10\%$, $M_{\text{total}}=6400$, $N_s=20$ and $\lambda_{\text{reg}}^2 $= 1.00e-3 yield an 6.44\% $E_{l^2}(S)$. }
    \label{fig:two_circle_nonconstant}
\end{figure}
\end{example}

\begin{example}[{One rectangle and one circle}]\label{ex:One rectangle and one circle}The spatial proximity of two sources poses a significant challenge for the inverse source problem. This difficulty arises because the fine,  high-frequency details of the small gap between them are smoothed away during wave propagation. In what follows,  we consider a numerical example with closely spaced sources
$$S(x_1, x_2)=\mathcal{X}_B-\mathcal{X}_C,$$  $B=\{(x_1, x_2)|\ (x_1- {0.71})^2+(x_2-0.5)^2 \leq r^2 ,  r=0.2\}$,  $C=\{(x_1, x_2)|\ 0.29\le x_1 \le 0.49,  0.3\le x_2 \le 0.7 \}$.\\
\noindent \emph{Experimental setup:}  $V_0=[0, 1]^2$,  $\Omega=[-0.5, 1.5]^2$. $k_{\text{min}}$=1,  $k_{\text{max}}=89$,  $N_s$=10,  $\delta=5\%$.\\
\noindent \emph{Hyperparameters settings:}\ The activation function for IRFM and IA-RFM is Tanh with $R_m=20$,  and for MA-RFM,  the basis functions are Tanh and sigmoid, corresponding to elliptical and rectangular shape functions  {detected by Algorithm \ref{alg: shape_detection}},  respectively. $M_1^{(i)}=M_2^{(i)}=0.5M_0$. 
\begin{align*}
& \phi_1^{(j)}(\boldsymbol{x};K_1^{(j)},\boldsymbol{\theta}_1^{(j)}) = \text{sigmoid}\left(\tilde{K}_1^{(j)} \cdot \left(1- \sum_{i=1}^2\left|\frac{x_i - c_{1,i}^{(j)}}{L_{1,i}^{(j)}}\right|^2\right)\right)
    ,\  j=1, 2, \cdots, M_1. \notag\\
    &\phi_{2}^{(j)}(\boldsymbol{x};K_2^{(j)},\boldsymbol{\theta}_2^{(j)})=\text{sigmoid}  \!\left(
    \hat{K}_\alpha^{(j)}
    \left(
    1-\max_{i\in\{1,2\}}
    \left|\frac{x_i-c_{2,i}^{(j)}}{L_{2,i}^{(j)}}\right|
    \right)
    \right),\ j=1, 2, \cdots, M_2. \notag
\end{align*}
 {For the phase 2, instead
of purely accumulating basis functions, we employ a refinement strategy referred to Remark 3. $M^{(i)}_1=M^{(i)}_2=0.5M_0$.}  $\epsilon_c=5\%$,  $\epsilon_{L}=5-10\%$. The initial mesh is $N_{x_1}=N_{x_2}=4$ with $n_{x_1}=n_{x_2}=3$,  $K_1^{(j)},K_2^{(j)}\sim U(1000,20000)$. \\

 {It is worth noting that in this instance, we take $\mathcal{Q}=\mathcal{Q}_{\text{grad}}\cap \mathcal{Q}_{\text{abs}}$ with $t_{\text{grad}}=t_{\text{abs}}=1/3$, in order to distinguish between two sources that are particularly close in distance yet differ significantly in value. Consistent with the results in Table \ref{tab:compare_one_circel_one_rec}, MA-RFM demonstrates superior performance even for this complex configuration with close proximity. While single phase methods saturate around 20\% error, MA-RFM effectively breaks this barrier achieving 15.27\% when $M=1600$.}

 { A noise experiment is conducted to test the stability and robustness of MA-RFM. Set $M_0=1600$, $M_1^{(1)}=M_2^{(1)}=2400$. The mesh evolution, $\mathcal{Q}_{\text{grad}}$ and MA-RFM reconstruction results with $\delta=5\%$ are illustrated in Figure \ref{process:one_circle_one_rec} and Figure \ref{fig:one_circle_one_rec_close}, respectively.}
  {Table \ref{tab:one_circle_rec_const} summarizes the results for different noise. Despite the complex geometry, the adaptive algorithm accurately retrieves the shape parameters and maintains a stable mesh size $n_{\text{integral}} \approx 7200$ and reduces the relative error from a saturated level of 23\% to 14\%--16\%.}  {See Tables \ref{tab:Comparison of IA-RFM and MA-RFM} and \ref{tab:extracted_parameters} for the corresponding stagewise comparison and detected basis information for the case $\delta=5\%$.}
\begin{table}[htbp]
    \centering
    \caption{Compare of IRFM and MA-RFM with $\delta=5\%$ for one rectangle and one circle.}
    \begin{tabular}{cccccc}
        \toprule
        Method & $M_{\text{total}}$ & $n_{\text{integral}}$ &  $\lambda^2$ & $E_{l^2}(S)$  &Time(s) \\
        \midrule
        \multirow{4}{*}{IRFM}
        &800 &$100^2$ &1.00e-4&25.79\%&4.8\\
        &1600&$100^2$ &1.00e-5&22.44\%&7.9\\
        &3200 &$100^2$ &1.00e-6&21.15\%&14.6\\
        &6400&$100^2$ &1.00e-6&20.25\%&34.9\\
        \hline
        \multirow{4}{*}{IA-RFM}
        &400 &$6381$ &1.00e-4&29.60\%&6.2\\
        &800&$7029$ &1.00e-4&25.84\%&11.5\\
        &1600 &$7245$ &1.00e-4&23.16\%&20.1\\
        &3200&$7245$ &1.00e-4&21.14\%&38.4\\
        \hline
        \multirow{4}{*}{MA-RFM}
        &800 &$6381$&1.00e-3&15.11\%&10.1\\
        &1600&$7029$&1.00e-4&15.27\%&32.5\\
        &3200 &$7245$&1.00e-4&15.04\%&66.2\\
        &6400&$7245$&1.00e-4&14.84\%&198.4\\
        \bottomrule
    \end{tabular}
    \label{tab:compare_one_circel_one_rec}
\end{table}

\begin{figure}[ht]
    \centering
    \includegraphics[width=\textwidth]{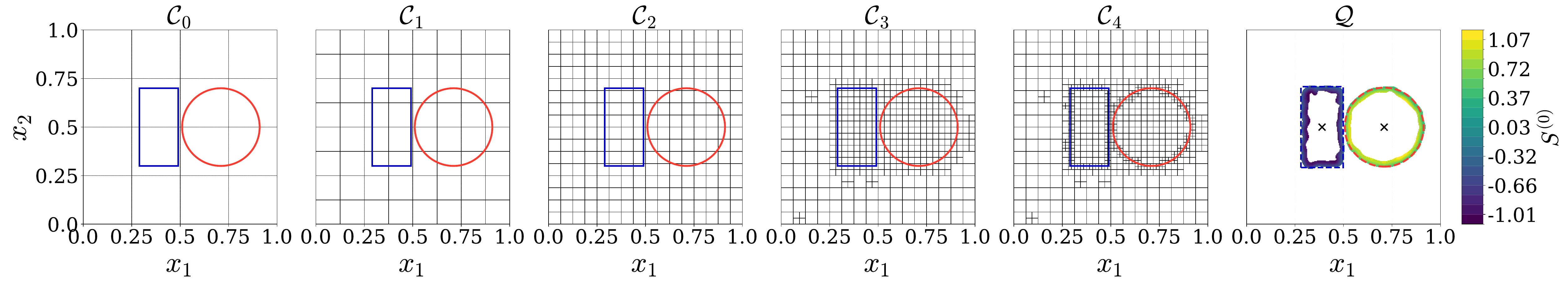}
    \caption{Example \ref{ex:One rectangle and one circle}:\ One rectangle and one circle: grid division diagram for IA-RFM with $\delta=5\%$, $M_0=2400$, $\text{Tanh}$, $R_m=2$. The last one shows the  $\mathcal{Q}=\mathcal{Q}_{\text{abs}}\cap \mathcal{Q}_{\text{grad}}$, with  {$t_{\text{grad}}=t_{\text{abs}}=1/3$}. The red and blue dashed lines indicate the detected boundary, while the “×” symbols denote the detected center obtained by Algorithm \ref{alg: shape_detection}.}
    \label{process:one_circle_one_rec}
\end{figure}
\begin{figure}[ht]
    \centering
    \begin{subfigure}[b]{0.32\textwidth}
    \includegraphics[width=\textwidth]{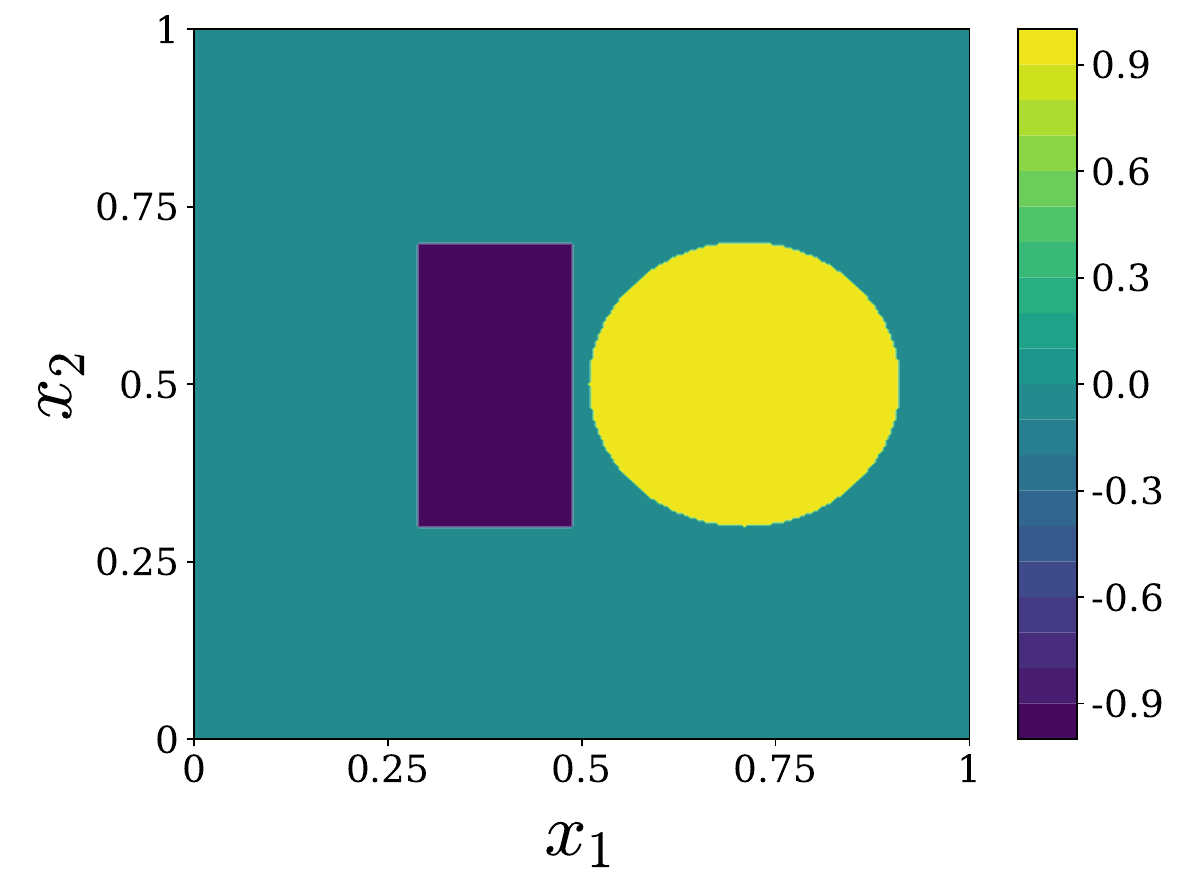}
        \caption{True source}
        %\label{S_ex}
    \end{subfigure}
    \begin{subfigure}[b]{0.32\textwidth}
    \includegraphics[width=\textwidth]{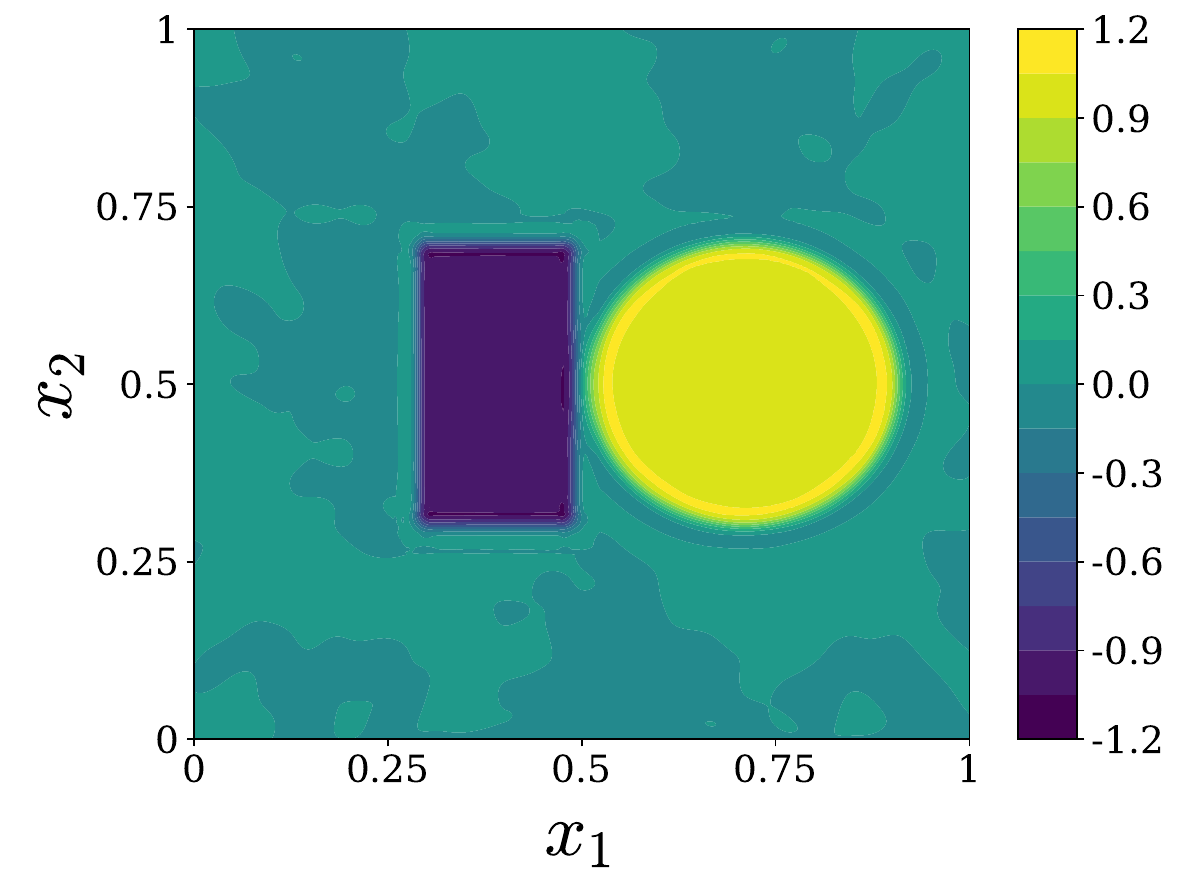}
        \caption{Reconstructed source}
    \end{subfigure}
    \begin{subfigure}[b]{0.32\textwidth}
    \includegraphics[width=\textwidth]{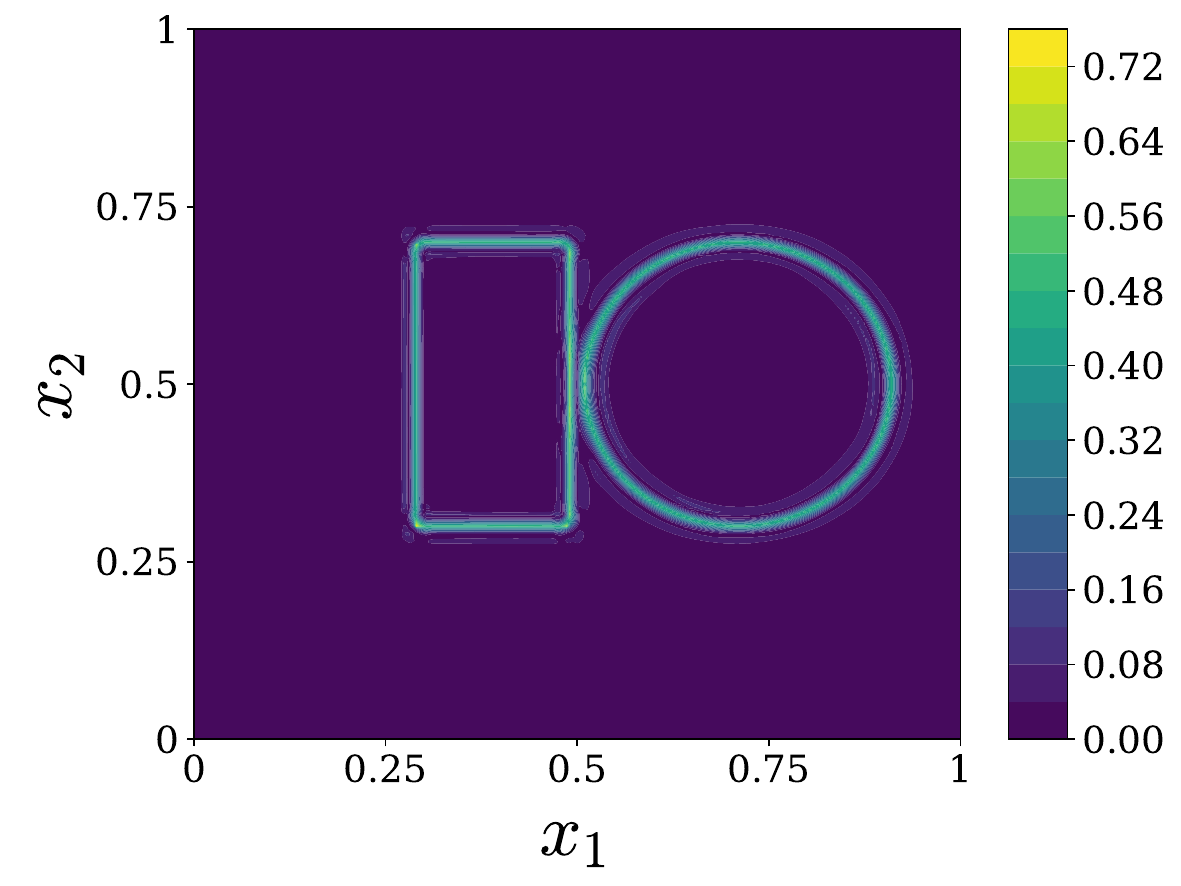}
        \caption{ {Pointwise absolute error}}
    \end{subfigure}
    \caption{Example \ref{ex:One rectangle and one circle}:\ One rectangle and one circle: Reconstruction with $\delta=5\%$, $M_0=1600$, $M_1=M_2=2400$, $N_s=10$ and $\lambda_{\text{reg}}^2 $= 1.00e-3 yielding 15.28\% $E_{l^2}(S)$ using MA-RFM.}
\label{fig:one_circle_one_rec_close}
\end{figure}
  \begin{table}[htbp]
      \centering
        \setlength{\tabcolsep}{3pt}
      \caption{The shape parameters and reconstruction errors for one rectangle and one circle.}
      \begin{tabular}{c|ccccc}
          \toprule
          $\delta$ & 0.5\% & 1\% & 5\% & 10\% & 20\% \\
          \midrule
          $n_{\text{integral}}$ & 7083 & 7461 & 7245 & 7245 & 7137 \\
$\hat{\boldsymbol c}_{1}$ & (3.90e-1,4.98e-1) & (3.90e-1,5.01e-1) & (3.90e-1,5.01e-1) & (3.90e-1,5.01e-1) & (3.89e-1,4.97e-1) \\
$\hat{\boldsymbol c}_{2}$ & (7.11e-1,4.99e-1) & (7.10e-1,5.00e-1) & (7.10e-1,5.00e-1) & (7.10e-1,4.99e-1) & (7.11e-1,4.99e-1) \\
$\hat{\boldsymbol L}_{1}$ & (2.17e-1,4.10e-1) & (2.17e-1,4.07e-1) & (2.20e-1,4.10e-1) & (2.13e-1,4.10e-1) & (2.13e-1,4.13e-1) \\
$\hat{\boldsymbol L}_{2}$ & (2.02e-1,2.02e-1) & (1.98e-1,2.03e-1) & (2.02e-1,2.05e-1) & (2.02e-1,2.02e-1) & (2.02e-1,2.00e-1) \\
          \midrule
          $E_{l^2}(S^{(0)})$ & 23.44\% & 23.68\% & 23.04\% & 22.69\% & 23.08\% \\
          $\lambda^2_{\text{MA-RFM}}$ & 1.00e-5 & 1.00e-5 & 1.00e-3 & 1.00e-3 & 1.00e-2 \\
          $E_{l^2}(S_{\text{final}})$ & 14.46\% & 14.84\% & 15.28\% & 15.34\% & 15.79\% \\
          \bottomrule
      \end{tabular}
      \label{tab:one_circle_rec_const}
  \end{table}

\end{example}

\begin{figure}[H]
    \centering
    \begin{subfigure}[b]{0.25\textwidth}
\includegraphics[width=\textwidth]{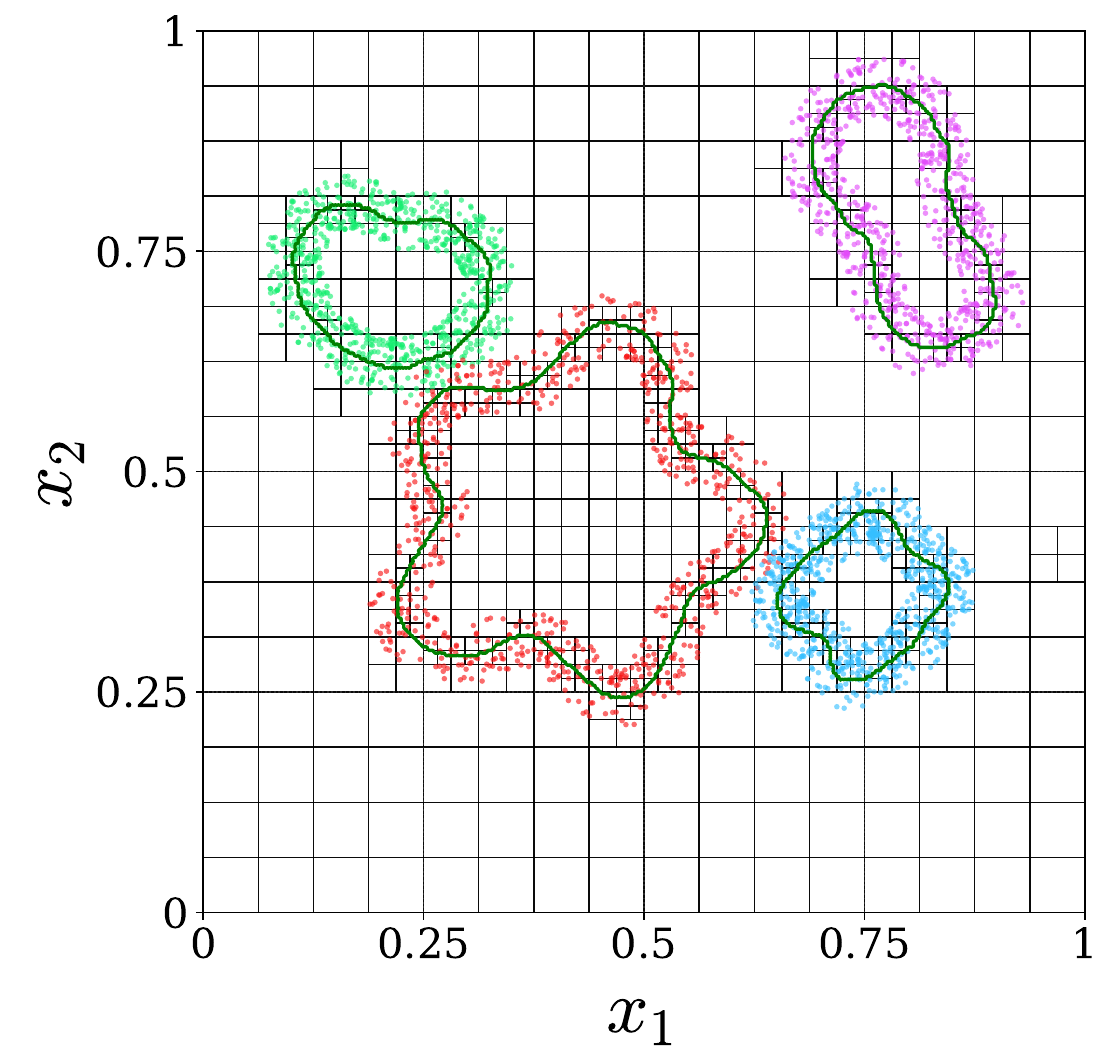}
        \caption{ $\mathcal{C}_{\text{final}}$ }
    \end{subfigure}
    \hfill
    \begin{subfigure}[b]{0.32\textwidth}
\includegraphics[width=\textwidth]{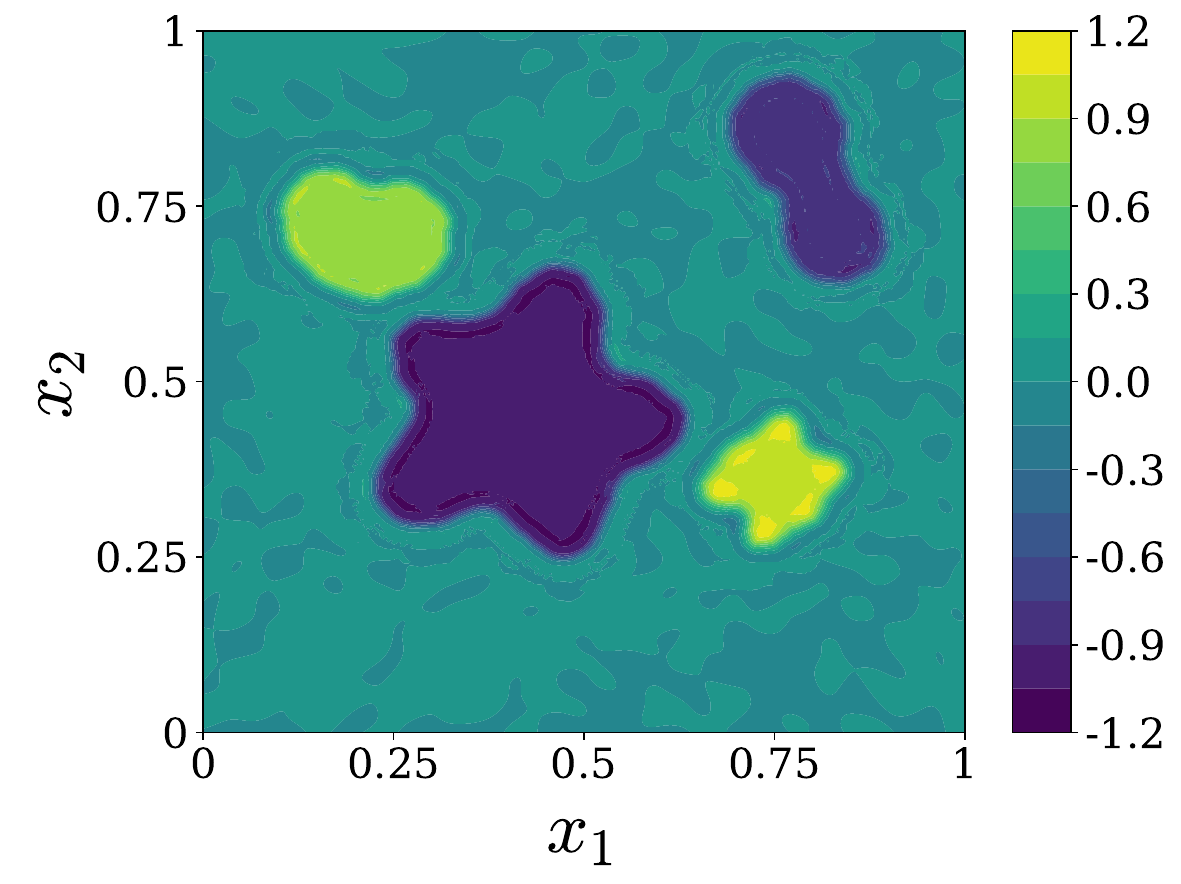}
        \caption{Reconstructed source}
    \end{subfigure}
    \hfill
    \begin{subfigure}[b]{0.32\textwidth}
\includegraphics[width=\textwidth]{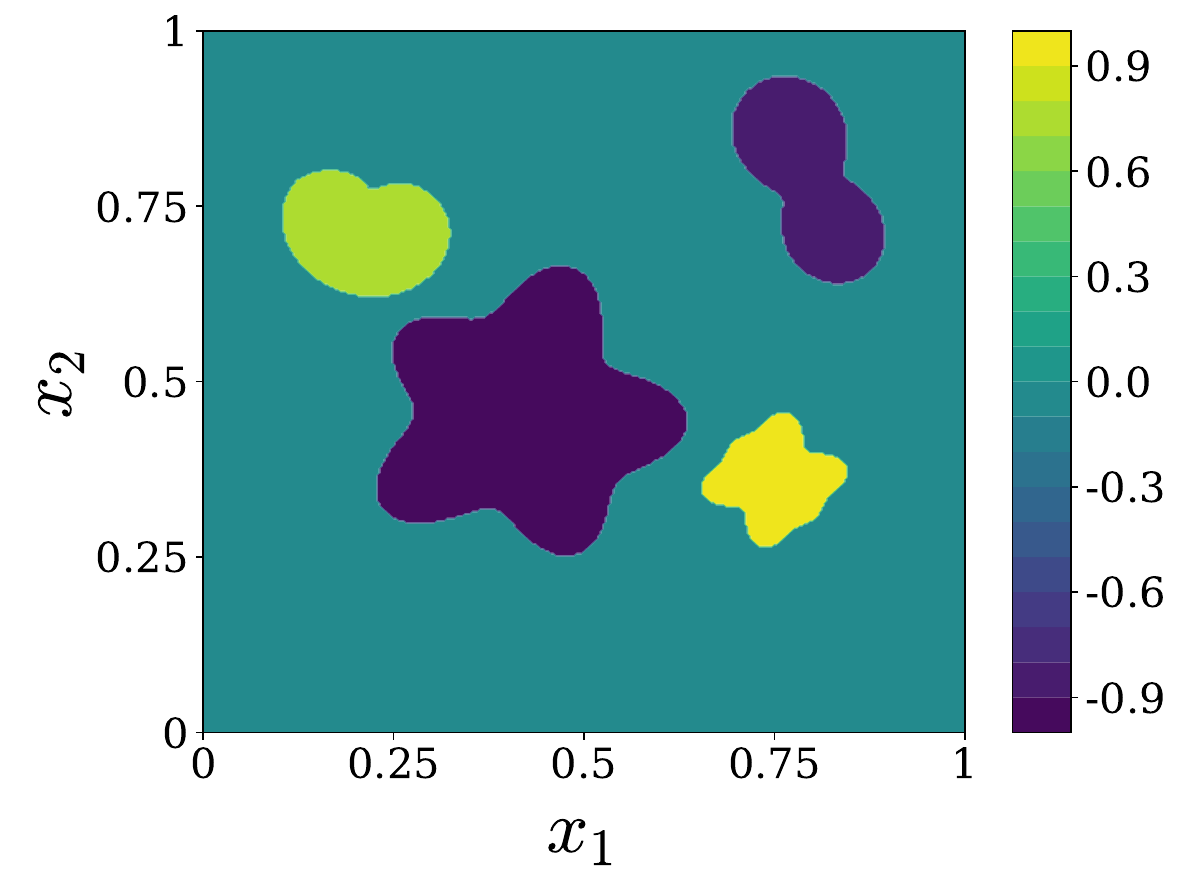}
        \caption{ {True source}}
        \label{ex:4.7:irregular}
    \end{subfigure}
    \caption{ {Example \ref{ex:Irregular source}:\ Irregular source: (a): Final adaptive mesh ($\mathcal{C}_{\text{final}}$) obtained via IA-RFM, yielding 25.05\%, with colored circle basis centers and the nominal boundary(green solid). (b) MA-RFM results with $\delta=5\%$, $M_0=1600$, $M_{\alpha,\text{general}}=1000$, $M_{\alpha,\text{circular}}=800$, $\lambda_{\text{reg}}^2$=1.00e-5 yield $17.84\% \ E_{l^2}(S)$.}}
    \label{fig:Irregular:detect_bound}
\end{figure}

\begin{example}[\emph{Composite kidney source}]\label{ex:Composite kidney source}Now consider complex geometries to test the generality of MA-RFM.
Consider the kidney-shaped line with the contour implicit function expression
\begin{equation}
    \label{hidden_function}
    ((x-x_0)^2+(y-y_0)^2-4a^2)^3=108a^4 (y-y_0)^2.
\end{equation}
Use the level set function $\psi(x, y)$ to represent regions,
$$
\psi(x, y)=((x-x_0)^2+(y-y_0)^2-4a^2)^3-108a^4 (y-y_0)^2,
$$
$\tau=\{(x,y)|\ \psi(x,y)\le0\}$,  introduce the radial parameter $r$,  parameterize the region to calculate the measurement data.
\begin{align}\label{para_area}
\left \{
\begin{aligned}
&x(s, \varphi)=r(3a \cos\varphi-a \cos 3\varphi),  \\
&y(s, \varphi)=r(3a \sin\varphi-a \sin 3\varphi),  \ r\in [0, 1],  \ \phi\in[0, 2\pi].
\end{aligned}
\right.
\end{align}
Consider a composite source that combines a continuous source and a discontinuous source.
\begin{align*}
%\label{composite}
S(x_1, x_2)&=S_{\text{kidney}}(x_1, x_2)+S_{\text{gauss}}(x_1, x_2),  \\
S_{\text{kidney}}(x_1, x_2)&=\mathcal{X}_{\{(x_1, x_2):\  \psi(x_1, x_2)\le0\}},  \\
S_{\text{gauss}}(x_1, x_2)&=1.2 \exp(-125((x_1-0.3)^2+(y-0.6)^2)).
\end{align*}
\noindent \emph{Experimental setup:}\ $V_0=[0, 1]^2$,   $\Omega=[-0.5, 1.5]^2$,  $x_0=0.6$,  $y_0=0.25$,  $a=0.05$. $k_{\text{min}}$=1,  $k_{\text{max}}=101$, $N_s$=20,  $\delta=5\%$.\\
\noindent \emph{Hyperparameter settings:}
The activation function for IA-RFM is Tanh, $R_m=20$,  $\epsilon_{\text{res}}=0.5\delta$. The basis function used in MA-RFM is sigmoid, comprising a general form  {detected by Algorithm \ref{alg: shape_detection}} and a supplementary circle basis function to compensate for the undesirable position of the detected boundary at the interface. 
    \begin{align*}
        \phi_{1, {\text{general}}}^{(j)}(\boldsymbol{x};K_1^{(j)},\theta^{(j)}_1) &= \text{sigmoid} \left(K_1^{(j)} \cdot d_1^{(j)}(\boldsymbol{x};\rho^{(j)}) \right),  \ \ j=1, \cdots, M_{1, {\text{general}}}. \\
        \phi_{1,\text{ {circular}}}^{(j)}(\boldsymbol{x};K_2^{(j)},\boldsymbol{\theta}_2^{(j)})&=\text{sigmoid}\left(K_1^{(j)}\cdot\left((r_1^{(j)})^2-\|\boldsymbol{x}-\boldsymbol{c}_{1}^{(j)}\|_2^2\right)\right), \ j=1, \cdots, M_{1, {\text{circular}}}.
    \end{align*}
$ {\rho^{(j)} \sim U(-10h, 0)}$, $K_1^{(j)} \sim U(1000,30000)$, $ {r_1^{(j)} \sim U(0, 10h)}$, $t_{\text{grad}}$=1/3,    $ {M_{1,\text{general}}}=M_0=1600$,  $ {M_{1,\text{circular}}=800}$,  {$\phi_2^{(j)}$ is the local circular basis function}. The initial mesh $N_{x_1}=N_{x_2}=4$ with $n_{x_1}=n_{x_2}=3$. \\

Figure \ref{fig:kidney:detect_bound} illustrates the $\mathcal{C}_{\text{final}}$ using IA-RFM and 
numerical scaled interface and real interface. Figure \ref{fig:kidney} shows the reconstruction using MA-RFM with $\delta=5\%$.  {The corresponding stagewise comparison and clusterwise detected parameters are reported in Tables \ref{tab:Comparison of IA-RFM and MA-RFM} and \ref{tab:extracted_parameters}, respectively.}
\begin{figure}[H]
    \centering
    \begin{subfigure}[t]{0.32\textwidth}
\includegraphics[width=\textwidth]{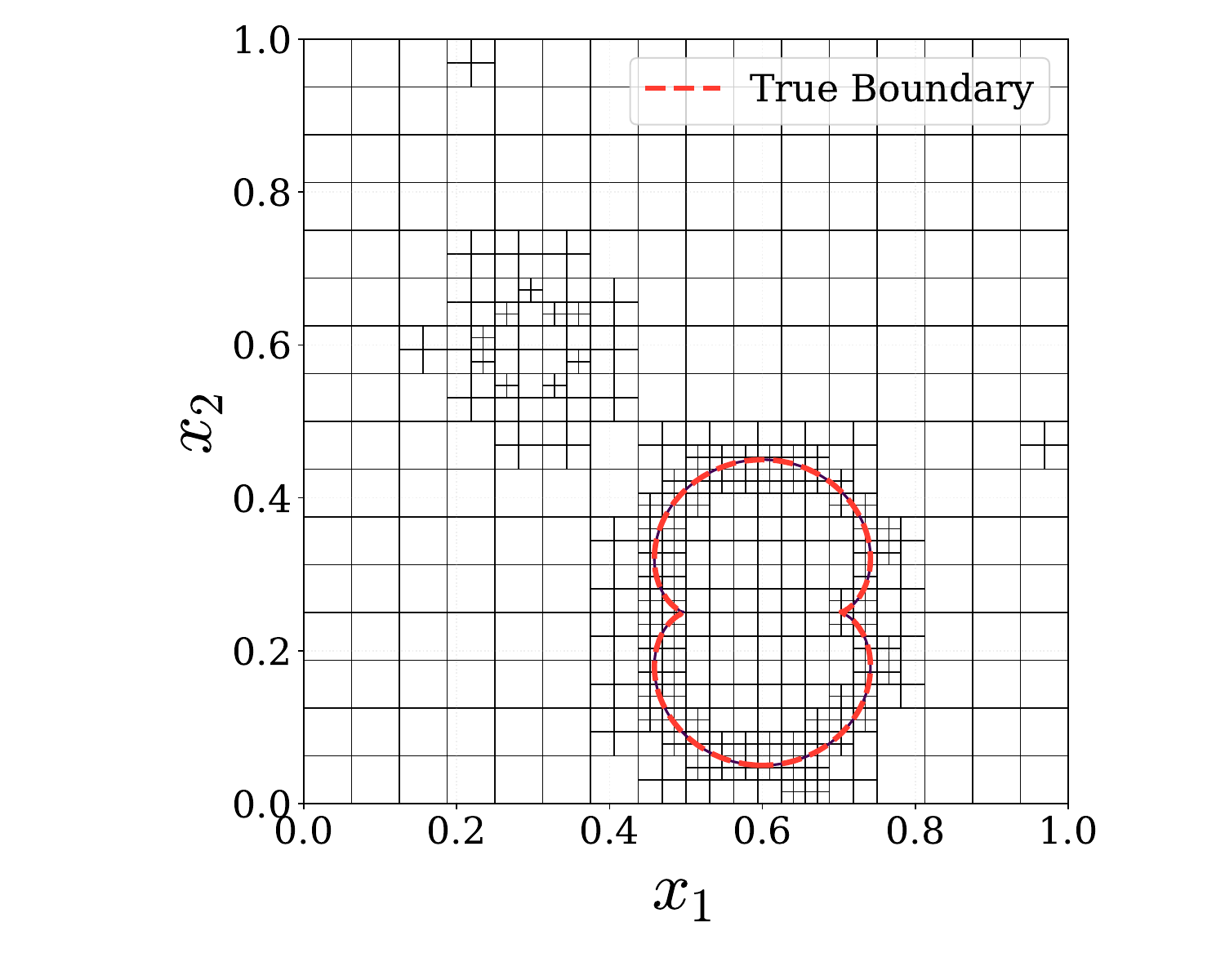}
        \caption{ {$\mathcal{C}_{\text{final}}$ }}
    \end{subfigure}
    \hfill
    \begin{subfigure}[t]{0.32\textwidth}
\includegraphics[width=\textwidth]{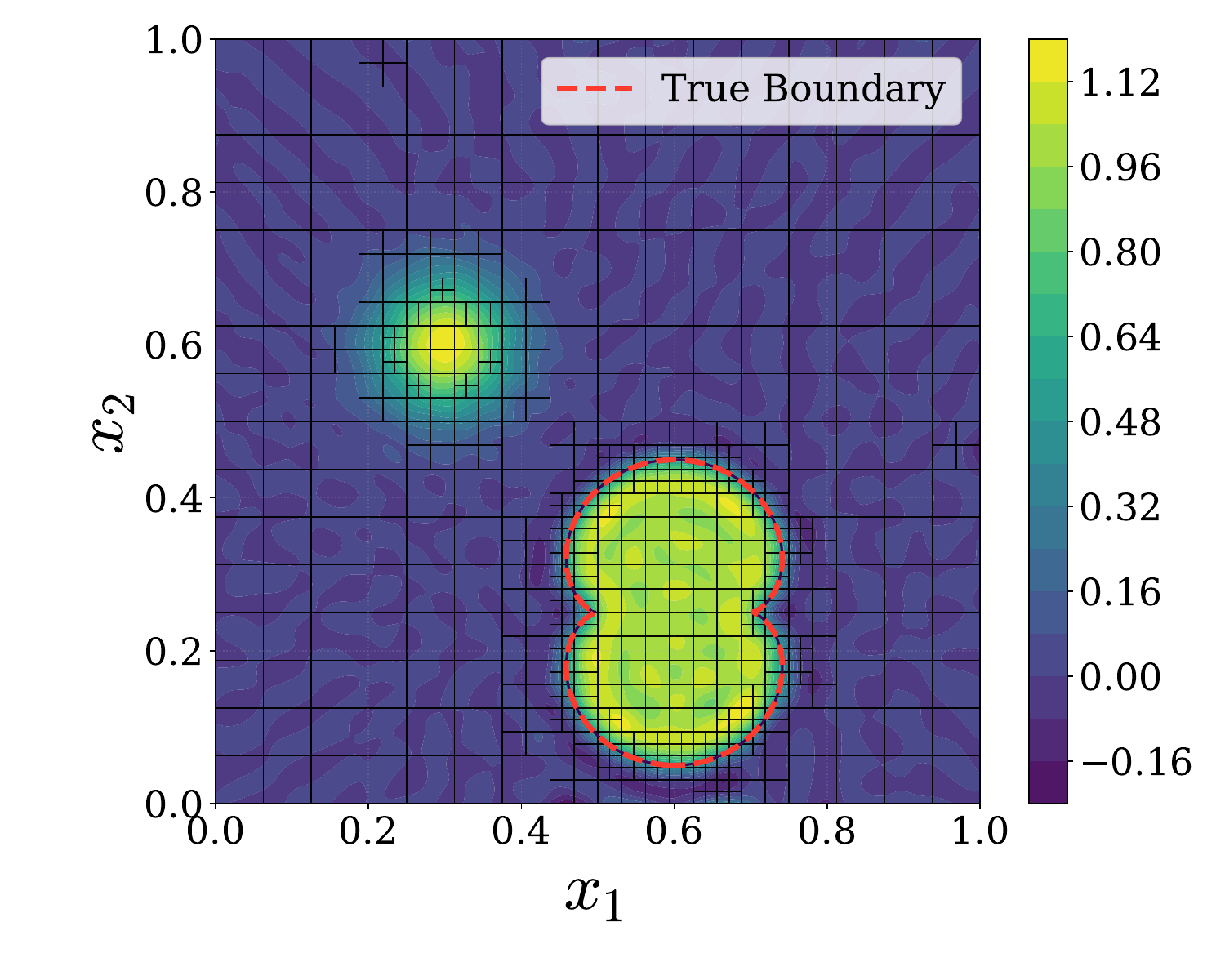}
        \caption{ {$S_{\text{M}}$}}
    \end{subfigure}
    \hfill
    \begin{subfigure}[t]{0.32\textwidth}
\includegraphics[width=\textwidth]{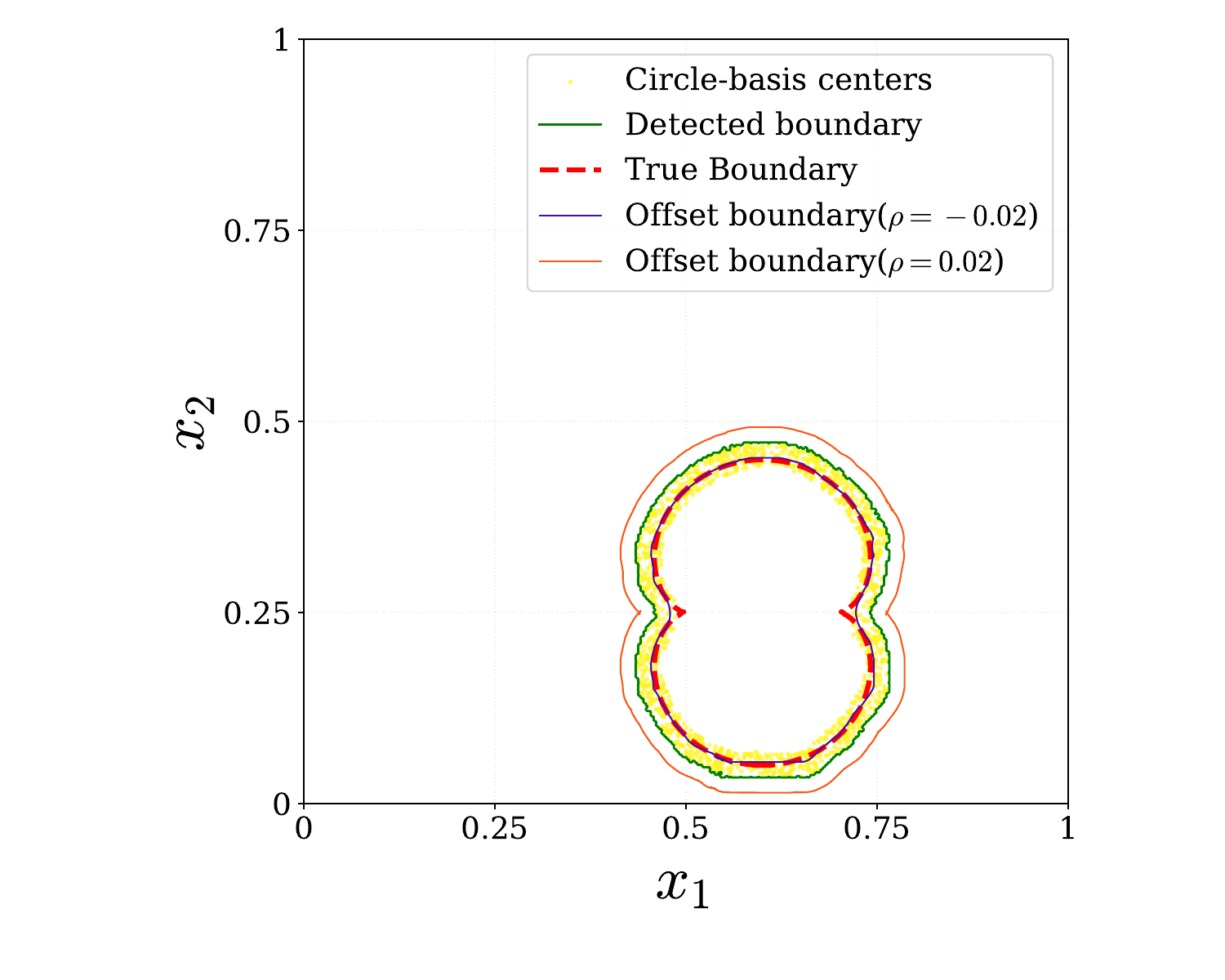}
        \caption{ {Numerical boundary}}
    \end{subfigure}
    \caption{Example \ref{ex:Composite kidney source}:\   {Composite kidney source}:  {(a)-(b) Final adaptive mesh ($\mathcal{C}_{\text{final}}$) and reconstruction obtained via IA-RFM, yielding a relative error of $E_{l^2}(S)=19.74\%$ with  {$n_{\text{integral}}=6354$}. (c) Numerical boundary reconstruction. The true boundary (red dashed) is compared with the nominal boundary extracted from $\mathcal{Q}_{\text{grad}}$ with $t_{\text{grad}}=1/3$ (green solid).   {The orange and purple contours represent the perturbed numerical level sets ($\rho=\pm 0.02$) with yellow circle basis centers.}}}
    \label{fig:kidney:detect_bound}
\end{figure}

\begin{figure}[ht]
    \centering
    \begin{subfigure}[b]{0.32\textwidth}
        \includegraphics[width=\textwidth]{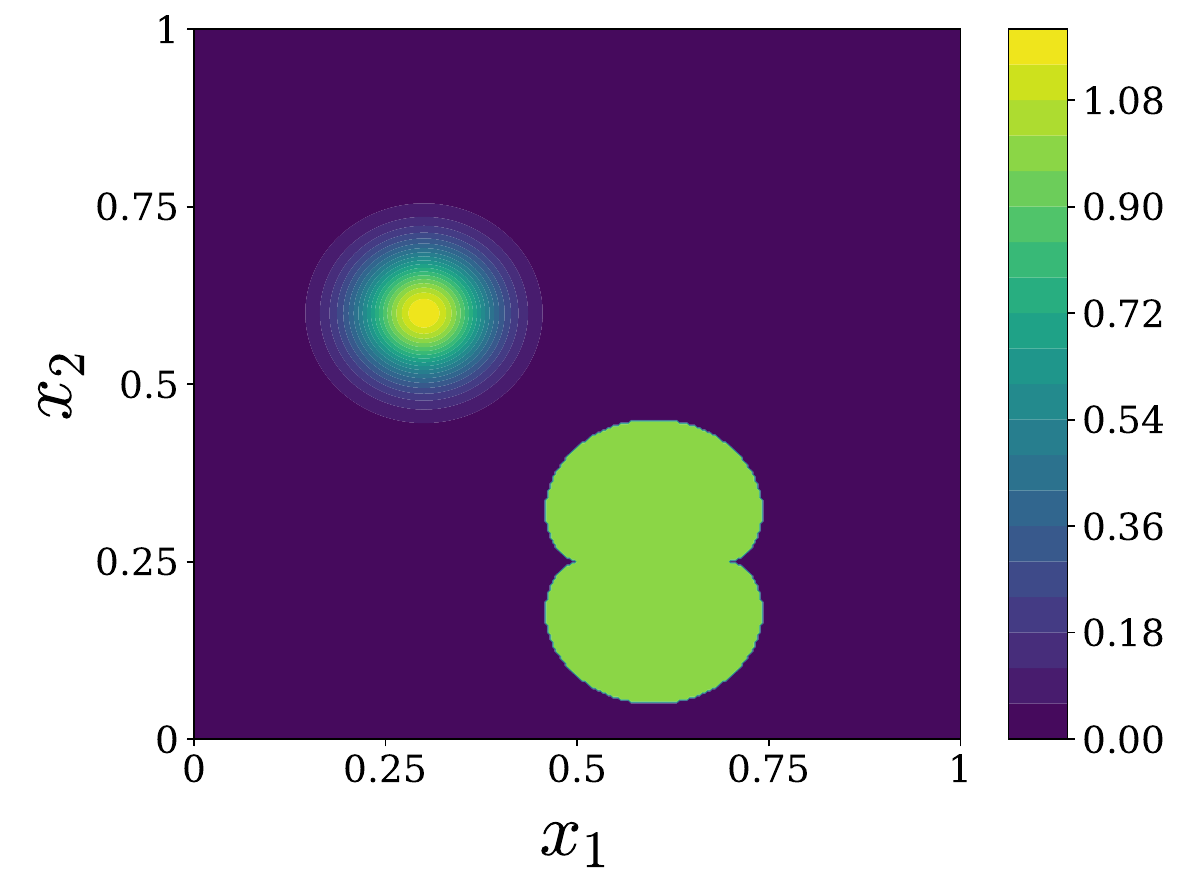}
        \caption{True source}
    \end{subfigure}
    \hfill
    \begin{subfigure}[b]{0.32\textwidth}
        \includegraphics[width=\textwidth]{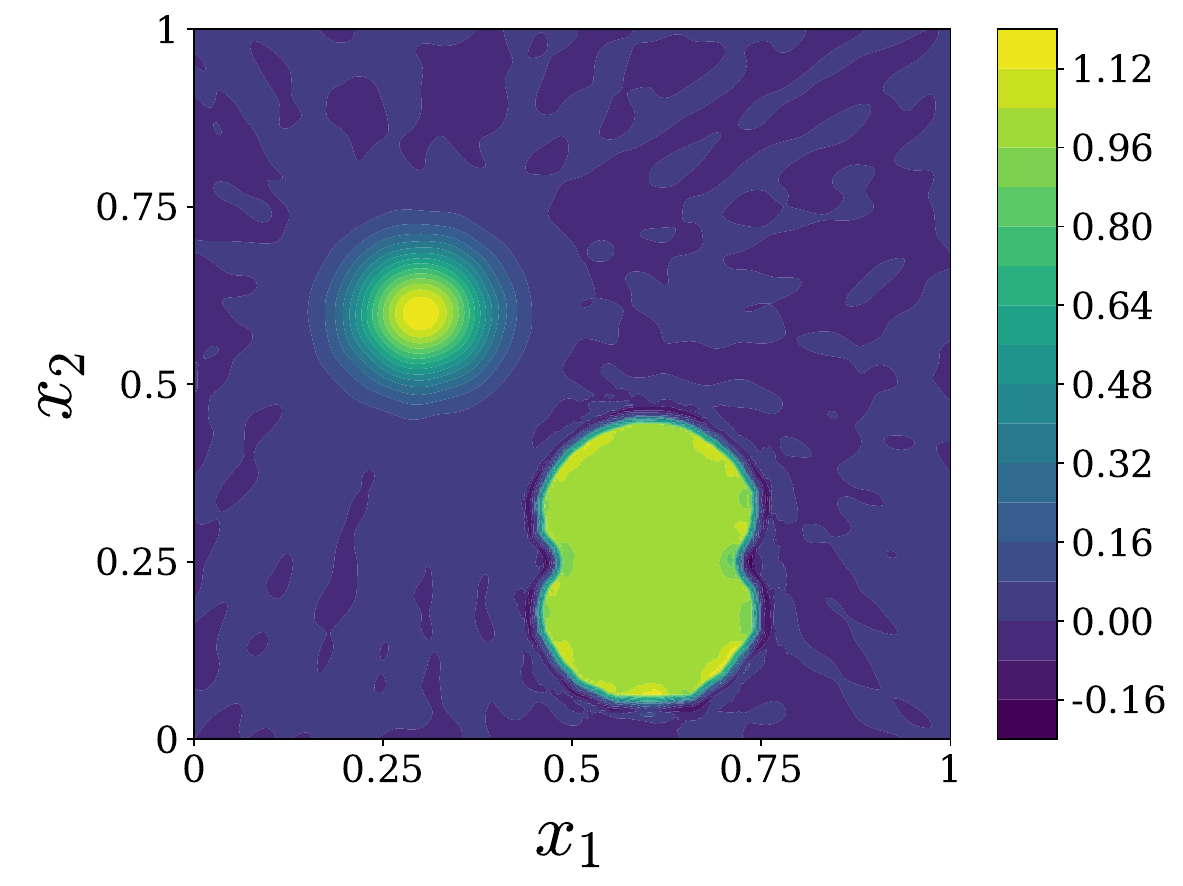}
        \caption{Reconstructed source}
    \end{subfigure}
    \hfill
    \begin{subfigure}[b]{0.32\textwidth}
\includegraphics[width=\textwidth]{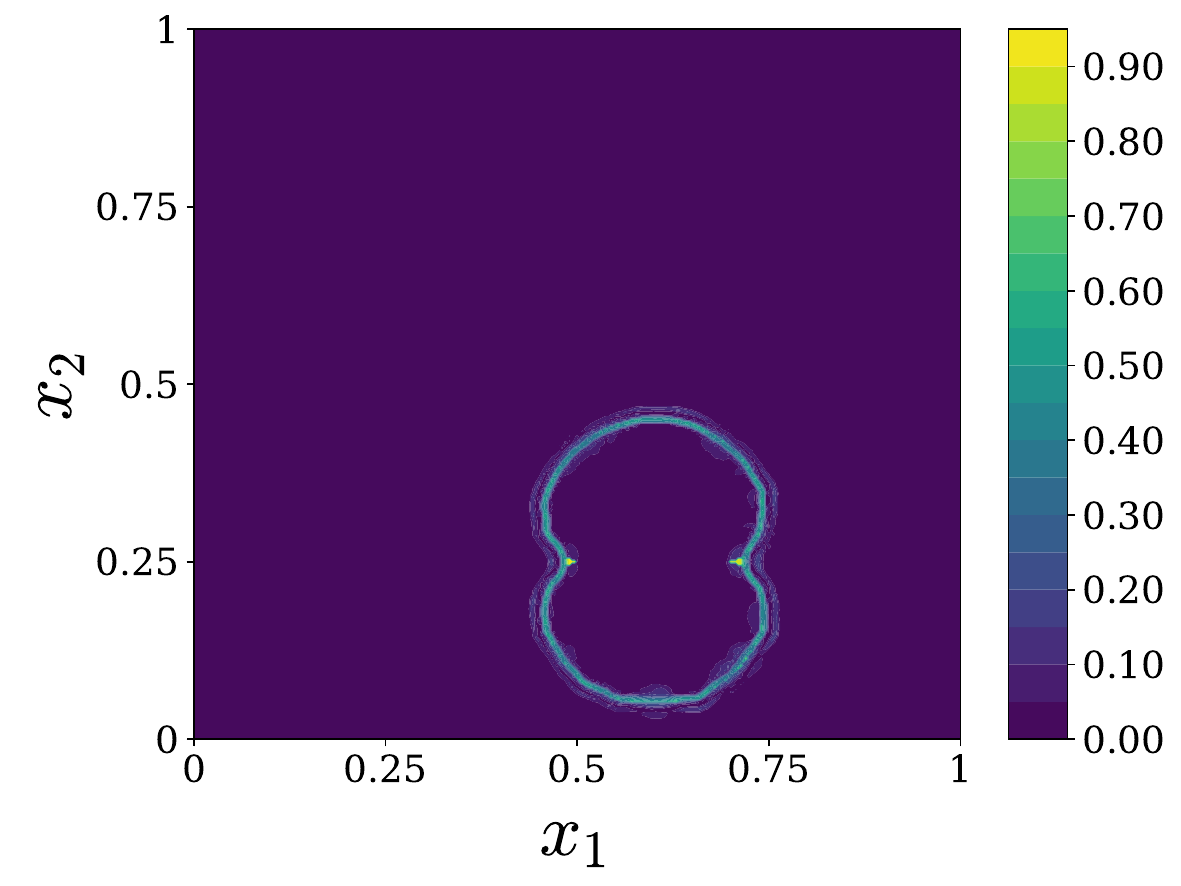}
        \caption{ {Pointwise absolute error}}
    \end{subfigure}
    \caption{Example \ref{ex:Composite kidney source}:\  {Composite kidney source}: MA-RFM results with $\delta=5\%$, $M_0=1600$,  $M_1=1600$,  $ {M_2=800}$, $\lambda_{\text{reg}}^2$ = 1.00e-3 yield 13.94\% $E_{l^2}(S)$.}
    \label{fig:kidney}
\end{figure}
\end{example}
\begin{example}[\emph{Irregular source}]\label{ex:Irregular source}
 {Next, we consider more general irregular sources with specific details given in Figure (\ref{ex:4.7:irregular}).}

\noindent  {\emph{Experimental setup:}\ $V_0=[0, 1]^2$,   $\Omega=[-0.5, 1.5]^2$. $k_{\text{min}}$=1,  $k_{\text{max}}=101$, $N_s$=20,  $\delta=5\%$.\\
\noindent \emph{Hyperparameter settings:}
The activation function for IA-RFM is Tanh, $R_m=20$,  $\epsilon_{\text{res}}=0.5\delta$. The basis function used in MA-RFM is sigmoid, comprising a general form  {detected by Algorithm \ref{alg: shape_detection}} and a supplementary circular basis function according to Scenario II. 
    \begin{align*}
        \phi_{\alpha,\text{general}}^{(j)}(\boldsymbol{x};K_\alpha^{(j)},\theta^{(j)}_\alpha) &= \text{sigmoid} \left(K_\alpha^{(j)} \cdot d_\alpha^{(j)}(\boldsymbol{x};\rho_{\alpha}^{(j)}) \right),  \ \ \alpha=1,\cdots,4,\ j=1, \cdots, M_{\alpha,\text{general}}, \\
        \phi_{\alpha,\text{circular}}^{(j)}(\boldsymbol{x};K_\alpha^{(j)},\boldsymbol{\theta}_\alpha^{(j)})&=\text{sigmoid}\left(K_\alpha^{(j)}\cdot\left((r_\alpha^{(j)})^2-\|\boldsymbol{x}-\boldsymbol{c}_{\alpha}^{(j)}\|_2^2\right)\right), \alpha=1,\cdots,4,\ j=1, \cdots, M_{\alpha,\text{circular}}.
    \end{align*}
$\rho_{\alpha}^{(j)} \sim U(-10h, 10h)$, $K_\alpha^{(j)} \sim U(1000,30000)$, $ {r_\alpha^{(j)} \sim U(0, 10h)}$, $t_{\text{grad}}=t_{\text{abs}}$=1/3,    $M_0=1600$, $M_{\alpha,\text{general}}=1000, M_{\alpha,\text{circular}}=800$. The initial mesh $N_{x_1}=N_{x_2}=4$ with $n_{x_1}=n_{x_2}=3$.\\}

 {Figure \ref{fig:Irregular:detect_bound} illustrates the $\mathcal{C}_{\text{final}}$ using IA-RFM, the centers of circle enhancement basis functions with colored scattering points, and the reconstruction using MA-RFM with $\delta=5\%$.}    {See Tables \ref{tab:Comparison of IA-RFM and MA-RFM} and \ref{tab:extracted_parameters} for the corresponding stagewise comparison and detected basis information.}\\
\end{example}

\begin{figure}[hbt]
    \centering
    \begin{subfigure}[b]{0.32\textwidth}
\includegraphics[width=\textwidth]{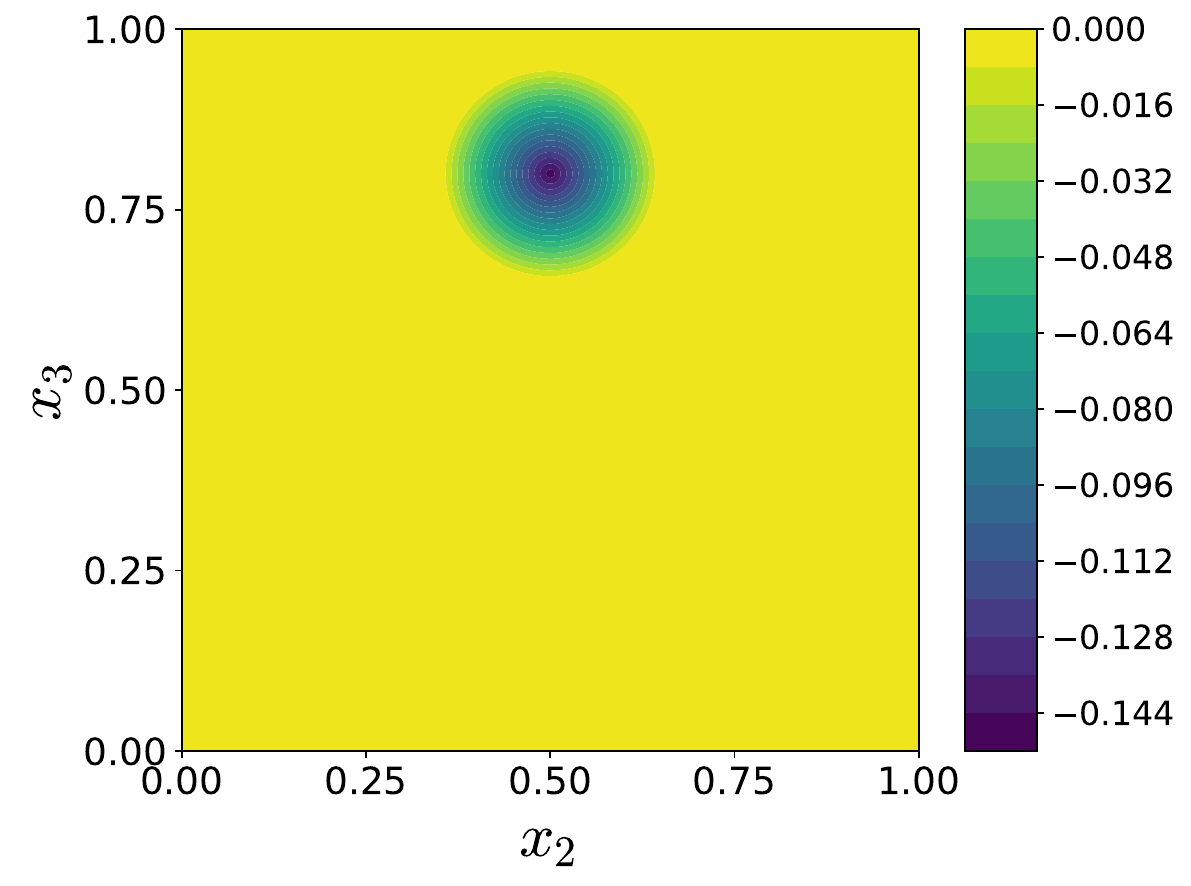}
    \caption{True source at $x_1=0.3$ }
    \end{subfigure}
    \hfill
    \begin{subfigure}[b]{0.32\textwidth}      \includegraphics[width=\textwidth]{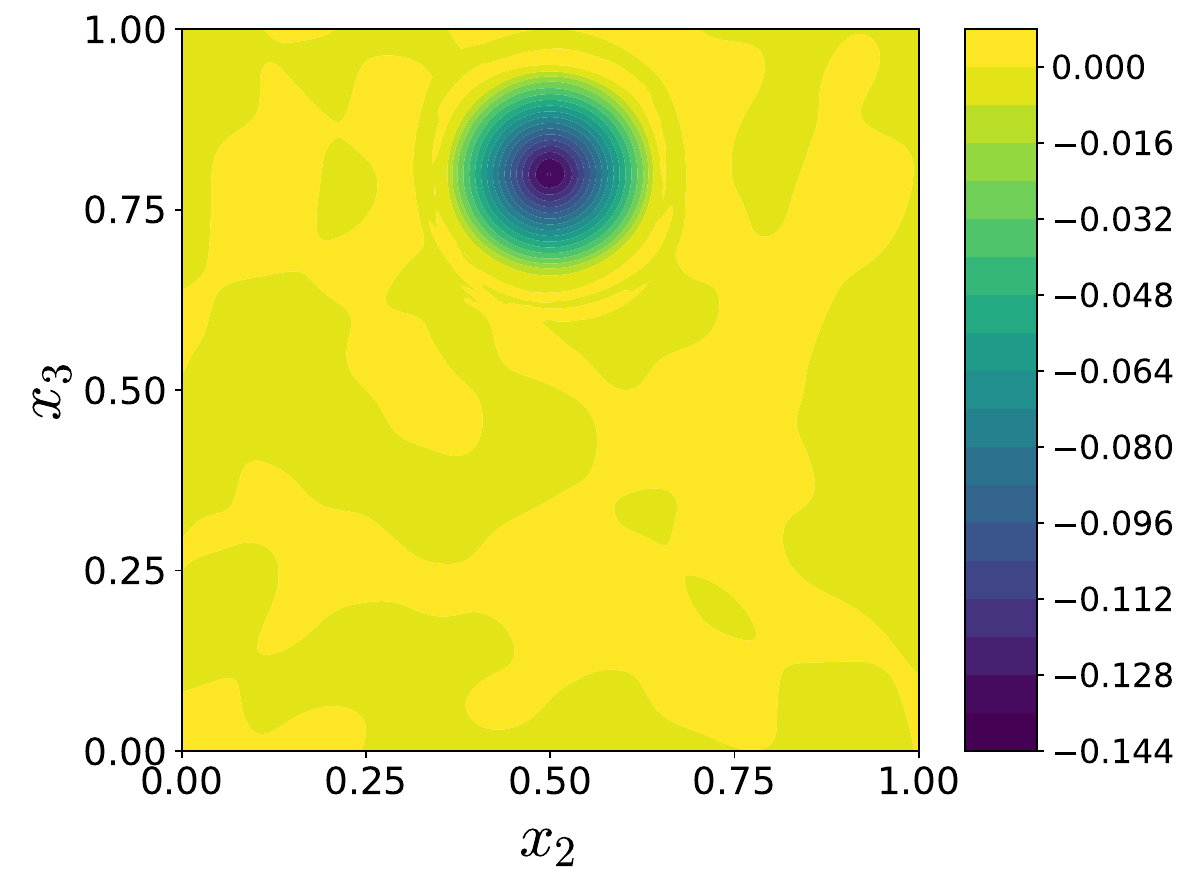}
        \caption{Reconstruction at $x_1=0.3$ }
    \end{subfigure}
    \hfill
    \begin{subfigure}[b]{0.32\textwidth}      \includegraphics[width=\textwidth]{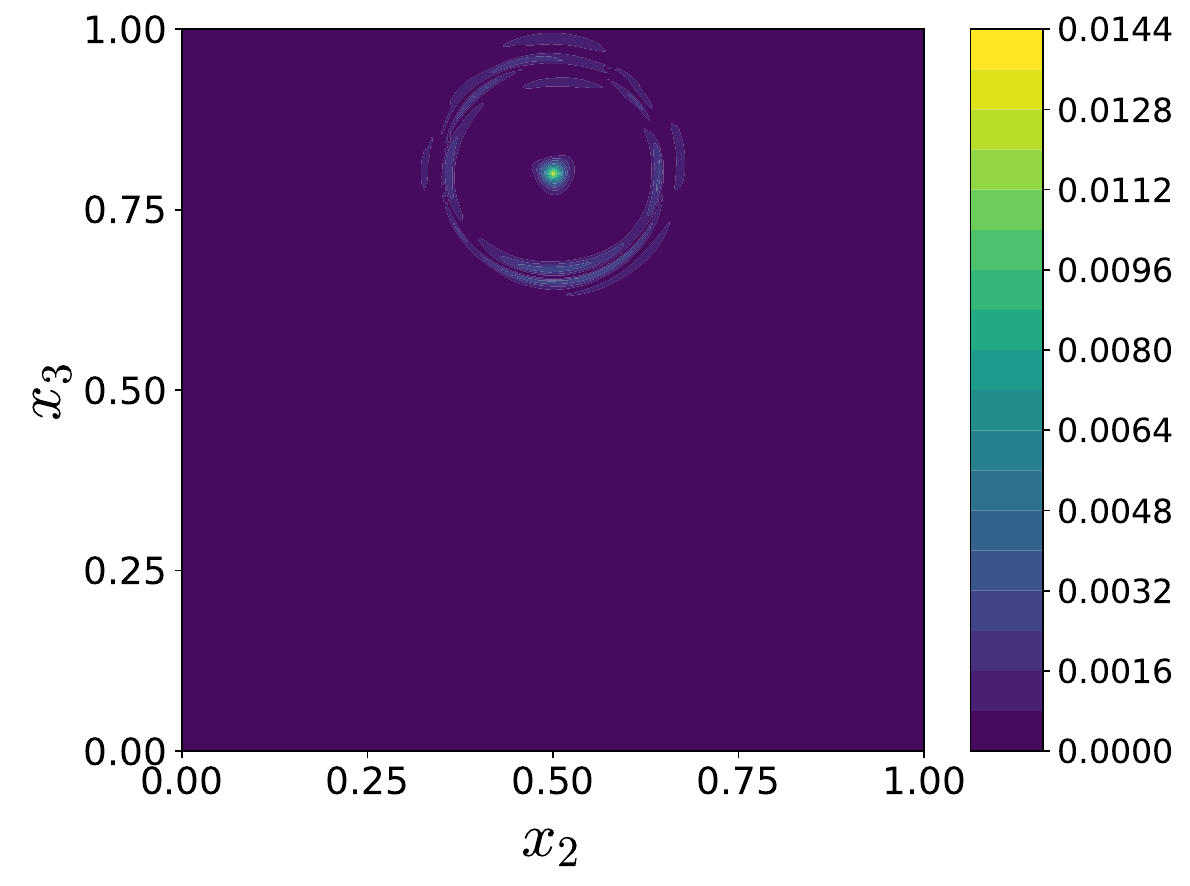}
        \caption{ {Pointwise absolute error}}
    \end{subfigure}
    \\[5pt]
    \begin{subfigure}[b]{0.32\textwidth}
\includegraphics[width=\textwidth]{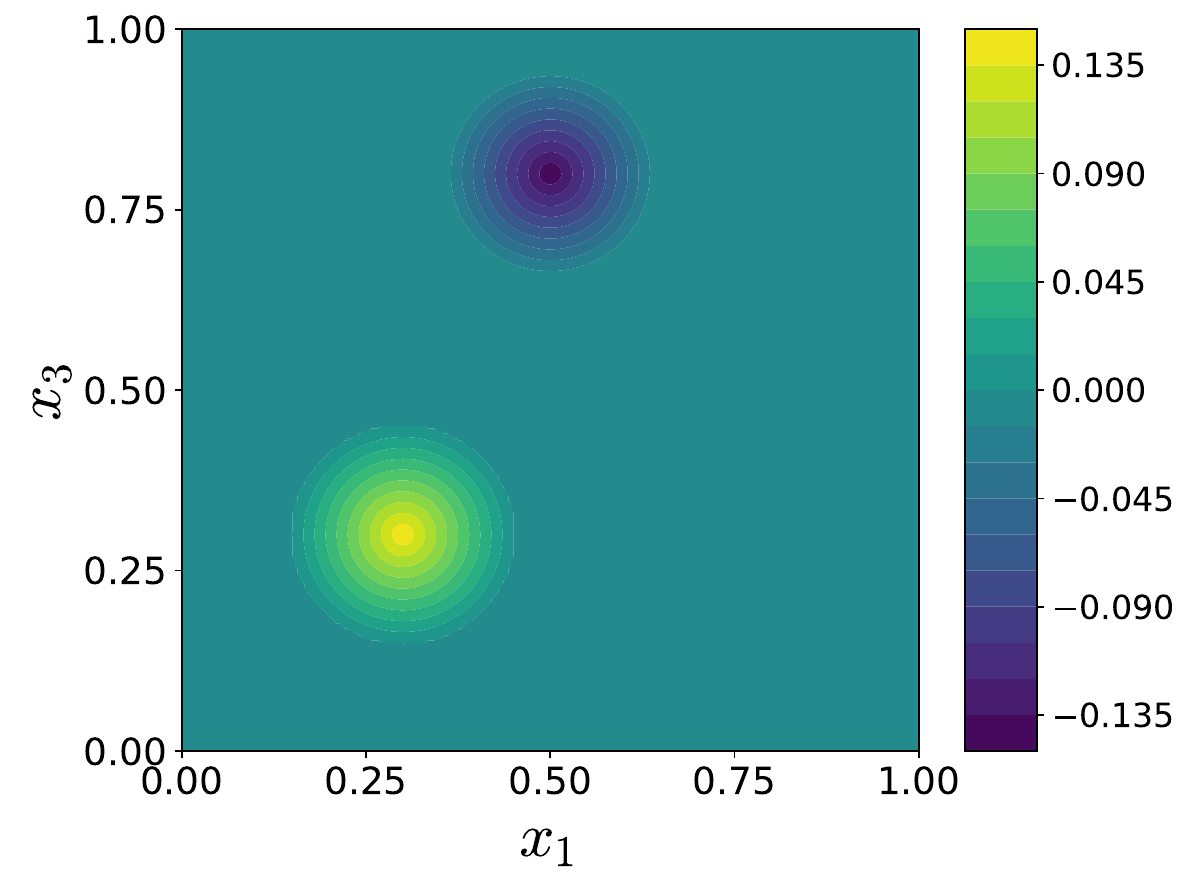}
        \caption{True source at $x_2=0.5$ }
    \end{subfigure}
    \hfill
    \begin{subfigure}[b]{0.32\textwidth} 
        \includegraphics[width=\textwidth]{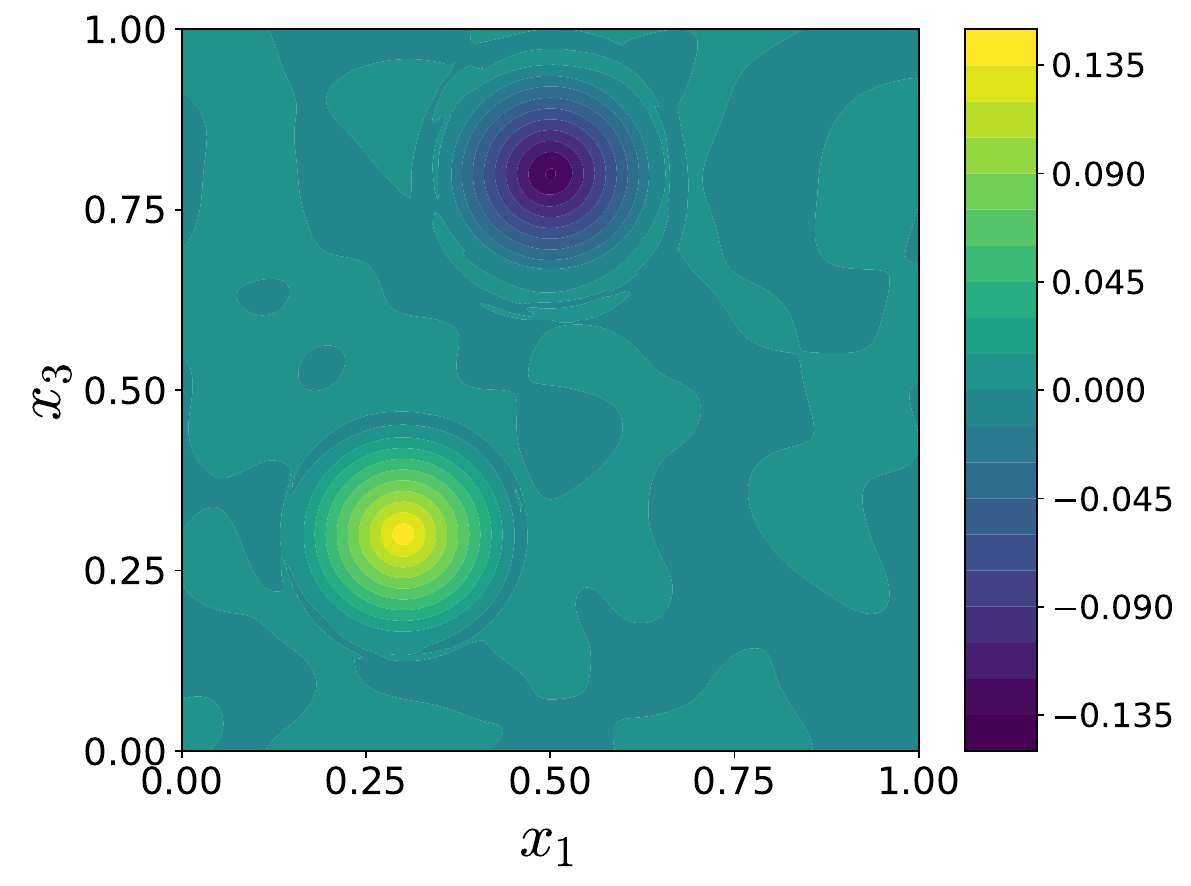}
        \caption{Reconstruction at $x_2=0.5$ }
    \end{subfigure}
    \hfill
    \begin{subfigure}[b]{0.32\textwidth}  
        \includegraphics[width=\textwidth]{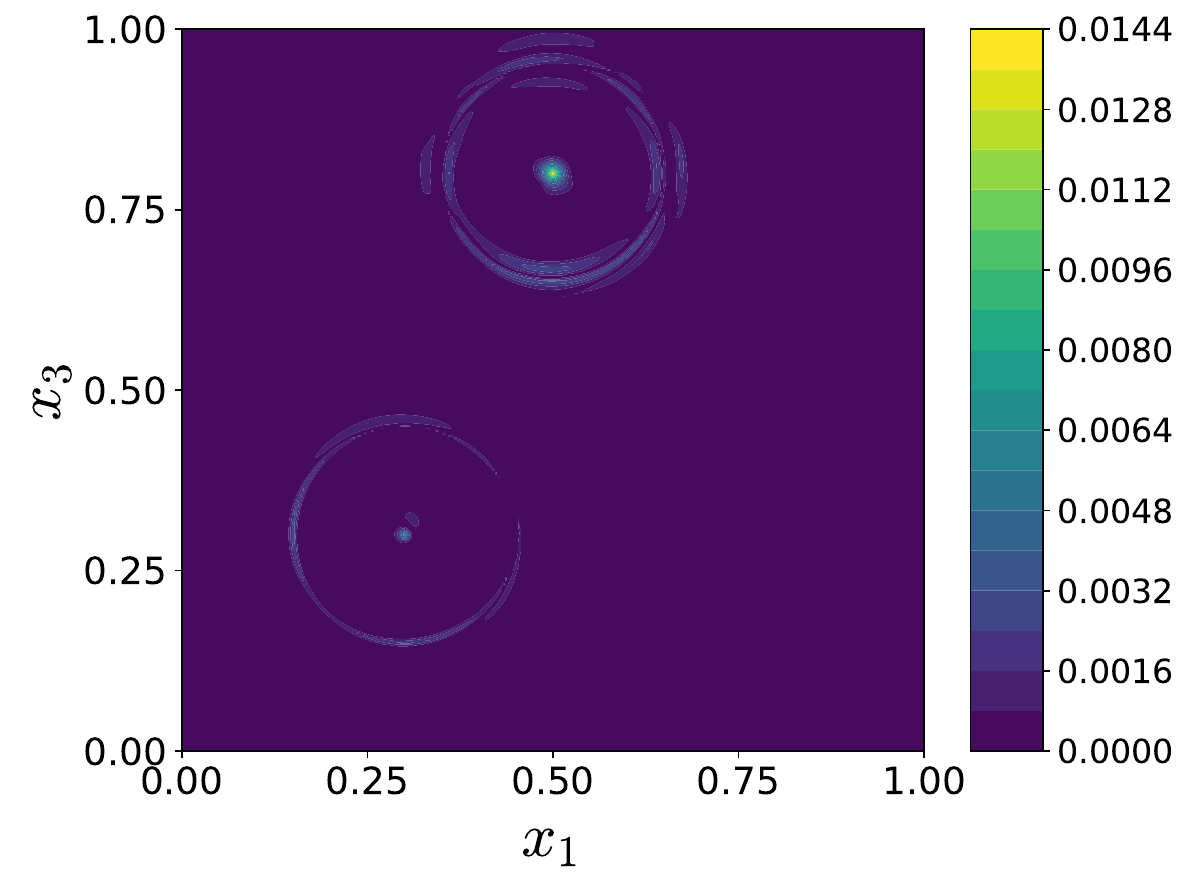}
        \caption{ {Pointwise absolute error}}
    \end{subfigure}
    \caption{Example \ref{ex:Three-Dimensional $C^0$ source}: Three-dimensional $C^0$ source: MA-RFM results with $\delta=5\%$, $M_0=3600$, $M_{\alpha,\text{ReLU}}=M_{\alpha,\text{exp}}=900$, $\lambda_{\text{reg}}^2$ = 1.00e-2 yield 1.99\% $E_{l^2}(S)$.}
    \label{fig:3D_S_num0_2D}
\end{figure}

\begin{example}[\emph{Three-dimensional $C^0$ source}]\label{ex:Three-Dimensional $C^0$ source}Consider $S(\boldsymbol{x})=S^{\boldsymbol{a}}(\boldsymbol{x})-S^{\boldsymbol{b}}(\boldsymbol{x})$
\begin{align*}
S^{\boldsymbol{y}}(\boldsymbol{x})=
\begin{cases}
0.15-r_{\boldsymbol{y}},&\mathrm{if}\quad r_{\boldsymbol{y}}<0.2, \\
\quad0,&\mathrm{if}\quad r_{\boldsymbol{y}}\geqslant0.2,
\end{cases}
\end{align*}
where $r_{\boldsymbol{y}}=\sqrt{(x_{1}-y_{1})^{2}+(x_{2}-y_{2})^{2}+(x_{3}-y_{3})^{2}}$, $\boldsymbol{a}=(0.3, 0.5, 0.3)$,  $\boldsymbol{b}=(0.5, 0.5, 0.8)$.

\begin{figure}[htb]
    \centering
    \begin{subfigure}[b]{0.27\textwidth}
    \includegraphics[width=\textwidth]{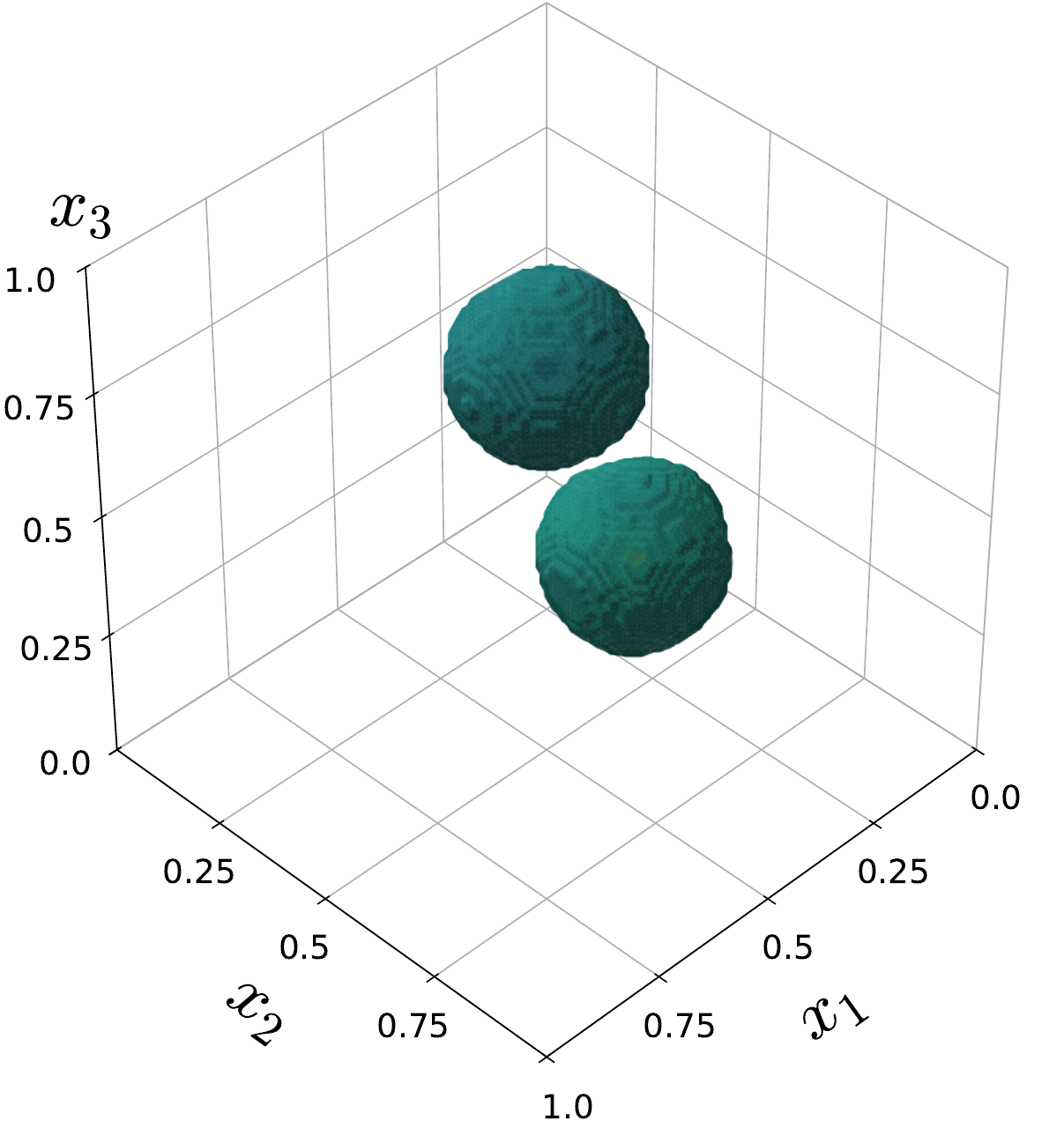}
        \caption{ }
    \end{subfigure}
    \hfill
     \begin{subfigure}[b]{0.3\textwidth}
        \includegraphics[width=\textwidth]{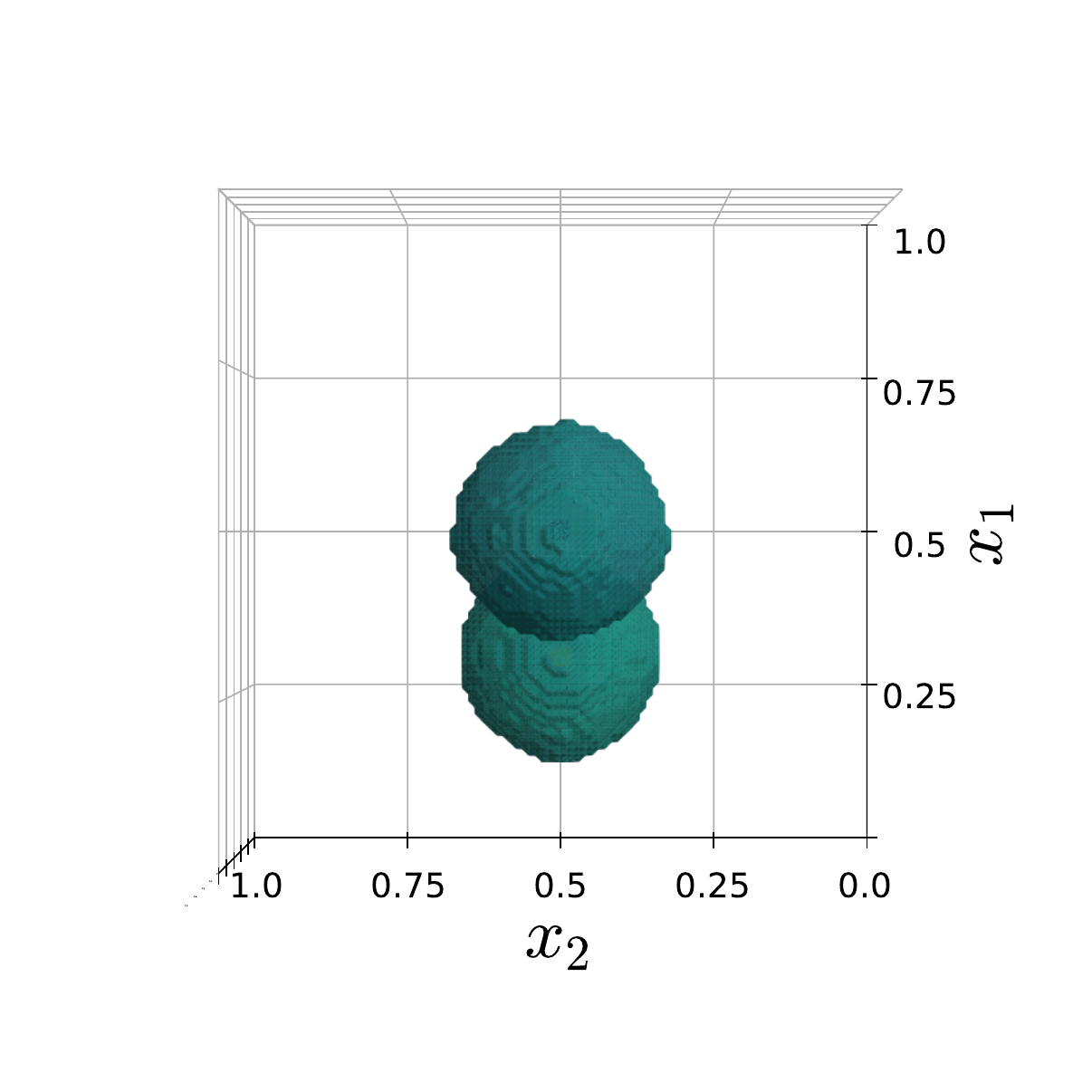}
        \caption{ }
    \end{subfigure}
    \hfill
    \begin{subfigure}[b]{0.3\textwidth}
        \includegraphics[width=\textwidth]{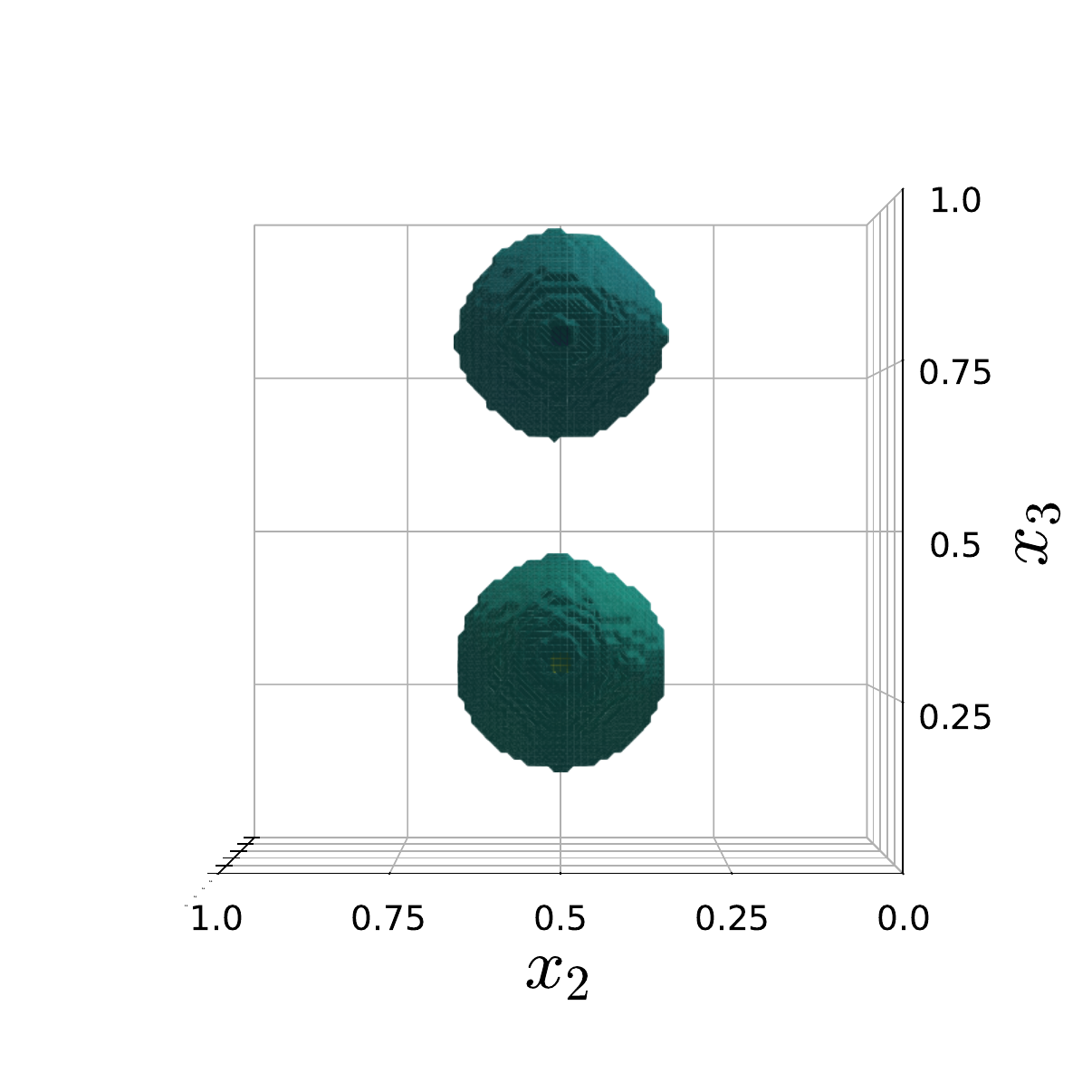}
        \caption{ }
    \end{subfigure}
    \hfill
    
    \begin{subfigure}[b]{0.275\textwidth}        \includegraphics[width=\textwidth]{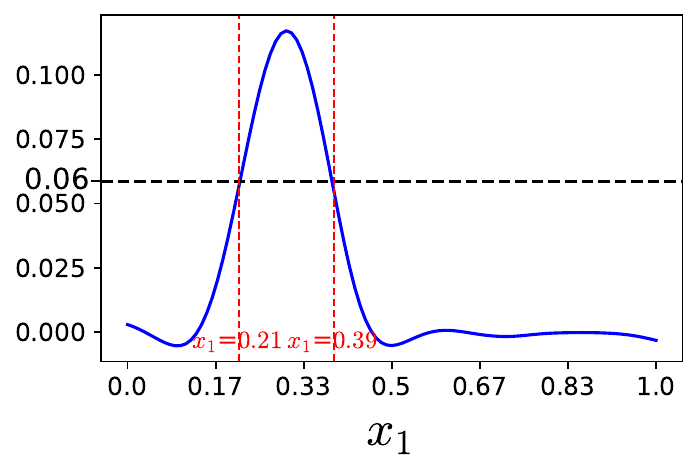}
        \caption{$S_{M}(t,0.5,0.30)$}
    \end{subfigure}
    \hfill
    \begin{subfigure}[b]{0.275\textwidth}
        \includegraphics[width=\textwidth]{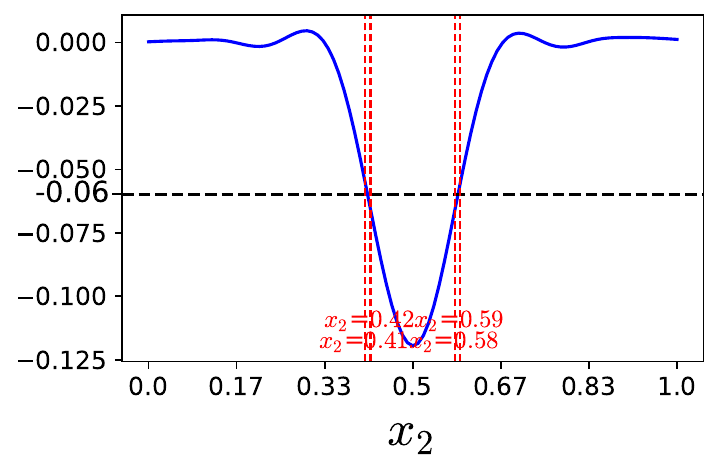}
        \caption{$S_{M}(0.5,t,0.79)$}
    \end{subfigure}
    \hfill
    \begin{subfigure}[b]{0.275\textwidth}
        \includegraphics[width=\textwidth]{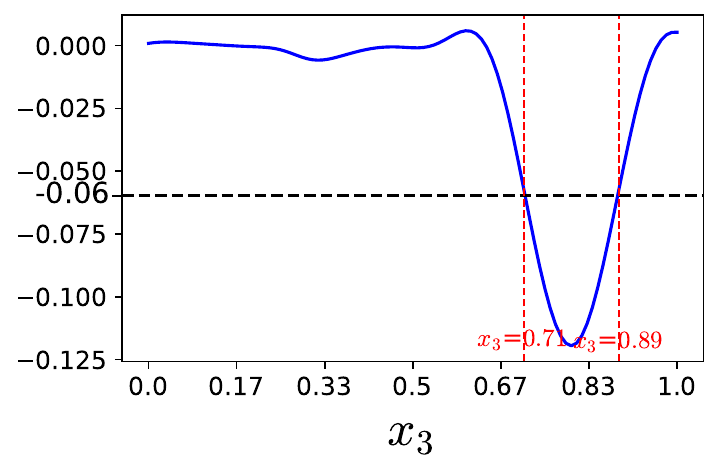}
        \caption{ $S_{M}(0.5,0.5,t)$}
    \end{subfigure}
    \hfill
    \caption{Example \ref{ex:Three-Dimensional $C^0$ source}: Three-dimensional $C^0$ source: (a),(b),(c) show the  $\mathcal{Q}=\mathcal{Q}_{\text{grad}}$ with $t_{\text{grad}}=1/3$ from different perspectives. 
    %The symbol "$\times$" represents the center  {$\hat{c}_1=(0.3000,0.5009,0.2991)$, $\hat{c}_2=(0.4991,0.4994,0.8000)$, and $\hat{L}_1=(0.155,0.150,0.160),\ \hat{L}_2=(0.160,0.160,0.160)$} detected by Algorithm \ref{alg: shape_detection}. 
    (d),(e),(f) show the numerical solution $g_i(t)$ for the slice $S_{M}^{(0)}$ (blue solid line).   {The horizontal black and vertical red lines indicate the $I_{\text{half}}$ and the corresponding position, respectively.}}
\label{fig:3D_detect_two_sources}
\end{figure}

\noindent \emph{Experimental setup:}\ $V_0=[0, 1]^3$, $\Omega=[-0.5, 1.5]^3$,   $k_{\text{min}}$=1,  $k_{\text{max}}=81$,   $N_s$=10,  $\delta=5\%$. 

\noindent \emph{Hyperparameter settings:}\ $V_0=[0, 1]^3$, $\Omega=[-0.5, 1.5]^3$,   $k_{\text{min}}$=1,  $k_{\text{max}}=81$,   $N_s$=10,  $\delta=5\%$. The activation function for IA-RFM is Tanh, $R_m=10$,  $\epsilon_{\text{res}}=0.05\delta$. The basis functions in MA-RFM are of a conical type,  using ReLU and Gaussian functions  {detected by Algorithm \ref{alg: shape_detection}} as follows: 
    \begin{align*}
    \phi_{\alpha,\text{ReLU}}^{(j)}(\boldsymbol{x};\boldsymbol{\theta}_\alpha^{(j)})&=\text{ReLU}\left(r_\alpha^{(j)}-\|\boldsymbol{x}-\boldsymbol{c}_{\alpha}^{(j)}\|_2\right)
    ,\ \alpha=1,2,\ j=1, 2, \cdots, M_{\alpha,\text{ReLU}}, \\
    \phi_{\alpha,\text{exp}}^{(j)}(\boldsymbol{x};\boldsymbol{\theta}_\alpha^{(j)})&= \exp\left(\tilde{v}_\alpha^{(j)} \cdot \left(1- \sum_{i=1}^3\left|\frac{x_i - c_{\alpha,i}^{(j)}}{L_{\alpha,i}^{(j)}}\right|^2\right)\right), \ \alpha=1,2,\  j=1, 2, \cdots, M_{\alpha,\text{exp}}.
    \end{align*}
 {The decay rate $v^{(j)}_\alpha$ is determined using the Full Width at Half Maximum (FWHM) method. Let $\boldsymbol{x}^* = (x_1^*, x_2^*, x_3^*)$ denote the global peak location of the initial solution magnitude $|S_{M}|$.  We define the 1-D slice function $g_i(t)$ by varying the $i$-th coordinate while fixing the other two coordinates at their peak values:
\begin{align*}
g_1(t):= S_{M}(t, x_2^*, x_3^*), \
g_2(t):= S_{M}(x_1^*, t, x_3^*), \
g_3(t):= S_{M}(x_1^*, x_2^*, t).
\end{align*}
The FWHM for each dimension $i \in \{1,2,3\}$ is then determined by:
\begin{align*}
    &t_{\text{max}} : = \mathop{\arg\max}_{t} |g_i(t)|,  \quad I_{\text{half}}=\frac{1}{2}g_i(t_{\text{max}}),\\&\text{FWHM}_i : = \sup\{t \mid g_i(t) \ge I_\text{{half}}\} - \inf\{t \mid g_i(t) \ge I_{\text{half}}\},\\
    &v_{\text{min}}=\frac{2.355^2}{2\cdot \max\limits_{i} \text{FWHM}_i}, \ v_{\text{max}}=\frac{2.355^2}{2\cdot \min\limits_{i} \text{FWHM}_i},
\end{align*}
}
$v^{(j)}_\alpha\sim U(0.5 v_{\text{min}}, 2 v_{\text{max}})$. $c^{(j)}_\alpha$ are selected based on the peak locations of $S_{M}^{(0)}$,  as illustrated in Figure \ref{fig:3D_detect_two_sources}. $M_0=3600$, $M_{\alpha,\text{ReLU}}=M_{\alpha,\text{exp}}=900$, $\tilde{v}^{(j)}_\alpha\sim U(0.185,0.887)$, $\epsilon_{L}= \epsilon_r=5\%$.  The initial mesh $N_{x_1}=N_{x_2}=N_{x_3}=4$ with $n_{x_1}=n_{x_2}=n_{x_3}=3$.

See Figure \ref{fig:3D_detect_two_sources} for the Cell division and detection details. And Figure \ref{fig:3D_S_num0_2D} shows the reconstruction using MA-RFM.  {Moreover, the corresponding stagewise comparison and clusterwise detected parameters are reported in Tables \ref{tab:Comparison of IA-RFM and MA-RFM} and \ref{tab:extracted_parameters}, respectively.}
\end{example}

\begin{example}[\emph{A doubly-connected source}]\label{ex:Three-Dimensional Donut}
Consider a 3-D ``donut'' segment source, which can be parameterized as follows:
\begin{align}
\label{para_Doughnut}
\left \{
\begin{aligned}
&x_1(u,v,w)=(R+wr\cos v)\cos u, &0\le u\le2 \pi,  \\
&x_2(u,v,w)=(R+wr\cos v)\sin u, &0\le v \le 2\pi,\\
&x_3(u,v,w)=w r\sin v, &0\le w\le 1.
\end{aligned}
\right.
\end{align} The source function is defined as 1 inside the donut and 0 outside with $r=0.15$ and $R=0.25$. 

\begin{figure}[H]
    \centering
    \begin{subfigure}[b]{0.27\textwidth}
        \includegraphics[width=\textwidth]{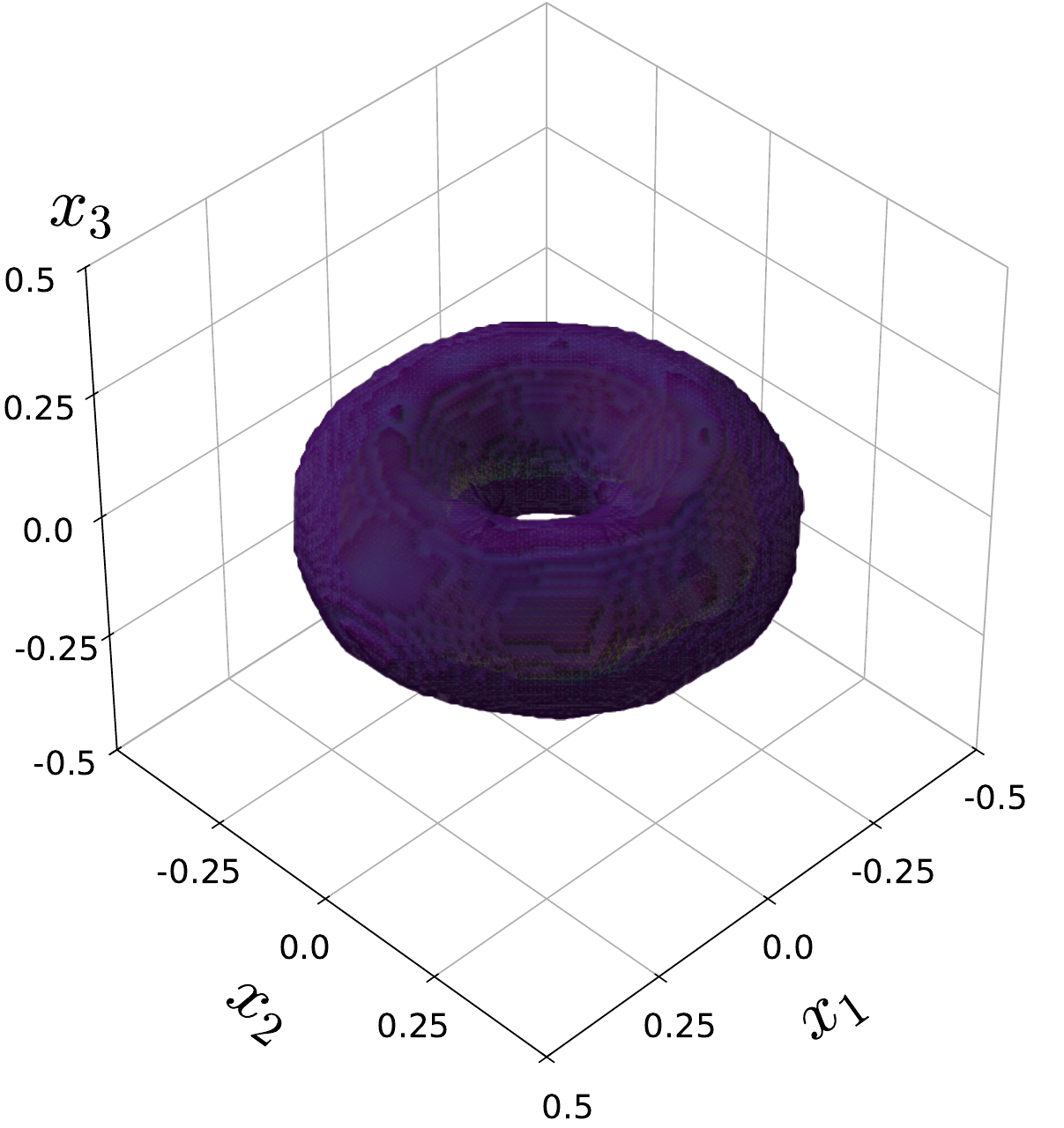}
        \caption{ }
        \label{Snum0_tily}
    \end{subfigure}
    \hspace{5em}
    % \begin{subfigure}[b]{0.29\textwidth}
    %     \includegraphics[width=\textwidth]{ex4.9_plot_3d_top.pdf}
    %     \caption{ }
    %     \label{Snum0_top}
    % \end{subfigure}
    % \vspace{2ex}
    % \begin{subfigure}[b]{0.29\textwidth}
    %     \includegraphics[width=\textwidth]{ex4.9_plot_3d_straight.pdf}
    %     \caption{ }
    %     \label{Snum0_straight}
    % \end{subfigure}
    \begin{subfigure}[b]{0.27\textwidth}
        \includegraphics[width=\textwidth]{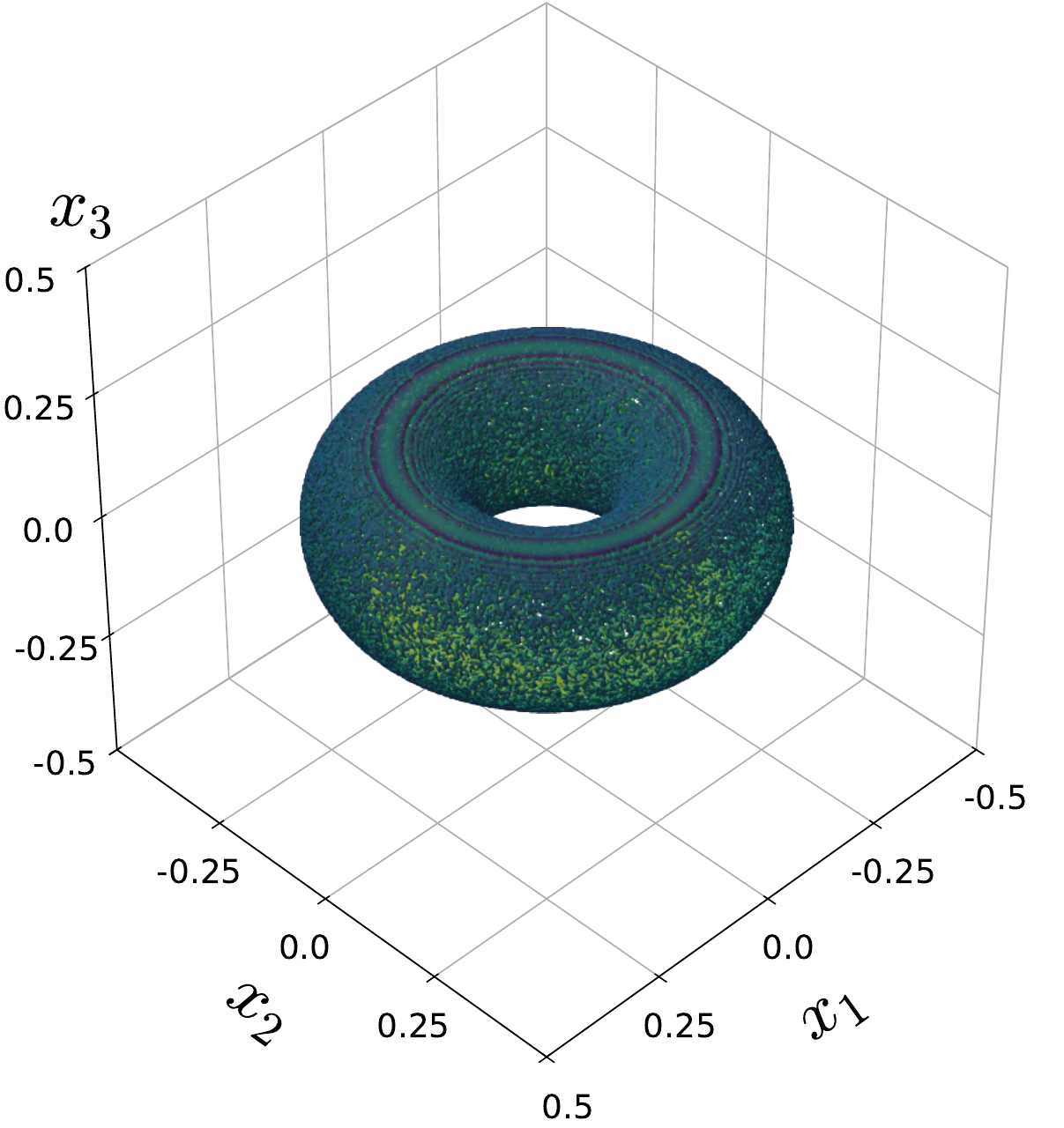}
        \caption{ }
        \label{Snum1_tily}
    \end{subfigure} \\[10pt]
    % \vspace{2ex}
    % \begin{subfigure}[b]{0.29\textwidth}
    %     \includegraphics[width=\textwidth]{ex4.9_plot_3d_top1.pdf}
    %     \caption{ }
    %     \label{Snum1_top}
    % \end{subfigure}
    % \vspace{2ex}
    % \begin{subfigure}[b]{0.29\textwidth}
    %     \includegraphics[width=\textwidth]{ex4.9_plot_3d_straight1.pdf}
    %     \caption{}
    %     \label{Snum1_straight}
    % \end{subfigure}
        \begin{subfigure}[b]{0.32\textwidth}
        \includegraphics[width=\textwidth]{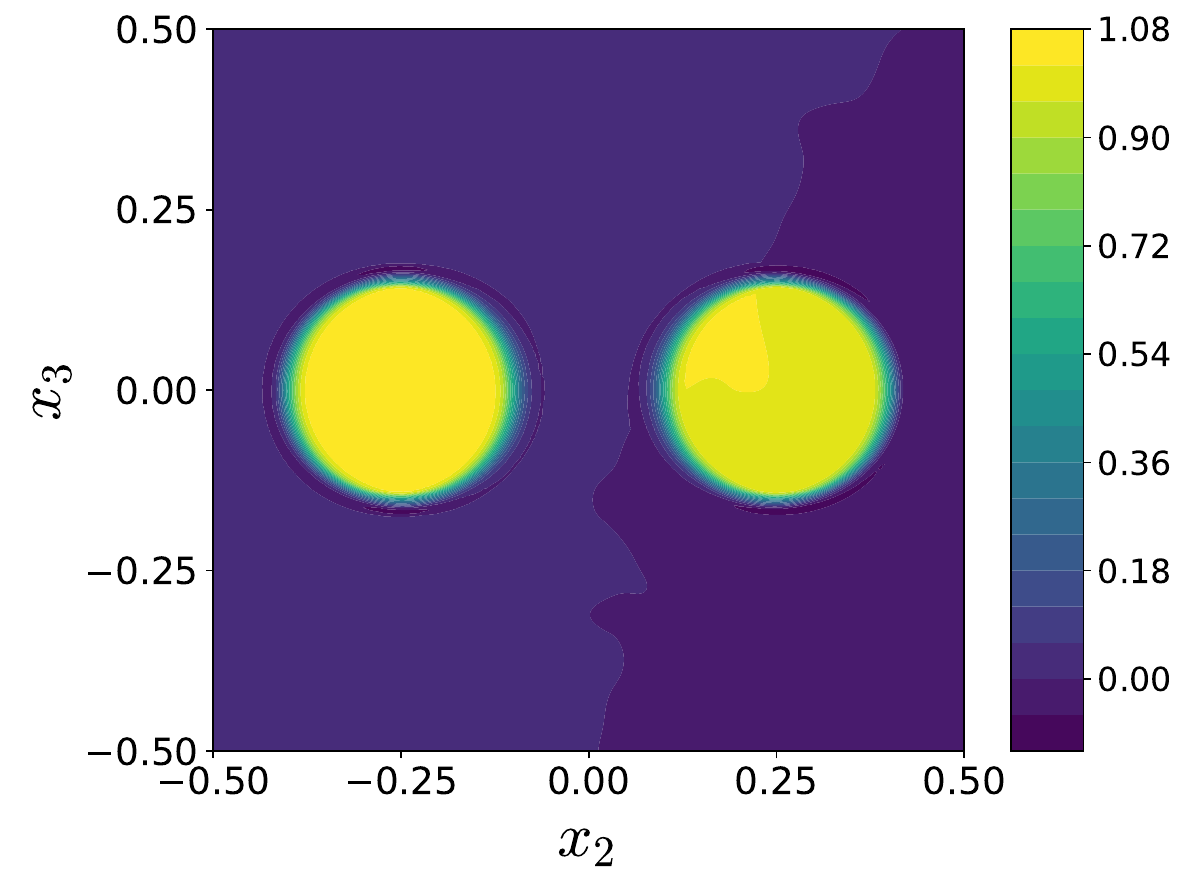}
        \caption{$S_M(0,x_2,x_3)$}
    \end{subfigure}
    \hfill
    \begin{subfigure}[b]{0.32\textwidth}
        \includegraphics[width=\textwidth]{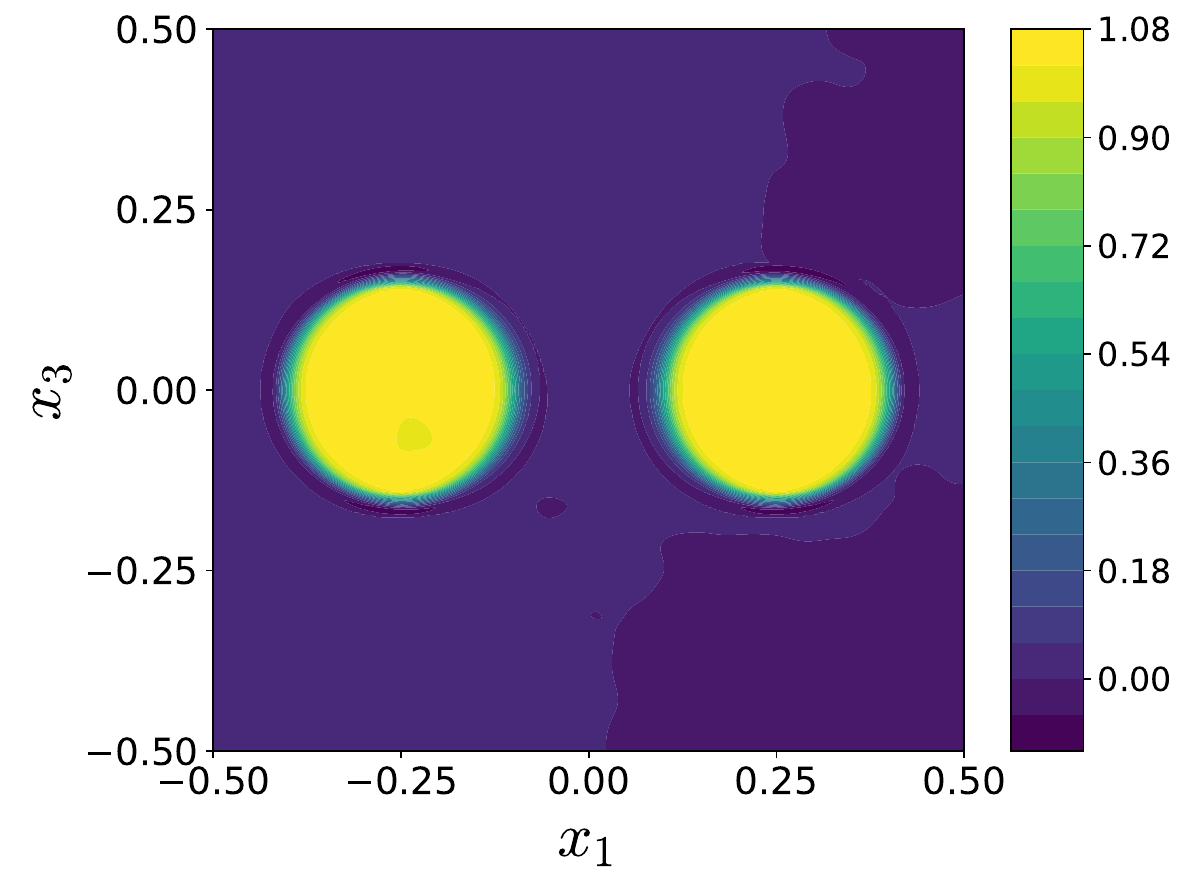}
        \caption{$S_M(x_1,0,x_3)$}
    \end{subfigure}
    \hfill
    \begin{subfigure}[b]{0.32\textwidth}
    \includegraphics[width=\textwidth]{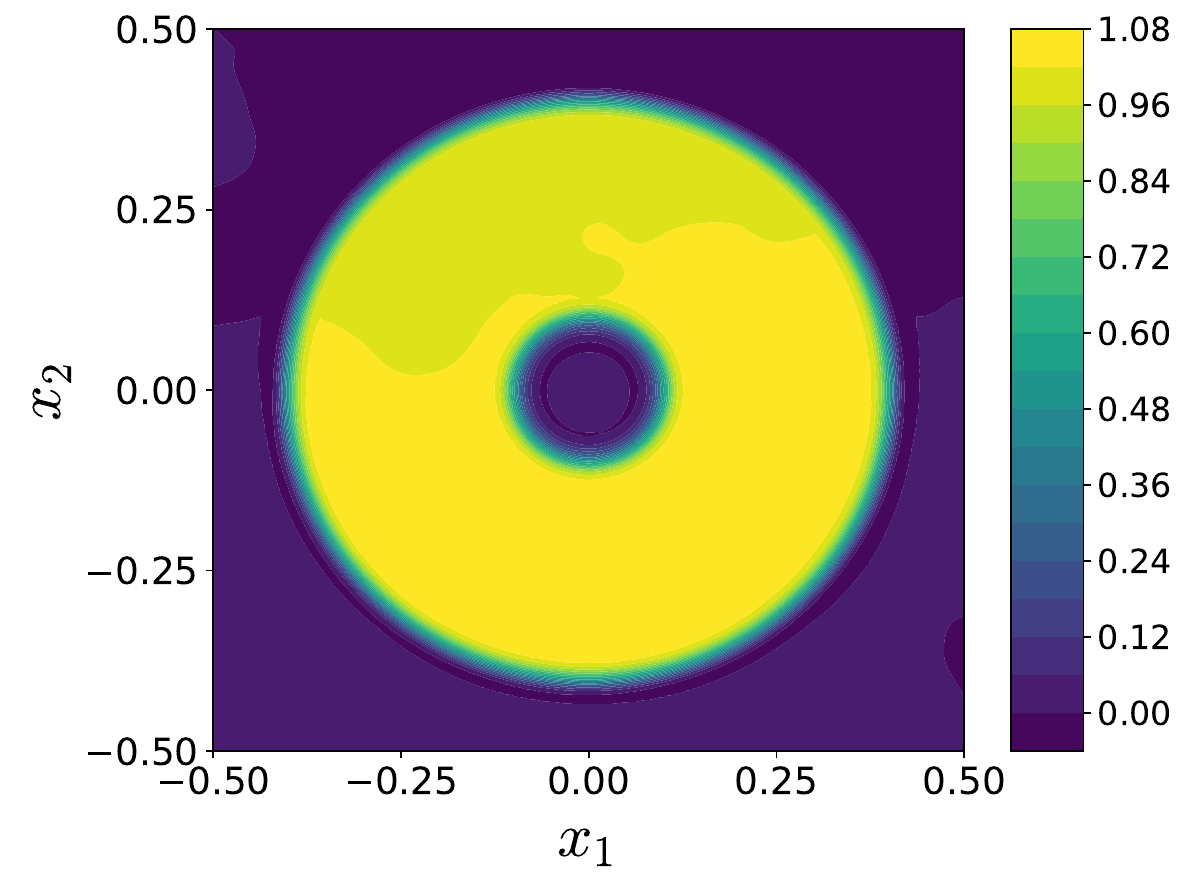}
        \caption{$S_M(x_1,x_2,0)$}
    \end{subfigure}
        \caption{Example \ref{ex:Three-Dimensional Donut}: Three-dimensional donut: (a) shows the $\mathcal{Q}=\mathcal{Q}_{\text{grad}}$ with $t_{\text{grad}}=1/3$. %from different perspectives.   
        %The symbol "$\times$" represents the center  {$\boldsymbol{c}$ =  [-6.73e-5, -2.68e-3, 1.53e-3]$^\top$}, and  {$R_{1}=0.2535,\ r_1=0.1489$} detected by Algorithm \ref{alg: shape_detection}. 
        (b) shows the reconstruction using MA-RFM, %from different perspectives, 
        and (c),(d),(e) show the 2-D cross section results with $M_0=3600$, $M_1=1600$, $\delta=5\%$ and $\lambda_{\text{reg}}^2$=1e-2, yielding $E_{l^2}(S)=15.41\%$.}
    \label{fig:3D_donut}
\end{figure}

\noindent \emph{Experimental setup:}\ $V_0=[-0.5, 0.5]^3$,  \, $\Omega=[-0.75, 0.75]^3$. $k_{\text{min}}$=1,  $k_{\text{max}}=81$,  $N_s$=10,  $\delta=5\%$. Dirichlet data only.

\noindent \emph{Hyperparameter settings:}\  The activation function for IA-RFM is Tanh, $R_m=20$. The basis function used in MA-RFM is sigmoid  and   the detected shape type is torus  {detected by Algorithm~\ref{alg: shape_detection}}. 
    \begin{align*}
    &d^{(j)}_1(\boldsymbol{x};\boldsymbol{\theta}_1^{(j)})=-\psi^{(j)}_1(\boldsymbol{x};\boldsymbol{\theta}_1^{(j)})=(r_1^{(j)})^2-\left(\left(\sqrt{(x_1-c_1)^2+(x_2-c_2)^2}-R^{(j)}_1\right)^2+(x_3-c_3)^2\right),\\
        &\phi_1^{(j)}
(\boldsymbol{x};K_1^{(j)},\boldsymbol{\theta}_1^{(j)})=\text{sigmoid}\left(K_1^{(j)}\cdot d^{(j)}_1(\boldsymbol{x};\boldsymbol{\theta}_1^{(j)})\right),\  j=1, \cdots, M_1.
    \end{align*}
     $M_0$=3600, $M_1=1600$.  {Figures \ref{Snum0_tily} 
     illustrates $\mathcal{Q}=\mathcal{Q}_{\text{grad}}\cap \mathcal{Q}_{\text{abs}}$ with $t_{\text{grad}}=t_{\text{abs}}=1/3$.}
 {$\epsilon_c=\epsilon_r=\epsilon_R=5\%, \boldsymbol{c}_1^{(j)} \sim  \prod_{l=1}^3 U\left( c_{1,l}-\epsilon_c |c_{1,l}|, c_{1,l}+\epsilon_c |c_{1,l}|\right).$ $r_1^{(j)}\sim U(r_1-\epsilon_r|r_1|,r_1+\epsilon_r|r_1|)$, $R_1^{(j)}\sim U(R_1-\epsilon_R|R_1|,R_1+\epsilon_R|R_1|)$,  $K_1^{(j)} \sim U(2000,50000)$. The initial mesh $N_{x_1}=N_{x_2}=N_{x_3}=4$ with $n_{x_1}=n_{x_2}=n_{x_3}=3$}.\\
 \end{example}
   {Figure \ref{fig:3D_donut} shows the details of two stages. See Tables \ref{tab:Comparison of IA-RFM and MA-RFM} and \ref{tab:extracted_parameters} for the corresponding stagewise comparison and detected basis information.}\\
\begin{example}[\emph{Limited aperture}]\label{ex:Limited aperture}In practical applications,  only partial data may be available. To simulate this condition,  we directly test the IA-RFM in a scenario where only partial measurement data is acquired. We consider the problem of reconstructing a source function consisting of four Gaussian peaks: 
\begin{align*}S(x_{1}, x_{2})&=\mathrm{e}^{-300\left(\left(x_{1}-0.15\right)^{2}+\left(x_{2}-0.15\right)^{2}\right)}+\mathrm{e}^{-300\left(\left(x_{1}+0.15\right)^{2}+\left(x_{2}-0.15\right)^{2}\right)}\\&+\mathrm{e}^{-300\left(\left(x_{1}+0.15\right)^{2}+\left(x_{2}+0.15\right)^{2}\right)}+\mathrm{e}^{-300\left(\left(x_{1}-0.15\right)^{2}+\left(x_{2}+0.15\right)^{2}\right)}.\end{align*}

\noindent\emph{Experimental setup:}\ $V_0=[-0.3, 0.3]\times[-0.3, 0.3]$,  \,  $\Omega=B_{0.55
    }(0, 0)$,  $\Gamma=\partial \Omega$,  $k_{\text{min}}$=1,  $k_{\text{max}}=101$, $\delta =1\%$.  The number of measurement points on each quarter-arc of the boundary is $N_s=25$. Dirichlet data only. \\
\noindent\emph{Hyperparameter settings:}\ The activation function for IA-RFM is $\sin$,  $R_m=20$,  $M_0=3200$.
The initial mesh $N_{x_1}=N_{x_2}=4$ with $n_{x_1}=n_{x_2}=3$.

\begin{figure}[H]
    \centering
    \begin{subfigure}[b]{0.32\textwidth}
        \includegraphics[width=\textwidth]{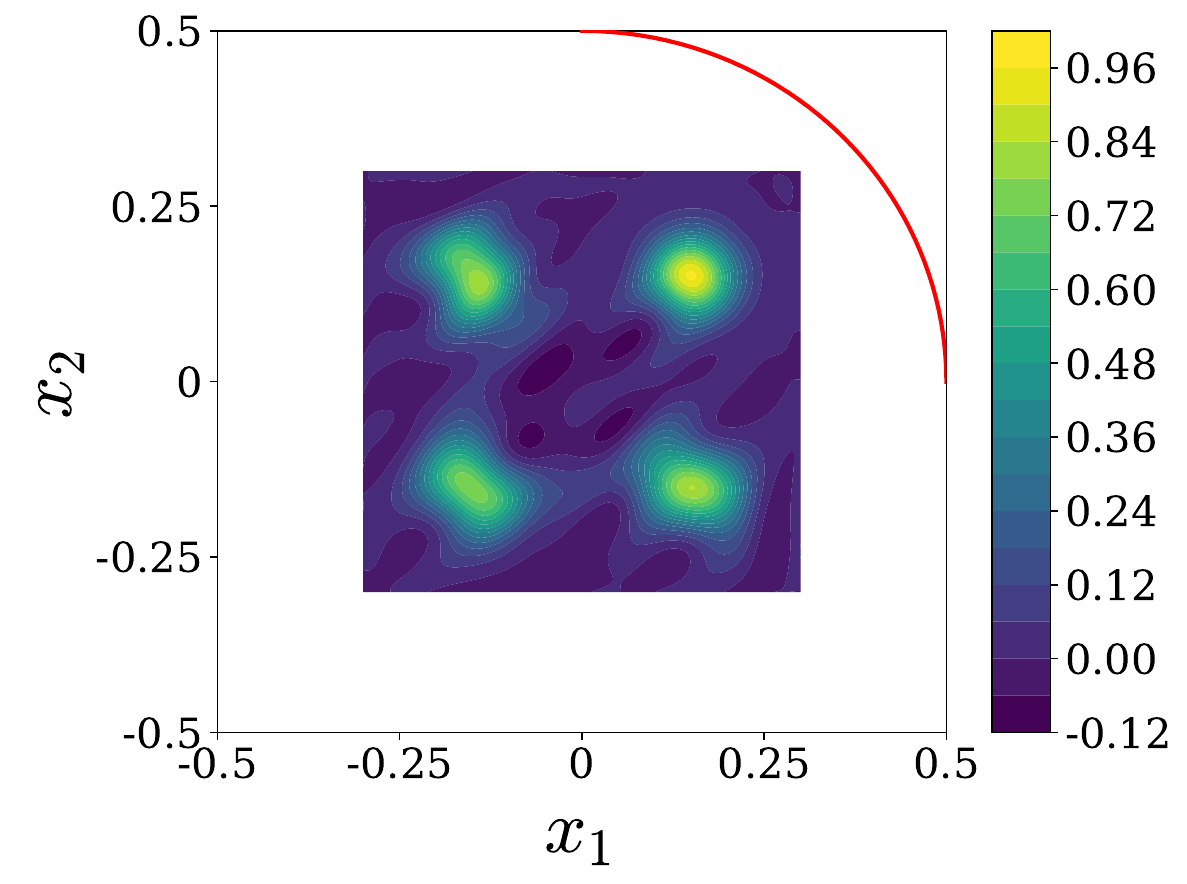}
        \caption{ }
    \end{subfigure}
    \begin{subfigure}[b]{0.32\textwidth}
        \includegraphics[width=\textwidth]{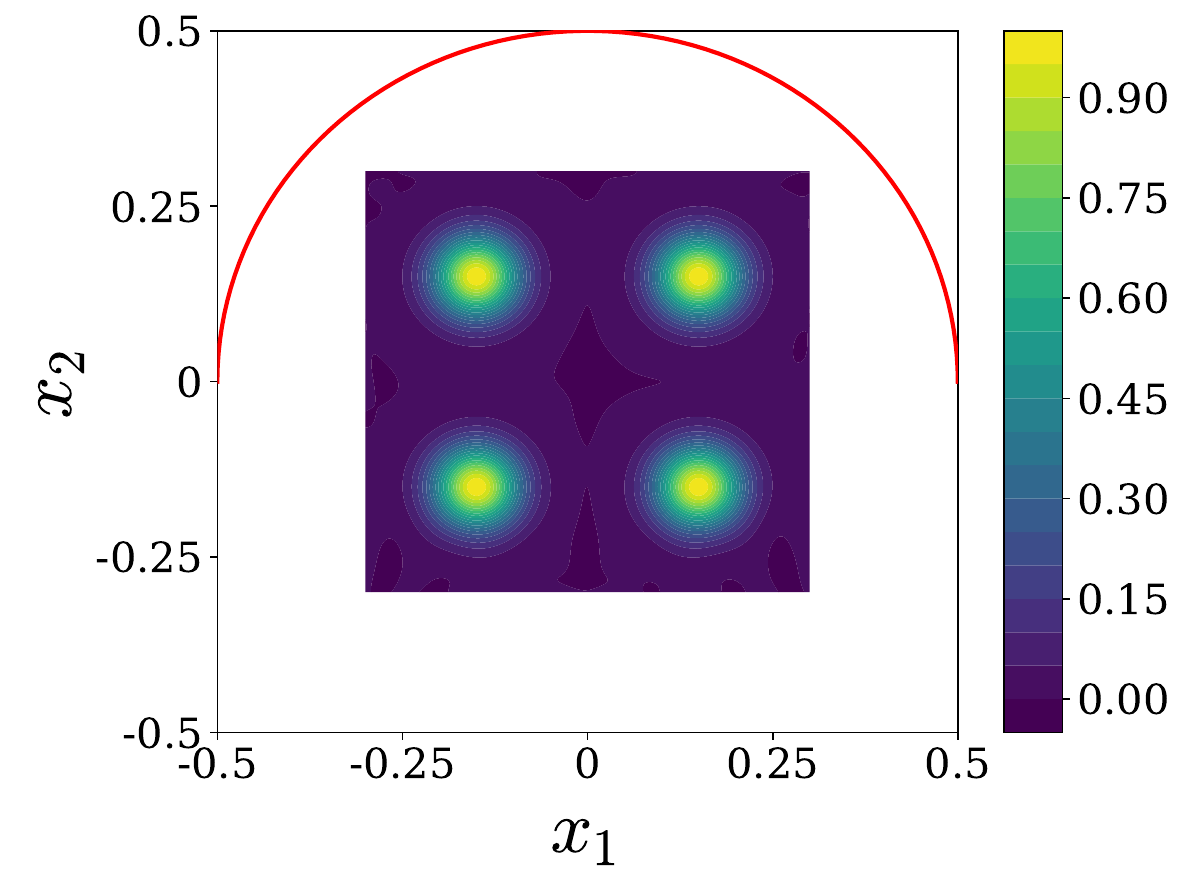}
        \caption{ }
    \end{subfigure}

    \begin{subfigure}[b]{0.32\textwidth}
        \includegraphics[width=\textwidth]{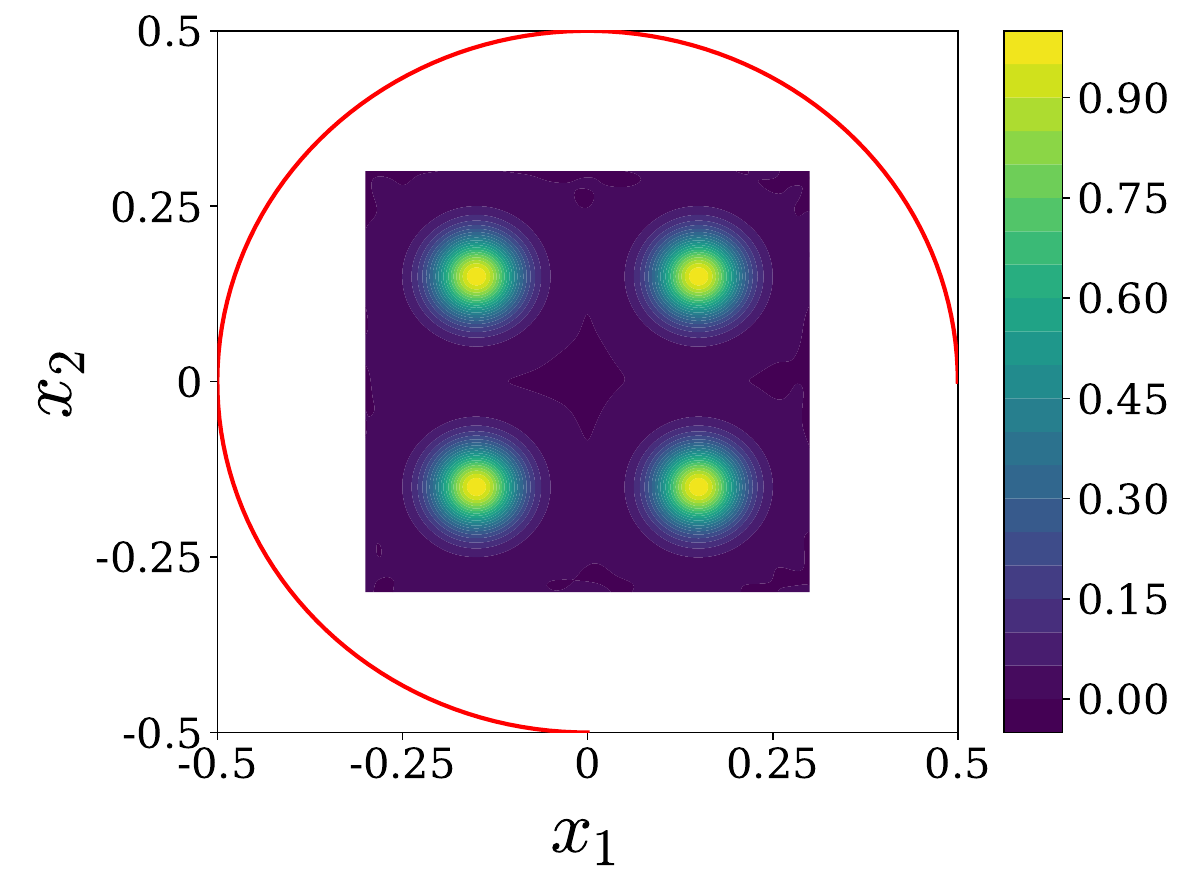}
        \caption{ }
    \end{subfigure}
    \begin{subfigure}[b]{0.32\textwidth}
        \includegraphics[width=\textwidth]{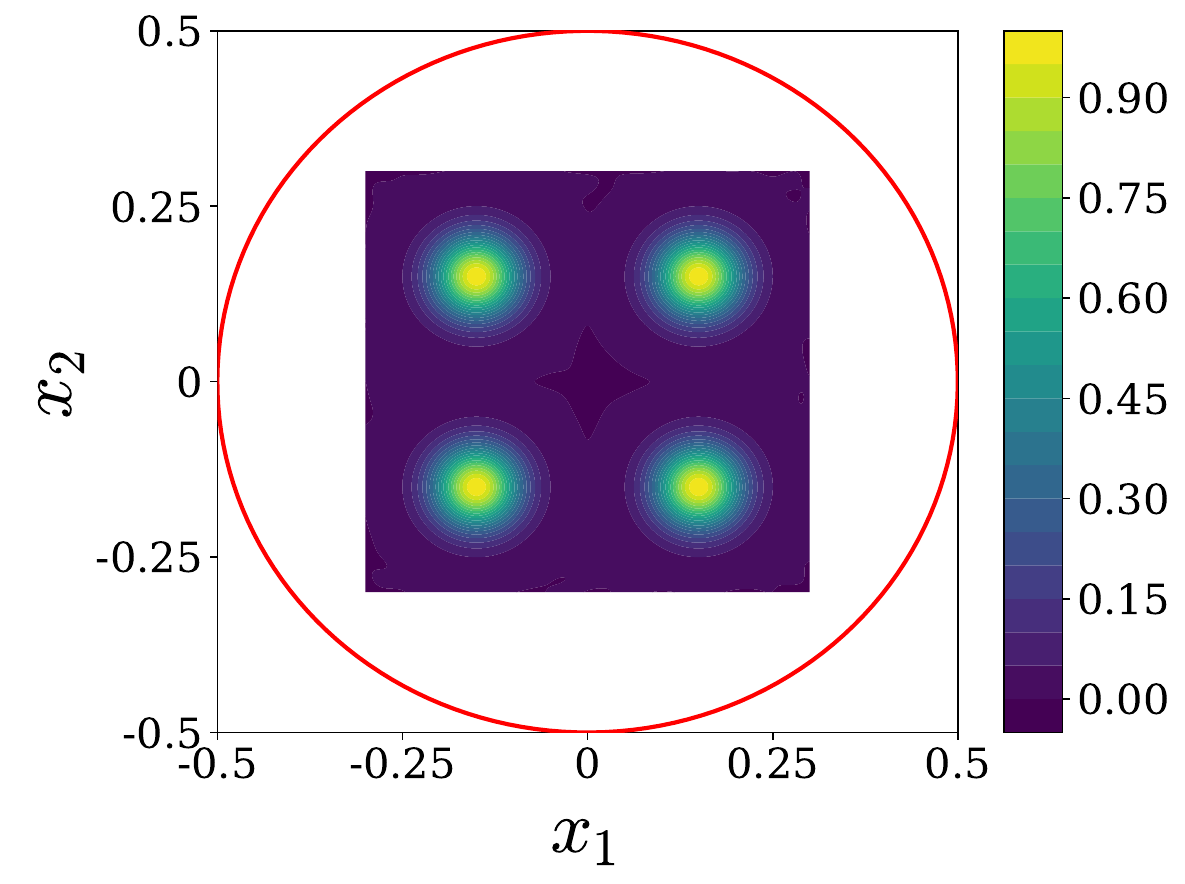}
        \caption{ }
    \end{subfigure}
    \caption{Example \ref{ex:Limited aperture}: Limited aperture: effect of the measurement aperture $\theta_{\text{max}}$ on the source reconstruction. (a) $\theta_{\text{max}}=\pi/2$,  (b) $\theta_{\text{max}}=\pi$,  (c) $\theta_{\text{max}}=3\pi/4$,  and (d) $\theta_{\text{max}}=2\pi$ with $\delta=1\%$. The measurement locations are indicated by the red outer arc.}
    \label{fig:limited_aperture}
\end{figure}
Table \ref{tab:limited aperture} and Figure \ref{fig:limited_aperture} show the $E_{l^2}(S)$ for different apertures with noise level $\delta=1\%$. It can be observed that when $\theta_{\text{max}} = \pi$,  the error is already very small. For $\theta_{\text{max}} = \pi/2$,  we can still roughly determine the locations of the sources. This result indicates that the proposed algorithm exhibits good performance and effectiveness under limited aperture conditions.
\begin{table}[H]
    \centering
    \caption{$E_{l^2}(S)$ with aperture $[0,  \theta_{\text{max}}]$  and $\delta=1\%$,  $M=3200$ using IA-RFM for limited aperture.}
    \begin{tabular}{ccccc}
        \toprule
        $\theta_{\text{max}}$ &2$\pi$& $3\pi/2$ & $\pi$ & $\pi/2$\\
        \midrule
        $n_{\text{integral}}$ &1872&1872&1899&2142\\
        $E_{l^2}(S)$&0.30\% & 0.33\% & 0.66\% & 24.47\% \\
        \bottomrule
    \end{tabular}
    \label{tab:limited aperture}
\end{table}

\end{example}
 
\section{Conclusion}
In conclusion, we have proposed a novel, efficient framework that successfully tackles the long-standing challenge of complex source geometries in the Helmholtz inverse source problem. The core of our method is a synergistic combination of spectral methods and neural networks. Specifically, we employ an integral equation formulation to satisfy the radiation condition as a hard constraint. Concurrently, we propose the MA-RFM, which uses the posterior information to design targeted basis functions of neural networks, enabling precise capture of local singular features of the source. Extensive numerical experiments conducted on a series of sources with complex situations validate the superior performance of our framework.  {The results show that the proposed method not only achieves high accuracy in all test cases but also reduces the number of integration points by two to three orders of magnitude compared to traditional uniform grid methods, significantly alleviating the matrix assembly burden while maintaining the same level of precision.} Furthermore, we believe that the underlying concept of morphology-based adaptive basis function enhancement holds significant promise for broader applications in other complex scientific and engineering computing problems.

\section*{Acknowledgments}
The work is partially supported by the NSFC Major Research Plan (Interpretable and General-purpose Next-generation Artificial Intelligence) No. 92370205, NSFC grants 12494543, 12425113, 
the Strategic Priority Research Program of the Chinese Academy of Sciences (grant no. XDA0480504), 
and the Key Laboratory of the Ministry of Education for Mathematical Foundations and Applications of Digital Technology, University of Science and Technology of China.
\section*{Data availability statement}
   {The code and data that support the findings of this study are available at:
  \url{https://github.com/Lydiaa66/MA-RFM}.}
\appendix
\section*{Appendix}
\addcontentsline{toc}{section}{Appendix}
\section{Proof of Theorem~\ref{the:Uniqueness of multi-frequency data}}

\begin{proof}
Assume $S = S_1 - S_2$ and $u = u_1 - u_2$. The source of $u$ is $S$,  and its support is contained in $\tau_1 \cup \tau_2$. In situation (a),  (b),  (c),   $u$ is a radiating solution to the homogeneous Helmholtz equation in the exterior domain $\mathbb{R}^d \backslash \overline{\Omega}$,  which implies $u(k_j,  \boldsymbol{x}) = 0$ for all $\boldsymbol{x} \in \mathbb{R}^d \backslash \overline{\Omega}$ from Theorem 9.10 in \cite{mclean2000elliptic}. And for (d),  we can deduce $u(k_j,  \boldsymbol{x}) = 0$ from Holmgren's uniqueness theorem \cite{hedenmalm2015Holmgren}. All of the situations imply that 
\begin{align}
\label{all_eng}
u(k_j, \boldsymbol{x})=0 \quad and \quad \partial_{\nu}u(k_j, \boldsymbol{x})=0,  \quad \forall \boldsymbol{x} \in \Gamma.
\end{align}
Now,  let $w_{k_j}$ be any solution to the homogeneous Helmholtz equation $\Delta w_{k_j} + k_j^2 w_{k_j} = 0$ in $\Omega$. Applying Green's formula, 
\begin{align}
\int_{\Omega} S(\boldsymbol{x}) w_{k_j}(\boldsymbol{x}) \mathrm{d}\boldsymbol{x} &= \int_{\Omega} (\Delta u + k_j^2 u) w_{k_j}(\boldsymbol{x}) \mathrm{d}\boldsymbol{x} \\
&= \int_{\Omega} u(\Delta w_{k_j} + k_j^2 w_{k_j}) \mathrm{d}\boldsymbol{x} + \int_{\Gamma} (w_{k_j} \partial_{\nu}u - u \partial_{\nu}w_{k_j}) \mathrm{d}\boldsymbol{x}.
\end{align}
Since $w_{k_j}$ is a solution to the homogeneous equation,  the first volume integral on the right-hand side is zero. Furthermore,  due to the boundary conditions in (\ref{all_eng}),  the boundary integral also vanishes. Therefore, 
$\int_{\Omega} S(\boldsymbol{x}) w_{k_j}(\boldsymbol{x}) \mathrm{d}\boldsymbol{x} = 0.$
Choose $w_{k_j}(\boldsymbol{x})$ to be a plane wave. For any $\boldsymbol{\xi} \in \mathbb{R}^d$ with magnitude $|\boldsymbol{\xi}| = k_j$,  $w_{k_j}(\boldsymbol{x}) = e^{-i\boldsymbol{\xi} \cdot \boldsymbol{x}}$ is a solution to $\Delta w_{k_j} + k_j^2 w_{k_j} = 0$. Substituting this into the integral,  we obtain the Fourier transform of $S$: 
\begin{align}
\hat{S}(\boldsymbol{\xi}) = \int_\Omega S(\boldsymbol{x})e^{-i\boldsymbol{\xi} \cdot \boldsymbol{x}}\mathrm{d}\boldsymbol{x} = 0,  \quad \text{for all } \boldsymbol{\xi} \in \mathbb{R}^d \text{ with } |\boldsymbol{\xi}|=k_j.
\end{align}
Since $\Omega$ is bounded,  the source $S$ has compact support. By the Paley-Wiener theorem,  its Fourier transform $\hat{S}(\boldsymbol{\xi})$ can be uniquely extended to an entire function on $\mathbb{C}^d$.
Consider the restriction of $\hat{S}(\boldsymbol{\xi})$ to a line, the function $f(z) = \hat{S}(z,  0,  \dots,  0)$ is an entire function of one complex variable. It is zero at the points $z = k_j$,  which have an accumulation point. From unique continuation for analytic functions,  $\hat{S}(\boldsymbol{\xi})$ is zero on the whole complex domain $\mathbb{C}^d$. Since the Fourier transform is identically zero and the transform is invertible,  the function $S(\boldsymbol{x})$ must be identically zero. Therefore,  $S_1(\boldsymbol{x}) = S_2(\boldsymbol{x})$.
\end{proof}
\section{Proof of Theorem \ref{Existence and Uniqueness}}
\begin{proof}
(a). Rewrite $\mathcal{L}_{\text{reg}}$:
\begin{align*}\mathcal{L}_{\text{reg}}(\boldsymbol{s}^{\delta})&=(\Psi_\mathrm{M}\boldsymbol{s}^{\delta}-U^\delta)^\top(\Psi_\mathrm{M}\boldsymbol{s}^{\delta}-U^\delta)+\lambda_{\text{reg}}^2 (\boldsymbol{s}^{\delta})^\top \boldsymbol{s}^{\delta}\\&=(\boldsymbol{s}^{\delta})^\top\Psi_\mathrm{M}^\top\Psi_\mathrm{M}\boldsymbol{s}^{\delta}-(\boldsymbol{s}^{\delta})^\top\Psi_\mathrm{M}^\top U^\delta-(U^\delta)^\top\Psi_\mathrm{M}\boldsymbol{s}^{\delta}+(U^\delta)^\top U^\delta+\lambda_{\text{reg}}^2(\boldsymbol{s}^{\delta})^\top \\&=(\boldsymbol{s}^{\delta})^\top(\Psi_\mathrm{M}^\top\Psi_\mathrm{M}+\lambda_{\text{reg}}^2I)\boldsymbol{s}^{\delta}-2(U^\delta)^\top\Psi_\mathrm{M}s+\|U^\delta\|_2^2.\end{align*}
Compute the Hessian matrix:
$$
H(\mathcal{L}_{\text{reg}})=2(\Psi_\mathrm{M}^\top\Psi_\mathrm{M}+\lambda_{\text{reg}}^2I) \succ 0.
$$
(b). Applying the triangle inequality,  
    \begin{align}
        \label{eq:err_decomp}
        \|\boldsymbol{s}^{\delta}-\boldsymbol{s}^{*}\|_2&=\|(\Psi_\text{M}^\top\Psi_\text{M}+\lambda_{\text{reg}}^2 I)^{-1}\Psi_\text{M}^\top U^{\delta}-\boldsymbol{s}^{*}\|_2 \notag\\
        &=\|(\Psi_\text{M}^\top\Psi_\text{M}+\lambda_{\text{reg}}^2 I)^{-1}[\Psi_\text{M}^{T}U^{\delta}-(\Psi_\text{M}^\top\Psi_\text{M}+\lambda_{\text{reg}}^2 I)\boldsymbol{s}^{*}]\|_2 \notag\\
        &=\|(\Psi_\text{M}^\top\Psi_\text{M}+\lambda_{\text{reg}}^2 I)^{-1}\Psi_\text{M}^\top(U^{\delta}-\Psi_\text{M}\boldsymbol{s}^{*})-\lambda_{\text{reg}}^2 (\Psi_\text{M}^\top\Psi_\text{M}+\lambda_{\text{reg}}^2 I)^{-1} \boldsymbol{s}^{*}\|_2 \notag\\
        &\leq \|(\Psi_\text{M}^\top\Psi_\text{M}+\lambda_{\text{reg}}^2 I)^{-1}\Psi_\text{M}^\top\|_2\cdot \|U^{\delta}-\Psi_\text{M}\boldsymbol{s}^{*}\|_2+\|\lambda_{\text{reg}}^2 (\Psi_\text{M}^\top\Psi_\text{M}+\lambda_{\text{reg}}^2 I)^{-1} \boldsymbol{s}^{*}\|_2.
    \end{align}
    Let $\Psi_\text{M}=U\Sigma V^\top$ be the singular value deposition of $\Psi_\text{M}$,  where $\sigma_i$ are the singular values of $\Psi_\text{M}$. For the first term,  we bound the operator norm: 
\begin{align*}
\|(\Psi_\text{M}^\top\Psi_\text{M}+\lambda_{\text{reg}}^2 I)^{-1}\Psi_\text{M}^\top\|_2& =\|(V\Sigma^\top\Sigma V^\top+ \lambda_{\text{reg}}^2 I)^{-1} V\Sigma^\top U^\top\|_2 \\
&=|V(\Sigma^\top\Sigma + \lambda_{\text{reg}}^2 I)^{-1}\Sigma^\top U^\top\|_2  \\
&= \max_i \frac{\sigma_i}{\sigma_i^2 + \lambda_{\text{reg}}^2} \le \frac{1}{2\lambda_{\text{reg}}}.
\end{align*}
The residual term $\|U^{\delta} - \Psi_\text{M} \boldsymbol{s}^{*}\|_2$ is bounded by the sum of data noise and model inconsistency: 
$$\|U^{\delta} - \Psi_\text{M} \boldsymbol{s}^{*}\|_2 \le \|U^{\delta} - U_{\text{true}}\|_2 + \|U_{\text{true}} - \Psi_\text{M} \boldsymbol{s}^{*}\|_2 \le \delta_{\text{all}} + \eta_\text{M}.$$
Thus,  the first term in \eqref{eq:err_decomp} is bounded by $\frac{1}{2\lambda_{\text{reg}}}(\delta_{\text{all}} + \eta_\text{M})$.
\begin{align*}
   \|(\Psi_\text{M}^\top\Psi_\text{M}+\lambda_{\text{reg}}^2 I)^{-1}\Psi_\text{M}^\top\|_2\cdot \|U^{\delta}-\Psi_\text{M}\boldsymbol{s}^{*}\|_2&\leq \frac{1}{2 \lambda_{\text{reg}}} (\delta_{\text{all}} + \eta_\text{M}).
\end{align*}
For the second term,  we use the source condition $\boldsymbol{s}^{*}=\Psi_\text{M}^{T}\Psi_\text{M}w$: 
\begin{align*}
   \|\lambda_{\text{reg}}^2 (\Psi_\text{M}^T\Psi_\text{M}+\lambda_{\text{reg}}^2 I)^{-1} \boldsymbol{s}^{*}\|&=\lambda_{\text{reg}}^2\|(V\Sigma^\top\Sigma V^\top+ \lambda_{\text{reg}}^2 I)^{-1} (V\Sigma^\top \Sigma V^\top)^{\nu}w\|_2\\
   &=\lambda_{\text{reg}}^2\|(V\Sigma^\top\Sigma V^\top+ \lambda_{\text{reg}}^2 I)^{-1} V(\Sigma^\top \Sigma )^{\nu}V^\top w\|_2\\
   &\leq \lambda_{\text{reg}}^2\|V(\Sigma^2+\lambda_{\text{reg}}^2 I)^{-1}\Sigma^{2\nu}V^\top\|_2\cdot\| w\|_2\\
   &\leq \max_i \frac{\lambda_{\text{reg}}^2 \sigma_i^{2\nu}}{\sigma_i^2+\lambda_{\text{reg}}^2}\|w\|_2\le \sup_{\sigma\ge0} \frac{\lambda_{\text{reg}}^2 \sigma^{\nu}}{\sigma+\lambda_{\text{reg}}^2}\|w\|_2.
\end{align*}
 By performing a change of variables with $\sigma = \lambda_{\text{reg}}^2 t$ (for $t \ge 0$),  $\frac{(\lambda_{\text{reg}}^2 t)^{\nu}}{t + 1} = \lambda_{\text{reg}}^{2\nu}  \frac{t^{\nu}}{t+1}$.For smoothness parameter $0 < \nu \le 1$,  $\frac{t^{\nu}}{t+1}$ is bounded for all $t \ge 0$. Therefore,  there exists a constant  $C_{\nu}$ such that
\begin{equation*}
    \sup_{\sigma\ge0} \frac{\lambda_{\text{reg}}^2 \sigma^{\nu}}{\sigma+\lambda_{\text{reg}}^2} = \lambda_{\text{reg}}^{2\nu}  \sup_{t \ge 0} \frac{t^{\nu}}{t+1} \le C_{\nu} \lambda_{\text{reg}}^{2\nu}.
\end{equation*}
Substituting this result back into the overall error estimate,  we obtain: 
\begin{align*}
    \|\boldsymbol{s}^{\delta}-\boldsymbol{s}^{*}\|_2\le \frac{1}{2 \lambda_{\text{reg}}}(\delta_{\text{all}}+\eta_\text{M})+ C_{\nu} \lambda_{\text{reg}}^{2\nu}\|w\|_2.
\end{align*}
The bound is minimized by selecting the regularization parameter $\lambda_{\text{reg}}$ according to the following a priori rule: 
$$\lambda_{\text{reg}}^2=\left(\frac{\delta_{\text{all}}+\eta_\text{M}}{4\nu C_{\nu}\|w\|_2}\right)^{\frac{2}{(2\nu+1)}}.$$
With this choice,  we establish the optimal convergence rate for the Tikhonov-regularized solution: 
$$\|\boldsymbol{s}^{\delta}-\boldsymbol{s}^{*}\|_2\leq \left[ (2\nu+1) \cdot (4\nu)^{-\frac{2\nu}{2\nu+1}} \cdot (C_{\nu}\|w\|_2)^{\frac{1}{2\nu+1}} \right] (\delta_{\text{all}}+\eta_\text{M})^{\frac{2\nu}{2\nu+1}}.$$
\end{proof}

\section{Additional numerical results}
\subsection{Torus residual}\label{app:torus_residual}
  {For a three-dimensional cluster \(C_\alpha=\{\boldsymbol{x}_j\}_{j=1}^{|C_\alpha|}\), we test the three coordinate directions as candidate torus axes. For each \(m\in\{1,2,3\}\), let
  \(\Pi_m:\mathbb{R}^3\to\mathbb{R}^2\) denote the orthogonal projection onto the plane perpendicular to the \(x_m\)-axis. Let \(\boldsymbol{c}_\alpha\in\mathbb{R}^3\) be the detected
  cluster center, and denote its projected center by
  \[
  \boldsymbol{c}_{\alpha,\perp}^{(m)}=\Pi_m(\boldsymbol{c}_\alpha).
  \]
  For each point \(\boldsymbol{x}_j\in C_\alpha\), we define its radial distance to the candidate axis and its axial offset by
  \[
  \rho_{\alpha,j}^{(m)}
  =
  \left\|
  \Pi_m(\boldsymbol{x}_j)-\boldsymbol{c}_{\alpha,\perp}^{(m)}
  \right\|_2,
  \qquad
  z_{\alpha,j}^{(m)}=x_{j,m}-c_{\alpha,m}.
  \]}

   {Next, for the projected point cloud \(\{\Pi_m(\boldsymbol{x}_j)\}_{j=1}^{|C_\alpha|}\), we estimate an annular structure in the projected plane. Whenever the outer and inner projected
  boundaries can be explicitly extracted, we compute their radial distances to the projected center \(\boldsymbol{c}_{\alpha,\perp}^{(m)}\) and define
  \(\rho_{\alpha,\mathrm{out}}^{(m)}\) and \(\rho_{\alpha,\mathrm{in}}^{(m)}\)
  as the corresponding median outer and inner radii, respectively. We then define
  \[
  R_\alpha^{(m)}
  =
  \frac{\rho_{\alpha,\mathrm{out}}^{(m)}+\rho_{\alpha,\mathrm{in}}^{(m)}}{2},
  \qquad
  r_\alpha^{(m)}
  =
  \frac{\rho_{\alpha,\mathrm{out}}^{(m)}-\rho_{\alpha,\mathrm{in}}^{(m)}}{2}.
  \]
  Here \(R_\alpha^{(m)}\) and \(r_\alpha^{(m)}\) are the corresponding major and minor radii of the candidate torus.}

   {To evaluate the quality of this candidate axis, we further compute an annulus-type fitting residual on the projected cross-section, and among the three coordinate directions we retain the
  one with the smallest projected residual. For this selected axis, the torus residual is finally defined in the full three-dimensional geometry by
  \[
  \mathcal{E}_{\mathrm{tori}}^{(\alpha)}(m)
  =
  \frac{1}{|C_\alpha|}
  \sum_{j=1}^{|C_\alpha|}
  \left|
  \frac{\sqrt{\bigl(\rho_{\alpha,j}^{(m)}-R_\alpha^{(m)}\bigr)^2+\bigl(z_{\alpha,j}^{(m)}\bigr)^2}}
  {r_\alpha^{(m)}}-1
  \right|,
  \]
  and we set
  \[
  \mathcal{E}_{\mathrm{tori}}^{(\alpha)}
  =
  \min_{m\in\{1,2,3\}}
  \mathcal{E}_{\mathrm{tori}}^{(\alpha)}(m)
  \]
  as the torus fitting error.}

\subsection{Quantitative comparison (IA-RFM vs. MA-RFM)}
  \begin{table}[H]
      \centering
      \caption{Comparison between the IA-RFM stage ($S^{(0)}$) and the MA-RFM stage ($S_{\text{final}}$).}
      \begin{tabular}{cccccc}
          \toprule
           & $\delta$ & $M_{\text{total}}$ & $E_{l^2}(S^{(0)})$ & $E_{l^2}(S_{\text{final}})$ & Error reduction \\
          \midrule
          Ex 4.3 & 5\% & 3200 & 20.03\% & 14.02\% & 30.00\% \\
          Ex 4.4 & 5\% & 6400 & 10.06\% & 6.04\% & 40.03\% \\
          Ex 4.5 & 5\% & 6400 & 23.04\% & 15.28\% & 33.69\% \\
          Ex 4.6 & 5\% & 4800 & 19.74\% & 13.94\% & 29.41\% \\
          Ex 4.7 & 5\% & 7200 & 25.05\% & 17.84\% & 28.79\% \\
          Ex 4.8 & 5\% & 7200 & 11.87\% & 1.99\% & 83.24\% \\
          Ex 4.9 & 5\% & 5200 & 36.44\% & 15.41\% & 57.72\% \\
          \bottomrule
      \end{tabular}
      \label{tab:Comparison of IA-RFM and MA-RFM}
  \end{table}

\subsection{Geometric parameters and basis type}
\begin{table}[H]
        \centering
        \caption{Extracted geometric parameters and statistical indicators with $\delta=5\%$ by
  Algorithm \ref{alg: shape_detection}.}
        \label{tab:extracted_parameters}
        \resizebox{\textwidth}{!}{%
        \begin{tabular}{cccccccccc}
            \toprule
            {\emph{Ex}} & Cluster $k$ & $\boldsymbol{c}_k$ & $\boldsymbol{L}_k$ &
            $\mathcal{E}_{\text{rect}}^{(k)}$ & $\mathcal{E}_{\text{ellip}}^{(k)}$ &
            $\mathcal{E}_{\text{torus}}^{(k)}$ & $\mathrm{CV}_k$ & $\mathcal{T}_k$ & $\sigma_k$ \\
            \midrule

            Ex 4.3 & 1 & (5.00e-01, 5.00e-01) & (2.03e-01, 2.03e-01) & 0.0906 &
            \textbf{0.0121} & - & 0.165 & \textbf{Ellipsoid} & Sigmoid \\
            \midrule

            \multirow{2}{*}{Ex 4.4}
              & 1 & (-6.04e-02, -2.86e-04) & (6.40e-02, 6.60e-02) & 0.0808 & \textbf{0.0345}
              & - & 0.710 & \textbf{Ellipsoid} & Exp \\
              & 2 & (8.04e-02, 3.07e-04) & (6.40e-02, 6.60e-02) & 0.0809 & \textbf{0.0349}
              & - & 0.710 & \textbf{Ellipsoid} & Exp \\
            \midrule

            \multirow{2}{*}{Ex 4.5}
              & 1 & (3.90e-01, 4.98e-01) & (1.08e-01, 2.05e-01) & \textbf{0.0231} & 0.1266
              & - & 0.252 & \textbf{Rectangle} & Sigmoid \\
              & 2 & (7.11e-01, 5.00e-01) & (2.02e-01, 2.02e-01) & 0.0882 & \textbf{0.0169}
              & - & 0.151 & \textbf{Ellipsoid} & Sigmoid \\
            \midrule

            Ex 4.6 & 1 & (6.01e-01, 2.51e-01) & - & 0.0698 & 0.0613 & - & 0.419 & \textbf{general}
  & Sigmoid \\
            \midrule

            \multirow{4}{*}{Ex 4.7}
              & 1 & (2.12e-01, 7.13e-01) & - & 0.1113 & 0.0675 & - & 0.190 & \textbf{general} &
  Sigmoid \\
              & 2 & (4.20e-01, 4.49e-01) & - & 0.1938 & 0.1211 & - & 0.183 & \textbf{general} &
  Sigmoid \\
              & 3 & (7.49e-01, 3.59e-01) & - & 0.1823 & 0.1102 & - & 0.253 & \textbf{general} &
  Sigmoid \\
              & 4 & (7.96e-01, 7.89e-01) & - & 0.2065 & 0.2152 & - & 0.193 & \textbf{general} &
  Sigmoid \\
            \midrule

            \multirow{2}{*}{Ex 4.8}
              & 1 & (3.00e-01, 5.01e-01, 2.99e-01) & (1.55e-01, 1.50e-01, 1.60e-01) &
              0.1575 & \textbf{0.0373} & 0.0938 & 0.739 & \textbf{Ellipsoid} & Exp \\
              & 2 & (4.99e-01, 4.99e-01, 8.00e-01) & (1.60e-01, 1.60e-01, 1.60e-01) &
              0.1575 & \textbf{0.0373} & 0.0938 & 0.739 & \textbf{Ellipsoid} & Exp \\
            \midrule

            Ex 4.9 & 1 &
            (-6.73e-05, -2.68e-03, 1.53e-03) &
            \begin{tabular}{@{}c@{}}
            $r_{1}$=1.49e-01 \\
            $R_{1}$=2.54e-01
            \end{tabular} &
            0.2498 & 0.1955 & \textbf{0.0332} & 0.252 & \emph{Tori} & Sigmoid \\
            \bottomrule
        \end{tabular}%
        }
  \end{table}
\bibliographystyle{unsrt}
\bibliography{main}
\end{document}